\documentclass[english]{article}
\usepackage[T1]{fontenc}
\usepackage[latin9]{inputenc}
\usepackage[letterpaper]{geometry}
\usepackage{xcolor}
\usepackage{babel}
\usepackage{array}
\usepackage{booktabs}
\usepackage{units}
\usepackage{mathrsfs}
\usepackage{mathtools}
\usepackage{amsmath}
\usepackage{amsthm}
\usepackage{amssymb}
\usepackage{setspace}
\usepackage{wasysym}
\usepackage[authoryear]{natbib}
\PassOptionsToPackage{normalem}{ulem}
\usepackage{ulem}
\usepackage[unicode=true,pdfusetitle,
 bookmarks=true,bookmarksnumbered=false,bookmarksopen=false,
 breaklinks=true,pdfborder={0 0 0},pdfborderstyle={},backref=false,colorlinks=true]
 {hyperref}
\hypersetup{
 linkcolor=blue, citecolor=blue}
\usepackage{breakurl}

\makeatletter

\providecommand{\tabularnewline}{\\}
\providecolor{lyxadded}{rgb}{0,0,1}
\providecolor{lyxdeleted}{rgb}{1,0,0}

\DeclareRobustCommand{\lyxsout}[1]{\ifx\\#1\else\sout{#1}\fi}

\numberwithin{equation}{section}
\theoremstyle{plain}
\newtheorem{assumption}{\protect\assumptionname}
\theoremstyle{plain}
\newtheorem{thm}{\protect\theoremname}
\theoremstyle{plain}
\newtheorem{lem}{\protect\lemmaname}

\usepackage{bbm}

\usepackage{scalerel,stackengine}
\stackMath
\newcommand\reallywidecheck[1]{%
\savestack{\tmpbox}{\stretchto{%
  \scaleto{%
    \scalerel*[\widthof{\ensuremath{#1}}]{\kern-.6pt\bigwedge\kern-.6pt}%
    {\rule[-\textheight/2]{1ex}{\textheight}}
  }{\textheight}%
}{0.5ex}}%
\stackon[1pt]{#1}{\scalebox{-1}{\tmpbox}}%
}
\date{}
\allowdisplaybreaks

\newtheoremstyle{remboldstyle}
  {}{}{}{}{\bfseries}{.}{.5em}{{\thmname{#1 }}{\thmnumber{#2}}{\thmnote{ (#3)}}}
\theoremstyle{remboldstyle}
\newtheorem{rembold}{Remark}[section]

\makeatletter
\renewenvironment{proof}[1][\proofname]{%
  \par\pushQED{\qed}\normalfont%
  \topsep6\p@\@plus6\p@\relax
  \trivlist\item[\hskip\labelsep\bfseries#1\@addpunct{.}]%
  \ignorespaces
}{%
  \popQED\endtrivlist\@endpefalse
}
\makeatother

\newtheoremstyle{ntnboldstyle}
  {}{}{}{}{\bfseries}{.}{.5em}{{\thmname{#1}}{\thmnote{(#2)}}}
\theoremstyle{ntnboldstyle}
\newtheorem{ntnbold}{Notation}

\makeatother

\providecommand{\assumptionname}{Assumption}
\providecommand{\lemmaname}{Lemma}
\providecommand{\theoremname}{Theorem}

\begin{document}

\title{\textbf{Monotonicity-Constrained Nonparametric Estimation and Inference
for First-Price Auctions}\thanks{This version: \today}}

\author{Jun Ma\thanks{School of Economics, Renmin University of China}\and
Vadim Marmer\thanks{Vancouver School of Economics, University of British Columbia}\and
Artyom Shneyerov\thanks{Department of Economics, Concordia University}\and
Pai Xu\thanks{School of Economics and Finance, University of Hong Kong}}
\maketitle
\begin{abstract}
We propose a new nonparametric estimator for first-price auctions
with independent private values that imposes the monotonicity constraint
on the estimated inverse bidding strategy. We show that our estimator
has a smaller asymptotic variance than that of Guerre, Perrigne and
Vuong\textquoteright s (2000) estimator. In addition to establishing
pointwise asymptotic normality of our estimator, we provide a bootstrap-based
approach to constructing uniform confidence bands for the density
function of latent valuations. 
\end{abstract}

\section{Introduction}

Shape restrictions on infinite-dimensional parameters have received
much attention in econometric research. The roles of shape restrictions
in the literature include facilitating identification, providing testable
implications and improving estimation and inference. See \citet{chetverikov2018econometrics}
for a recent review. 

This paper focuses on the first-price sealed-bid auction model with
symmetric bidders, which is the same as that studied in \citet[GPV, hereafter]{vuong2000one}.
GPV's estimation strategy uses a nonparametrically estimated inverse
bidding function to generate pseudo valuations. However, the true
bidding strategy must be strictly increasing and the plug-in nonparametric
estimator of GPV ignores this shape restriction. Imposing such a constraint
at the estimation stage nonparametrically is interesting. For this
purpose, we may use the methods in the monotone nonparametric regression
literature.\footnote{See \citet{henderson2009imposing} for a comprehensive survey of the
literature on monotone nonparametric regression.} Some of these methods can be easily adapted to producing constrained
estimators for the inverse bidding strategy in the auction setting.
E.g., \citet{henderson_2012_JOE} take the constrained re-weighting
approach pioneered by \citet{hall2001nonparametric}, which was originally
used to build a monotone estimator for the conditional expectation
function. \citet{luo2017integrated} impose monotonicity by using
the greatest convex minorant of the integrated quantile function of
values.

In this paper, we pursue a different approach. We investigate the
asymptotic properties of a new monotonicity-constrained estimator
based on the smooth rearrangement approach of \citet{dette2006simple}
and propose a uniform confidence band around this monotonicity-constrained
nonparametric estimator. We show that this rearrangement-based monotonicity-constrained
estimator is asymptotically normal with an asymptotic variance smaller
than the unconstrained estimator as in GPV. Since the asymptotic variance
plays an important role in determining the width of a uniform confidence
band in large samples, the fact that the rearrangement-based estimator
has a smaller asymptotic variance will result in sharper inference.
Our estimator also has substantial computational advantage over \citet{henderson_2012_JOE}.\footnote{See \citet{dette2006comparative} for simulation studies that compare
the rearrangement and reweighting approaches to monotone regression.
\citet{dette2006comparative} notice that the rearrangement approach
has computational advantage since the reweighting approach requires
solving constrained optimization.} 

As a by-product, our method also produces a simple estimator for the
true bidding function. Note that GPV's procedure is based on the inverse-bidding
strategy, which has a simple form. On the other hand, the bidding
function has an integral expression that depends on the unknown distribution
of latent valuations and constructing its direct plug-in type estimator
would be cumbersome. Our simple estimator of the bidding function
can be of interest on its own, as it can be used in practical applications
for computing counterfactual bids. 

We compare the finite sample performances of the confidence bands
based on the rearrangement-based and the unconstrained estimators
in Monte Carlo experiments. We find that the confidence band based
on the rearrangement-based estimator tends to be narrower without
sacrificing the coverage accuracy.

The literature on structural econometrics of auctions is vast. See
\citet{gentry2018structural} for a recent review; see also the reviews
of the literature in \citet{athey2007nonparametric} and \citet{hendricks2007empirical}.
In their seminal paper, GPV demonstrate nonparametric identification
of the first-price auction model with independent private values,
and propose two-step nonparametric estimation of the density of latent
valuations. This paper improves on the GPV estimator by incorporating
the monotonicity constraint in the nonparametric estimation procedure. 

In a recent paper, \citet[MMS, hereafter]{ma2016inference} describe
the asymptotic distribution of the GPV estimator and propose a valid
bootstrap procedure based on the GPV estimator. They also propose
a procedure for constructing uniform confidence bands for the density
of latent valuations. This paper builds on their results.

The GPV estimator has been widely used in the empirical literature.
For examples of applications, see the literature review in \citet{athey2007nonparametric}
and \citet{hendricks2007empirical}. The GPV approach has been also
utilized in models with risk aversion (\citealp{guerre2009nonparametric},
\citealp{zincenko2018nonparametric}), unobserved heterogeneity (\citealp{krasnokutskaya2011identification}),
bidder asymmetry and affiliated values (\citealp{li2002sea}), common
values (\citealp{haile2003nonparametric}, \citealp{hendricks2003empirical}),
and entry (\citealp{li2009entry}, \citealp{marmer2013model}, \citealp{gentry2014identification}).

The recent related literature includes \citet{liuvuong2013}, who
propose a test for the monotonicity of the bidding function, \citet{liu2017nonparametric},
who propose a procedure for comparing valuation distributions, \citet{Marmer_Shneyerov_Quantile_Auctions},
\citet{luo2017integrated}, \citet{gimenes2017econometrics}, who
propose quantile based methods in the context of auctions. Our paper
is also related to the econometrics literature on two-step nonparametric
estimation. See, e.g., \citet{mammen2012nonparametric}. 

The rest of the paper is organized as follows. Section \ref{sec:The-Auction-Model}
introduces the empirical auction model studied in this paper and its
estimation technique, including the GPV estimator and a new monotonicity-constrained
estimator. In Section \ref{sec:Asymptotic-Properties}, we show that
the new estimator is asymptotically normal with a smaller asymptotic
variance, compared to the unconstrained estimator. Section \ref{sec:Inference}
provides an estimator for the asymptotic variance and a uniform confidence
band around the new monotonicity-constrained estimator. We extend
the proposed estimation and inference method to an auction model with
observed auction heterogeneity in Section \ref{sec:Observed-Heterogeneity}.
Section \ref{sec:Monte-Carlo-Simulations} reports Monte Carlo simulation
results. Proofs are collected in the appendix. 

\begin{ntnbold} ``$a\coloneqq b$'' is understood as ``$a$ is
defined by $b$''. ``$a\eqqcolon b$'' is understood as ``$b$
is defined by $a$''. $\mathbbm{1}\left(\cdot\right)$ denotes the
indicator function, and we also denote $\mathbbm{1}_{A}\coloneqq\mathbbm{1}\left(\cdot\in A\right)$.
Let $\ell^{\infty}\left(A\right)$ be the class of bounded functions
defined on $A$. For any $f\in\ell^{\infty}\left(A\right)$, let $\left\Vert f\right\Vert _{A}\coloneqq\underset{x\in A}{\mathrm{sup}}\left|f\left(x\right)\right|$
be the sup-norm.\end{ntnbold}

\section{The Auction Model and Estimation\label{sec:The-Auction-Model}}

In this section, we consider an auction model for homogeneous goods
and with a fixed number of bidders. A model with observed covariates
capturing auction-specific heterogeneity will be considered in Section
\ref{sec:Observed-Heterogeneity}. The econometrician observes bids
from $L$ auctions, with a fixed number of bidders in each auction:
\begin{equation}
\left\{ B_{il}:i=1,\ldots,N,\,l=1,\ldots,L\right\} .\label{eq:original sample}
\end{equation}
Bidders' valuations
\[
\left\{ V_{il}:i=1,\ldots,N,l=1,\ldots,L\right\} 
\]
are not unobservable to the econometrician. We assume the distribution
of the valuations satisfy the following assumption.
\begin{assumption}[\textbf{Data Generating Process}]
\textup{\label{assu:DGP}(a). The unobserved valuations
\[
\left\{ V_{il}:i=1,\ldots,N,l=1,\ldots,L\right\} 
\]
are i.i.d. with PDF $f$ and CDF $F$. (b). $f$ is strictly positive
and bounded away from zero on its support, a compact interval $\left[\underline{v},\overline{v}\right]\subseteq\mathbb{R}_{+}$,
and is twice continuously differentiable on $\left(\underline{v},\overline{v}\right)$. }
\end{assumption}
Assumption \ref{assu:DGP}(a) assumes that the bidders are symmetric
and the auctions are identical. Assumption \ref{assu:DGP} is similar
to Assumptions A1 and A2 of GPV and Assumption 1 of MMS. The object
of interest is the PDF of the valuations at interior points of $\left[\underline{v},\overline{v}\right]$.
Suppose that $v_{l}>\underline{v}$, $v_{u}<\overline{v}$ and $I\coloneqq\left[v_{l},v_{u}\right]$
is an inner closed sub-interval of $\left[\underline{v},\overline{v}\right]$. 

We assume that the observed bids are generated from the valuations
and by the Bayesian Nash equilibrium (BNE) bidding strategy:
\begin{equation}
B_{il}=s\left(V_{il}\right)\coloneqq V_{il}-\frac{1}{F\left(V_{il}\right)^{N-1}}\int_{\underline{v}}^{V_{il}}F\left(u\right)^{N-1}\mathrm{d}u.\label{eq:bidding function}
\end{equation}
The BNE requires that the bidding strategy $s$ is strictly increasing.
Moreover, GPV show that $s$ is at least three times continuously
differentiable on $\left(\underline{v},\overline{v}\right)$. The
inverse of the BNE bidding strategy can be written as
\begin{equation}
\xi\left(b\right)\coloneqq s^{-1}\left(b\right)=b+\frac{1}{N-1}\frac{G\left(b\right)}{g\left(b\right)},\label{eq: inverse bid strat}
\end{equation}
where $G$ and $g$ are CDF and PDF of the bids, respectively. Let
$\overline{b}\coloneqq s\left(\overline{v}\right)$ and $\underline{b}\coloneqq s\left(\underline{v}\right)$
denote the boundaries of support for the observed i.i.d. bids. GPV
show that $g$ is three-times continuously differentiable and also
bounded away from zero on its support $\left[\underline{b},\overline{b}\right]$:
\begin{equation}
\underline{C}_{g}\coloneqq\underset{b\in\left[\underline{b},\overline{b}\right]}{\mathrm{inf}}g\left(b\right)>0.\label{eq:density of bids bounded  from 0}
\end{equation}

Let $\widehat{G}$ be the empirical CDF of the bids:
\[
\widehat{G}\left(b\right)\coloneqq\frac{1}{N\cdot L}\sum_{i,l}\mathbbm{1}\left(B_{il}\leq b\right)
\]
 and $\widehat{g}$ be the kernel density estimator of $g$:
\begin{equation}
\widehat{g}\left(b\right)\coloneqq\frac{1}{N\cdot L}\sum_{i,l}\frac{1}{h_{g}}K_{g}\left(\frac{B_{il}-b}{h_{g}}\right)\label{eq:kernel density g}
\end{equation}
with some bandwidth $h_{g}>0$ and kernel $K_{g}$. Therefore, the
plug-in nonparametric estimator of the inverse bidding strategy $\xi$
is
\begin{equation}
\widehat{\xi}\left(b\right)\coloneqq b+\frac{1}{N-1}\frac{\widehat{G}\left(b\right)}{\widehat{g}\left(b\right)}.\label{eq:plug in estimator of ksi}
\end{equation}
 For each $B_{il}$, we construct a pseudo valuation by $\widehat{V}_{il}\coloneqq\widehat{\xi}\left(B_{il}\right)$.
There is a boundary bias issue when ordinary kernel density estimator
as in (\ref{eq:kernel density g}) is used. Thus, the pseudo valuations
corresponding to bids in boundary regions are contaminated. GPV propose
to trim off bids that lie in $\left[\widehat{\underline{b}},\widehat{\underline{b}}+h_{g}\right)\cup\left(\widehat{\overline{b}}-h_{g},\widehat{\overline{b}}\right]$,
where
\begin{align*}
\widehat{\overline{b}}\coloneqq & \mathrm{max}\left\{ B_{il}:i=1,...,N,l=1,...,L\right\} \\
\widehat{\underline{b}}\coloneqq & \mathrm{min}\left\{ B_{il}:i=1,...,N,l=1,...,L\right\} .
\end{align*}

We modify the kernel density estimator $\widehat{g}\left(b\right)$
in the boundary region $b\in\left[\underline{b},\underline{b}+h_{g}\right)\cup\left(\overline{b}-h_{g},\overline{b}\right]$
to avoid boundary bias and trimming.\footnote{See \citet{hickman_hubbard_2014_JAE} for more discussion on why we
should avoid trimming when estimating the auction model using GPV
approach.} Another concern is that, in order to remove more bias, we need to
make use of the fact that the bid density $g$ is smoother than the
valuation density $f$ and use a higher-order kernel when estimating
the inverse bidding strategy.\footnote{See Remark 2.3 of MMS.} A
suitable choice is the local quadratic minimum contrast estimator
(MCE). See \citet[Chapter 11.3]{bickel2015mathematical}.\footnote{See \citet{jales2017optimal} and \citet{Ma_EL_MC} for more recent
applications of MCE in econometrics.} Similar to the local polynomial regression, the local quadratic MCE
automatically adapts to the boundary so that the rate of the bias
is the same in the boundary region and the interior.\footnote{Note that the MCE requires knowledge of the locations of the endpoints
$\underline{b}$ and $\overline{b}$. Since the estimators $\widehat{\underline{b}}$
and $\widehat{\overline{b}}$ are super-consistent: $\widehat{\underline{b}}=\underline{b}+O_{p}\left(\mathrm{log}\left(L\right)/L\right)$
and $\widehat{\overline{b}}=\overline{b}+O_{p}\left(\mathrm{log}\left(L\right)/L\right)$,
we can replace the unknown endpoints in the MCE with these estimators
without affecting the validity of the asymptotic results.} The local quadratic MCE coincides with the kernel density estimator
(\ref{eq:kernel density g}) with a fourth-order kernel, and therefore
we achieve desired bias removal in the interior region $\left[\underline{b}+h_{g},\overline{b}-h_{g}\right]$.
The GPV estimator is 
\[
\widehat{f}_{GPV}\left(v\right)\coloneqq\frac{1}{N\cdot L}\sum_{i,l}\frac{1}{h_{f}}K_{f}\left(\frac{\widehat{V}_{il}-v}{h_{f}}\right),
\]
for some bandwidth $h_{f}>0$ and kernel $K_{f}$. We note again that
trimming can be avoided. 

Another important observation is that the plug-in estimator (\ref{eq:plug in estimator of ksi})
may not be monotone in finite samples, although its population counterpart
$\xi$ is strictly increasing under the assumption that the empirical
auction model is correctly-specified. We apply smooth rearrangement
to build a new monotonicity-constrained estimator of $\xi$ on the
plug-in estimator $\widehat{\xi}$. Define 
\begin{equation}
\widehat{s}\left(t\right)\coloneqq\int_{\widehat{\underline{b}}}^{\widehat{\overline{b}}}\int_{-\infty}^{t}\frac{1}{h_{r}}K_{r}\left(\frac{\widehat{\xi}\left(b\right)-u}{h_{r}}\right)\mathrm{d}u\mathrm{d}b+\widehat{\underline{b}},\;t\in\mathbb{R},\label{eq:rearrangement 1}
\end{equation}
for some bandwidth $h_{r}>0$ and (second-order) kernel function $K_{r}$.
Denote $\widetilde{K}_{r}\left(u\right)\coloneqq\int_{-\infty}^{u}K_{r}\left(t\right)\mathrm{d}t$.
Alternatively we can write
\begin{equation}
\widehat{s}\left(t\right)=\int_{\widehat{\underline{b}}}^{\widehat{\overline{b}}}\widetilde{K}_{r}\left(\frac{t-\widehat{\xi}\left(b\right)}{h_{r}}\right)\mathrm{d}b+\widehat{\underline{b}},\;t\in\mathbb{R}.\label{eq:rearrangement 2}
\end{equation}
(\ref{eq:rearrangement 2}) contains less challenge in computation
as the expression of $\widetilde{K}_{r}$ is always available for
standard kernel functions. 

It is clear that $\widehat{s}$ is increasing on $\mathbb{R}$ and
\[
\widehat{s}\left(t\right)=\begin{cases}
\widehat{\overline{b}} & \textrm{if \ensuremath{t\leq\underset{b\in\left[\widehat{\underline{b}},\widehat{\overline{b}}\right]}{\mathrm{inf}}\widehat{\xi}\left(b\right)-h_{r}}}\\
\widehat{\underline{b}} & \textrm{if \ensuremath{t\geq\underset{b\in\left[\widehat{\underline{b}},\widehat{\overline{b}}\right]}{\mathrm{sup}}\widehat{\xi}\left(b\right)+h_{r}}}
\end{cases}
\]
It is easy to see that $\widehat{s}$ can be viewed as an estimator
of the bidding function $s$ (see Lemma \ref{lem:Lemma 2}). Note
that this estimator is of interest on its own, as the bidding function
$s$ is an important structural object. Moreover, constructing an
estimator of $s$ directly from its expression in (\ref{eq:bidding function})
can be cumbersome.

Let $\widehat{s}^{-1}$ denote the pseudo inverse of $\widehat{s}$:
\[
\widehat{s}^{-1}\left(b\right)\coloneqq\mathrm{inf}\left\{ u\in\mathbb{R}:\widehat{s}\left(u\right)\geq b\right\} .
\]
$\widehat{s}^{-1}$ is a rearrangement-based estimator of $\xi$ with
the monotonicity constraint imposed. A new modified GPV estimation
procedure now can be proposed: First, we construct $\widehat{\xi}$,
the plug-in nonparametric estimator of the inverse bidding strategy
$\xi$. To avoid trimming, we use the local quadratic MCE instead
of the ordinary kernel density estimator. Then, we construct the monotonicity-imposed
estimator of the inverse bidding strategy: $\widehat{s}^{-1}$ and
generate monotonicity-constrained pseudo valuations $\widehat{V}_{il}^{\dagger}\coloneqq\widehat{s}^{-1}\left(B_{il}\right)$
for $i=1,...,N$, $l=1,...,L$. A new estimator, the rearrangement-based
GPV (RGPV), is
\[
\widehat{f}_{RGPV}\left(v\right)\coloneqq\frac{1}{N\cdot L}\sum_{i,l}\frac{1}{h_{f}}K_{f}\left(\frac{\widehat{V}_{il}^{\dagger}-v}{h_{f}}\right).
\]

The RGPV estimation procedure is computationally more involved than
the standard GPV procedure. When implementing it in practice, the
integral in (\ref{eq:rearrangement 2}) can be approximated by an
upper Riemann sum. Let $M\in\mathbb{N}$ be a very large number. Let
$d\coloneqq\left(\widehat{\overline{b}}-\widehat{\underline{b}}\right)/M$.
We can approximate $\int_{\widehat{\underline{b}}}^{\widehat{\overline{b}}}\widetilde{K}_{r}\left(\left(t-\widehat{\xi}\left(b\right)\right)/h_{r}\right)\mathrm{d}b$
by
\[
\sum_{i=1}^{M}\widetilde{K}_{r}\left(\frac{t-\widehat{\xi}\left(\widehat{\underline{b}}+i\cdot d\right)}{h_{r}}\right)d.
\]
However, unlike the estimator in \citet{henderson_2012_JOE}, it is
not required to solve a constrained optimization problem. Thus the
RGPV estimation procedure is computationally less demanding than \citet{henderson_2012_JOE}'s
estimation procedure, which is based on constrained reweighting. 

An alternative, and closely related, approach is to impose the monotonicity
restriction through the ``non-smooth'' rearrangement. Instead of
using (\ref{eq:rearrangement 1}), we can define 
\[
\widehat{s}_{0}\left(t\right)\coloneqq\int_{\widehat{\underline{b}}}^{\widehat{\overline{b}}}\mathbbm{1}\left(\widehat{\xi}\left(b\right)\leq t\right)\mathrm{d}b+\widehat{\underline{b}},\;t\in\mathbb{R},
\]
and use $\widehat{s}_{0}^{-1}$ as a monotonicity-constrained estimator
of $\xi$ to generated pseudo valuations. If the empirical auction
model is correctly specified so that $\xi$ is strictly increasing,
$\widehat{s}_{0}^{-1}$ is always an improvement over the plug-in
estimator $\widehat{\xi}$, in the sense that $\widehat{s}_{0}^{-1}$
has a strictly smaller (finite-sample) integrated mean square error
whenever $\widehat{\xi}$ is not monotonic. See \citealp[Proposition 1]{chernozhukov2009improving}.
However, for the structural auction model, the parameter of interest
is the density $f$. It is unclear whether the density estimator based
on pseudo valuations generated by $\widehat{s}_{0}^{-1}$ is an improvement
over the unconstrained GPV estimator theoretically. In this paper,
we focus on smooth rearrangement and in the next section, we show
that based on the RGPV estimator we could potentially achieve sharper
inference in large samples, which can be regarded as theoretical advantage
over the unconstrained GPV estimator.

\section{Asymptotic Properties\label{sec:Asymptotic-Properties}}

The following assumptions are imposed on the kernel functions and
the bandwidths, respectively.
\begin{assumption}[\textbf{Kernel}]
\label{assu:Assumption 2 Kernel}\textup{ $K_{f}$ is a probability
density function that is symmetric around 0, compactly supported on
$[-1,1]$ and has at least two Lipschitz continuous derivatives on
$\mathbb{R}$. Moreover, $K_{r}=K_{f}$ and the same kernel function
is used in the local quadratic MCE of the bid density. }
\end{assumption}
\begin{assumption}[\textbf{Bandwidth}]
\textup{\label{assu:bandwidth}Let $h$ be a sequence $\left\{ h_{L}\right\} _{L=1}^{\infty}$
satisfying $h=L^{-\gamma}$ for $1/7\leq\gamma\leq1/3$. $h_{f}=\lambda_{f}h$,
$h_{g}=\lambda_{g}h$ and $h_{r}=\lambda_{r}h$ for some positive
constants $\lambda_{f}$, $\lambda_{g}$ and $\lambda_{r}$.}
\end{assumption}
Assumption \ref{assu:Assumption 2 Kernel} implies that the kernel
functions are of second order. When the kernel used in the local quadratic
MCE of the bid density is second-order, at the interior points, the
MCE is the same as the ordinary kernel density estimator (\ref{eq:kernel density g})
with $K_{g}$ being fourth-order. Assumption \ref{assu:bandwidth}
is similar to Assumption 3 in MMS. 

It is shown in MMS that under assumptions \ref{assu:DGP}-\ref{assu:bandwidth},
\begin{equation}
\left(Lh_{f}^{2}h_{g}\right)^{1/2}\left(\widehat{f}_{GPV}\left(v\right)-f\left(v\right)\right)\rightarrow_{d}\mathrm{N}\left(0,\mathrm{V}_{GPV}\left(v\right)\right),\label{eq:GPV asymptotic normality}
\end{equation}
where
\begin{equation}
\mathrm{V}_{GPV}\left(v\right)\coloneqq\frac{1}{N\left(N-1\right)^{2}}\frac{F\left(v\right)^{2}f\left(v\right)^{2}}{g\left(s\left(v\right)\right)^{3}}\int\left\{ \int K_{f}'\left(u\right)K_{g}\left(w-\frac{\lambda_{f}}{\lambda_{g}}s'\left(v\right)u\right)\mathrm{d}u\right\} ^{2}\mathrm{d}w.\label{eq:asymptotic variance GPV}
\end{equation}
MMS show that the asymptotic variance (\ref{eq:asymptotic variance GPV})
can be consistently estimated by some estimator $\widehat{\mathrm{V}}_{GPV}\left(v\right)$
and an asymptotically valid confidence interval for $f\left(v\right)$
can be constructed:
\begin{equation}
\left[\widehat{f}_{GPV}\left(v\right)-z_{1-\alpha/2}\sqrt{\frac{\widehat{\mathrm{V}}_{GPV}\left(v\right)}{Lh_{f}^{2}h_{g}}},\widehat{f}_{GPV}\left(v\right)+z_{1-\alpha/2}\sqrt{\frac{\widehat{\mathrm{V}}_{GPV}\left(v\right)}{Lh_{f}^{2}h_{g}}}\right],\label{eq:GPV confidence interval}
\end{equation}
where $z_{1-\alpha/2}$ denotes the $1-\alpha/2$ quantile of the
standard normal distribution. Note that $\widehat{\mathrm{V}}_{GPV}\left(v\right)$
and its probabilistic limit $\mathrm{V}_{GPV}\left(v\right)$ play
important roles in determining the length of the confidence interval.
Furthermore, a bootstrap uniform confidence band is given by 
\[
CB_{GPV}\left(v\right)\coloneqq\left[\widehat{f}_{GPV}\left(v\right)-\zeta_{GPV,\alpha}\sqrt{\frac{\widehat{\mathrm{V}}_{GPV}\left(v\right)}{Lh_{f}^{2}h_{g}}},\widehat{f}_{GPV}\left(v\right)+\zeta_{GPV,\alpha}\sqrt{\frac{\widehat{\mathrm{V}}_{GPV}\left(v\right)}{Lh_{f}^{2}h_{g}}}\right],\textrm{ for \ensuremath{v\in I}},
\]
where $\zeta_{GPV,\alpha}$ is some bootstrap critical value. It can
be shown that 
\[
\mathrm{P}\left[f\left(v\right)\in CB_{GPV}\left(v\right),\textrm{ for all \ensuremath{v\in I}}\right]\rightarrow1-\alpha,\textrm{ as \ensuremath{L\uparrow\infty}}.
\]

Next, we show that a similar asymptotic normality result holds for
the RGPV estimator, but with a smaller asymptotic variance. The proof
uses the same arguments as in the proof of Theorem 2.1 in MMS. First,
we derive the following asymptotic representation:
\begin{equation}
\widehat{f}_{RGPV}\left(v\right)-f\left(v\right)=\frac{1}{\left(N-1\right)}\cdot\frac{1}{\left(N\cdot L\right)^{2}}\sum_{i,l}\sum_{j,k}\mathcal{M}\left(B_{il},B_{jk};v\right)+o_{p}\left(\left(Lh^{3}\right)^{-1/2}\right),\label{eq:f_RGPV_hat - f}
\end{equation}
where the remainder term is uniform in $v\in I$ . Moreover, 
\[
\mathcal{M}\left(b',b;v\right)\coloneqq-\frac{1}{h_{f}^{2}}K_{f}'\left(\frac{\xi\left(b'\right)-v}{h_{f}}\right)\xi'\left(b'\right)\int_{\underline{b}}^{\overline{b}}\frac{1}{h_{r}}K_{r}\left(\frac{\xi\left(b'\right)-\xi\left(u\right)}{h_{r}}\right)\frac{G\left(u\right)}{g\left(u\right)^{2}}\left(\frac{1}{h_{g}}K_{g}\left(\frac{b-u}{h_{g}}\right)-g\left(u\right)\right)\mathrm{d}u.
\]
Note that this ``kernel'' is different from that of the unconstrained
GPV estimator. See Equation (2.6) of MMS.

Define 
\begin{align}
 & \mathcal{M}_{1}\left(b;v\right)\coloneqq\int\mathcal{M}\left(b,b';v\right)\mathrm{d}G\left(b'\right),\nonumber \\
 & \mathcal{M}_{2}\left(b;v\right)\coloneqq\int\mathcal{M}\left(b',b;v\right)\mathrm{d}G\left(b'\right)\textrm{ and }\mu_{\mathcal{M}}\left(v\right)\coloneqq\int\int\mathcal{M}\left(b,b';v\right)\mathrm{d}G\left(b\right)\mathrm{d}G\left(b'\right).\label{eq:definition of R_1 R_2 miu}
\end{align}
Note that $\mu_{\mathcal{M}}\left(v\right)=\mathrm{E}\left[\mathcal{M}_{1}\left(B_{11};v\right)\right]=\mathrm{E}\left[\mathcal{M}_{2}\left(B_{11};v\right)\right]$.
Applying Hoeffding decomposition to the leading term in (\ref{eq:f_RGPV_hat - f})
and techniques from the theories of empirical processes and U processes,
we can show that
\[
\widehat{f}_{RGPV}\left(v\right)-f\left(v\right)=\frac{1}{N-1}\cdot\frac{1}{N\cdot L}\sum_{i,l}\left(\mathcal{M}_{2}\left(B_{il};v\right)-\mu_{\mathcal{M}}\left(v\right)\right)+o_{p}\left(\left(Lh^{3}\right)^{-1/2}\right),
\]
where the remainder term is uniform in $v\in I$. $\mathcal{M}_{2}\left(B_{il};v\right)-\mu_{\mathcal{M}}\left(v\right)$,
$i=1,...,N$, $l=1,...,L$ are independent, zero-mean but dependent
on the bandwidths. This term can be shown asymptotically normal. 
\begin{thm}
\textup{\label{thm:normality}Suppose that Assumptions \ref{assu:DGP}
- \ref{assu:bandwidth} are satisfied and $v\in\left(\underline{v},\overline{v}\right)$.
Also assume that $\lambda_{r}=\lambda_{f}$. Then,
\begin{equation}
\left(Lh_{f}^{2}h_{g}\right)^{1/2}\left(\widehat{f}_{RGPV}\left(v\right)-f\left(v\right)\right)\rightarrow_{d}\mathrm{N}\left(0,\mathrm{V}_{RGPV}\left(v\right)\right),\label{eq:statement of the main theorem}
\end{equation}
where
\begin{equation}
\mathrm{V}_{RGPV}\left(v\right)\coloneqq\frac{1}{N\left(N-1\right)^{2}}\frac{F\left(v\right)^{2}f\left(v\right)^{2}}{g\left(s\left(v\right)\right)^{3}}\int\left\{ \int\int K_{f}'\left(u\right)K_{r}\left(u-z\right)K_{g}\left(w-\frac{\lambda_{f}}{\lambda_{g}}s'\left(v\right)z\right)\mathrm{d}z\mathrm{d}u\right\} ^{2}\mathrm{d}w.\label{eq:asymptotic covariance}
\end{equation}
Moreover,
\begin{equation}
\mathrm{V}_{RGPV}\left(v\right)\leq\mathrm{V}_{GPV}\left(v\right)\label{eq:variance inequality}
\end{equation}
for all $v\in\left(\underline{v},\overline{v}\right)$. }
\end{thm}
\begin{rembold}In the proof of Theorem \ref{thm:normality}, we show
that 
\begin{align}
 & \mathrm{E}\left[\frac{Lh_{f}^{2}h_{g}}{\left(N-1\right)^{2}}\left(\frac{1}{N\cdot L}\sum_{i,l}\left(\mathcal{M}_{2}\left(B_{il};v\right)-\mu_{\mathcal{M}}\left(v\right)\right)\right)^{2}\right]\nonumber \\
= & \frac{1}{N\left(N-1\right)^{2}h_{f}^{2}h_{g}}\int\left\{ \int K_{f}'\left(\frac{\xi\left(b'\right)-v}{h_{f}}\right)\xi'\left(b'\right)\int_{\underline{b}}^{\overline{b}}\frac{1}{h_{r}}K_{r}\left(\frac{\xi\left(b'\right)-\xi\left(u\right)}{h_{r}}\right)\frac{G\left(u\right)}{g\left(u\right)^{2}}K_{g}\left(\frac{b-u}{h_{g}}\right)\mathrm{d}u\mathrm{d}G\left(b'\right)\right\} ^{2}\mathrm{d}G\left(b\right)\nonumber \\
 & +O\left(h^{3}\right)\nonumber \\
\eqqcolon & \mathrm{V}_{\mathcal{M}}\left(v\right)+O\left(h^{3}\right),\label{eq:R_2 average variance}
\end{align}
where the remainder term is uniform in $v\in I$. The asymptotic variance
is the limit of the leading term of (\ref{eq:R_2 average variance})
as $h\downarrow0$:
\begin{align*}
 & \underset{h\downarrow0}{\mathrm{lim}}\frac{1}{h_{f}^{2}h_{g}}\int\left\{ \int K_{f}'\left(\frac{\xi\left(b'\right)-v}{h_{f}}\right)\xi'\left(b'\right)\int_{\underline{b}}^{\overline{b}}\frac{1}{h_{r}}K_{r}\left(\frac{\xi\left(b'\right)-\xi\left(u\right)}{h_{r}}\right)\frac{G\left(u\right)}{g\left(u\right)^{2}}K_{g}\left(\frac{b-u}{h_{g}}\right)\mathrm{d}u\mathrm{d}G\left(b'\right)\right\} ^{2}\mathrm{d}G\left(b\right)\\
= & \frac{F\left(v\right)^{2}f\left(v\right)^{2}}{g\left(s\left(v\right)\right)^{3}}\int\left\{ \int\int K_{f}'\left(u\right)K_{r}\left(u-z\right)K_{g}\left(w-\frac{\lambda_{f}}{\lambda_{g}}s'\left(v\right)z\right)\mathrm{d}z\mathrm{d}u\right\} ^{2}\mathrm{d}w.
\end{align*}
A consistent estimator of the asymptotic variance $\mathrm{V}_{RGPV}\left(v\right)$
can be derived based on the sample analogue of $\mathrm{V}_{\mathcal{M}}\left(v\right)$.
\end{rembold}

\begin{rembold}The proof of Theorem \ref{thm:normality} also incorporates
the bias term: 
\[
\left(Lh_{f}^{2}h_{g}\right)^{1/2}\left(\widehat{f}_{RGPV}\left(v\right)-f\left(v\right)-\iota\left(v\right)\right)\rightarrow_{d}\mathrm{N}\left(0,\mathrm{V}_{RGPV}\left(v\right)\right),
\]
where 
\[
\iota\left(v\right)\coloneqq\frac{1}{2}f''\left(v\right)\left(\int K_{f}\left(u\right)u^{2}\mathrm{d}u\right)h_{f}^{2}+\frac{1}{2}\frac{\left(s'''\left(v\right)f\left(v\right)+s''\left(v\right)f'\left(v\right)\right)s'\left(v\right)-s''\left(v\right)^{2}f\left(v\right)}{s'\left(v\right)^{2}}\left(\int K_{r}\left(u\right)u^{2}\mathrm{d}u\right)h_{r}^{2}.
\]
A comparison of $\iota\left(v\right)$ with the bias given in Remark
2.3 of MMS shows that the smooth rearrangement incurs additional bias.
For inference, we take the ``under-smoothing'' approach to select
sufficiently small bandwidths so that these bias terms become negligible.\end{rembold}

\begin{rembold}\label{numerical }MMS show that the GPV estimator
has a smaller asymptotic variance than the quantile-based estimator
of \citet{Marmer_Shneyerov_Quantile_Auctions}. The proof of (\ref{eq:variance inequality})
uses similar arguments. It is easy to show that the inequality (\ref{eq:variance inequality})
is strict for all $v\in\left(\underline{v},\overline{v}\right)$,
if the kernel function satisfies $K'\left(u\right)<0$ for all $u\in\left(0,1\right)$
and $K'\left(u\right)>0$ for all $u\in\left(-1,0\right)$. 

Suppose that the valuations are drawn from the family of distributions
\[
F\left(v\right)=\begin{cases}
0, & v<0\\
v^{\theta}, & 0\leq v\leq1\\
1, & v>1
\end{cases}
\]
supported on $\left[0,1\right]$ with some parameter $\theta>0$.
The Bayesian Nash equilibrium bidding strategy in this example is
\[
s\left(v\right)=\left(1-\frac{1}{\theta\left(N-1\right)+1}\right)v,
\]
which is linear in $v$. Thus, the ratio $\mathrm{V}_{GPV}\left(v\right)/\mathrm{V}_{RGPV}\left(v\right)$
is independent from $v$, by the definitions of $\mathrm{V}_{RGPV}\left(v\right)$
and $\mathrm{V}_{GPV}\left(v\right)$. In this example, we choose
the triweight kernel
\[
K\left(u\right)=\frac{35}{32}\left(1-u^{2}\right)^{3}\mathbbm{1}\left(\left|u\right|\leq1\right).
\]
We can analytically evaluate the multi-dimensional integrals in (\ref{eq:asymptotic variance GPV})
and (\ref{eq:asymptotic covariance}) and calculate $\mathrm{V}_{GPV}\left(v\right)/\mathrm{V}_{RGPV}\left(v\right)$
in this example. We experiment with different combinations of $\left(\theta,N\right)$,
and find that the ratio $\mathrm{V}_{GPV}\left(v\right)/\mathrm{V}_{RGPV}\left(v\right)$
can be quite large in my cases. For instance, in the case of $\left(\theta,N\right)=\left(1,5\right)$,
$\mathrm{V}_{GPV}\left(v\right)/\mathrm{V}_{RGPV}\left(v\right)$
is approximately 1.587.\end{rembold}

\section{Inference\label{sec:Inference}}

If the asymptotic variance $\mathrm{V}_{RGPV}(v)$ in (\ref{eq:asymptotic covariance})
can be consistently estimated by some estimator $\widehat{\mathrm{V}}_{RGPV}\left(v\right)$,
Theorem \ref{thm:normality} shows that we can construct an asymptotically
valid confidence interval for $f\left(v\right)$:
\begin{equation}
\left[\widehat{f}_{RGPV}\left(v\right)-z_{1-\alpha/2}\sqrt{\frac{\widehat{\mathrm{V}}_{RGPV}\left(v\right)}{Lh_{f}^{2}h_{g}}},\widehat{f}_{RGPV}\left(v\right)+z_{1-\alpha/2}\sqrt{\frac{\widehat{\mathrm{V}}_{RGPV}\left(v\right)}{Lh_{f}^{2}h_{g}}}\right].\label{eq:RGPV confidence interval}
\end{equation}
As shown in (\ref{eq:variance inequality}), our rearrangement-based
estimator has a smaller asymptotic variance than the GPV estimator.
Therefore, the confidence intervals based on our estimator in (\ref{eq:RGPV confidence interval})
should be shorter than those based on the GPV estimator in (\ref{eq:GPV confidence interval})
in large samples. 

An estimator of $\mathrm{V}_{RGPV}\left(v\right)$ is derived based
on the sample analogue of $\mathrm{V}_{\mathcal{M}}\left(v\right)$,
(see (\ref{eq:R_2 average variance})):
\begin{align*}
\widehat{\mathrm{V}}_{RGPV}\left(v\right)\coloneqq & \frac{1}{N\left(N-1\right)^{2}h_{f}^{2}h_{g}}\frac{1}{\left(N\cdot L\right)\left(N\cdot L-1\right)\left(N\cdot L-2\right)}\sum_{i,l}\sum_{\left(j,k\right)\neq\left(i,l\right)}\sum_{\left(j',k'\right)\neq\left(i,l\right),\,\left(j',k'\right)\neq\left(j,k\right)}\eta_{il,jk}(v)\eta_{il,j'k'}(v),\\
\eta_{il,jk}(v)\coloneqq & K_{f}'\left(\frac{\widehat{V}_{jk}^{\dagger}-v}{h_{f}}\right)\frac{1}{\widehat{s}'\left(\widehat{V}_{jk}^{\dagger}\right)}\int_{\widehat{\underline{b}}}^{\widehat{\overline{b}}}\frac{1}{h_{r}}K_{r}\left(\frac{\widehat{V}_{jk}^{\dagger}-\widehat{\xi}\left(u\right)}{h_{r}}\right)\frac{\widehat{G}\left(u\right)}{\widehat{g}\left(u\right)^{2}}K_{g}\left(\frac{B_{il}-u}{h_{g}}\right)\mathrm{d}u,
\end{align*}
where we use the fact $\xi'\left(b\right)=1/s'\left(\xi\left(b\right)\right)$
and
\[
\widehat{s}'\left(v\right)\coloneqq\int_{\widehat{\underline{b}}}^{\widehat{\overline{b}}}\frac{1}{h_{r}}K_{r}\left(\frac{\widehat{\xi}\left(b\right)-v}{h_{r}}\right)\mathrm{d}b.
\]
The integral can be approximated by an upper Riemann sum in practice.
The following result provides uniform consistency of the variance
estimator and also an estimate of its uniform rate of convergence.
The proof uses the same arguments as in the proof of Theorem 3.1 of
MMS. 
\begin{thm}
\textup{\label{thm:variance estimator}Suppose that Assumptions \ref{assu:DGP}
- \ref{assu:bandwidth} are satisfied. Then,
\[
\underset{v\in I}{\mathrm{sup}}\left|\widehat{\mathrm{V}}_{RGPV}\left(v\right)-\mathrm{V}_{\mathcal{M}}\left(v\right)\right|=O_{p}\left(\left(\frac{\mathrm{log}\left(L\right)}{Lh^{3}}\right)^{1/2}+h^{2}\right).
\]
}
\end{thm}
An alternative approach to constructing pointwise confidence intervals
is based on bootstrapping. Let
\begin{equation}
\left\{ B_{il}^{*}:i=1,\ldots,N,l=1,\ldots,L\right\} \label{eq:bootstrap sample}
\end{equation}
denote a set of independent random variables drawn from the original
sample (\ref{eq:original sample}) with replacement. $\widehat{G}^{*}$
and $\widehat{g}^{*}$ denote the bootstrap analogues of $\widehat{G}$
and $\widehat{g}$ respectively. Let $\widehat{\xi}^{*}$ be the bootstrap
analogue of $\widehat{\xi}$. $\widehat{\xi}^{*}$ is defined by replacing
$\widehat{G}$ and $\widehat{g}$ with $\widehat{G}^{*}$ and $\widehat{g}^{*}$.
We further define 
\[
\widehat{s}^{*}\left(t\right)\coloneqq\int_{\widehat{\underline{b}}}^{\widehat{\overline{b}}}\int_{-\infty}^{t}\frac{1}{h_{r}}K_{r}\left(\frac{\widehat{\xi}^{*}\left(b\right)-u}{h_{r}}\right)\mathrm{d}u\mathrm{d}b+\widehat{\underline{b}},\;t\in\mathbb{R}
\]
and bootstrap analogues of $\widehat{V}_{il}^{\dagger}$, denoted
by $\widehat{V}_{il}^{\dagger*}$, $i=1,\ldots,N$, $l=1,\ldots,L$
by using the pseudo inverse of $\widehat{s}^{*}$. Lastly, we construct
a bootstrap analogue of $\widehat{f}_{RGPV}$:
\[
\widehat{f}_{RGPV}^{*}\left(v\right)\coloneqq\frac{1}{N\cdot L}\sum_{i,l}\frac{1}{h_{f}}K_{f}\left(\frac{\widehat{V}_{il}^{\dagger*}-v}{h_{f}}\right).
\]
The percentile bootstrap pointwise confidence interval for $f\left(v\right)$
is 
\[
\left[q_{\alpha/2}^{*}\left(v\right),q_{1-\alpha/2}^{*}\left(v\right)\right],
\]
where $q_{\tau}^{*}\left(v\right)$ is the $\tau-$th quantile of
the conditional distribution of $\widehat{f}_{RGPV}^{*}\left(v\right)$
given the original sample.

The pointwise inference results for $f(v)$ described above can be
extended for the inference on the optimal reserve price, as the latter
is a function of the density at the reserve price. Such an extension
is discussed in Section 8 in MMS. Similarly, a function of the density
$(1-F(v))/f(v)$ is of interest in applications as it represents the
markup of the bidder with value $v$. Again, since relatively to GPV's
our rearrangement estimator has a smaller asymptotic variance, basing
inference on optimal reserve price or the markup on our estimator
would result in more powerful tests and shorter confidence intervals.

Next, we show that a bootstrap-based uniform confidence band for $\left\{ f\left(v\right):v\in I\right\} $
centered at the rearrangement-based estimator can be constructed by
using intermediate Gaussian approximation pioneered by \citet{chernozhukov2014gaussian,chernozhukov2014anti,chernozhukov2016empirical}.
Consider the following bootstrap process
\begin{equation}
Z^{*}\left(v\right)\coloneqq\frac{\widehat{f}_{RGPV}^{*}\left(v\right)-\widehat{f}_{RGPV}\left(v\right)}{\left(Lh_{f}^{2}h_{g}\right)^{-1/2}\widehat{\mathrm{V}}_{RGPV}\left(v\right)^{1/2}},\,v\in I.\label{eq:Z_star definition}
\end{equation}
Let $\mathrm{P}^{*}\left[\cdot\right]$ denote the probability conditional
on the original sample and
\[
\zeta_{RGPV,\alpha}\coloneqq\mathrm{inf}\left\{ z\in\mathbb{R}:\mathrm{P}^{*}\left[\left\Vert Z^{*}\right\Vert _{I}\leq z\right]\geq1-\alpha\right\} 
\]
be the $\left(1-\alpha\right)$-quantile of the conditional distribution
of $\left\Vert Z^{*}\right\Vert _{I}$ given the original sample.
A uniform confidence band around the rearrangement-based estimator
is

\[
CB_{RGPV}\left(v\right)\coloneqq\left[\widehat{f}_{RGPV}\left(v\right)-\zeta_{RGPV,\alpha}\sqrt{\frac{\widehat{\mathrm{V}}_{RGPV}\left(v\right)}{Lh_{f}^{2}h_{g}}},\,\widehat{f}_{RGPV}\left(v\right)+\zeta_{RGPV,\alpha}\sqrt{\frac{\widehat{\mathrm{V}}_{RGPV}\left(v\right)}{Lh_{f}^{2}h_{g}}}\right],\,\textrm{for \ensuremath{v\in I}}.
\]
The following theorem establishes the asymptotic validity of $CB_{RGPV}$.
Its proof uses the same arguments as in the proof of Corollary 4.3
of MMS. 
\begin{thm}
\textup{\label{thm:uniform confidence band}Suppose that Assumptions
\ref{assu:DGP} - \ref{assu:bandwidth} are satisfied. Then,}
\end{thm}
\[
\mathrm{P}\left[f\left(v\right)\in CB_{RGPV}\left(v\right),\textrm{ for all \ensuremath{v\in I}}\right]\rightarrow1-\alpha,\textrm{ as \ensuremath{L\uparrow\infty}}.
\]

\section{Auction-Specific Heterogeneity\label{sec:Observed-Heterogeneity}}

In previous sections, we focused on the case of identical auctions
with a fixed number of bidders. In this section, we consider auction
models with auction-specific heterogeneity. The econometrician observes
data from $L$ auctions. Let $\boldsymbol{X}_{l}$ denote the $d$-dimensional
relevant characteristics for the object in the $l$-th auction. Let
$N_{l}$ denote the number of bidders in the $l$-th auction. Let
$B_{il}$ denote the bid submitted by the $i$-th bidder in the $l$-th
auction. The data observed by the econometrician is given by
\[
\left\{ \left(B_{il},\boldsymbol{X}_{l},N_{l}\right):i=1,...,N_{l},\,l=1,...,L\right\} .
\]
Unobserved bidders' valuations are denoted by
\[
\left\{ V_{il}:i=1,...,N_{l},\,l=1,...,L\right\} .
\]

Assume that $\left\{ \left(\boldsymbol{X}_{l},N_{l}\right):l=1,...,L\right\} $
are i.i.d. and for each $l=1,...,L$, given $\boldsymbol{X}_{l}=\boldsymbol{x}$
and $N_{l}=n$, the valuations $\left\{ V_{il}:i=1,...,n\right\} $
are i.i.d. with conditional PDF $f\left(\cdot|\boldsymbol{x}\right)$.
We follow the literature and assume that the valuations and the number
of bidders $N_{l}$ are conditionally independent given the observed
characteristics $\boldsymbol{X}_{l}$.\footnote{For a test of this assumption, see \citet{liu2017nonparametric}.}
Also assume that the conditional probability mass function of $N_{l}$
given $\boldsymbol{X}_{l}$ has a known support $\left\{ \underline{n},...,\overline{n}\right\} $.

The observed bid $B_{il}$ is assumed from the Bayesian Nash equilibrium
bidding for risk-neutral bidder $i$ submitted in the $l$-th auction.
Let $G\left(\cdot|\boldsymbol{x},n\right)$ denote the conditional
CDF of $B_{il}$ given $\boldsymbol{X}_{l}=\boldsymbol{x}$ and $N_{l}=n$.
Let $g\left(\cdot|\boldsymbol{x},n\right)$ be the conditional PDF.
The inverse bidding strategy in this context becomes
\begin{equation}
V_{il}=\xi\left(B_{il},\boldsymbol{X}_{l},N_{l}\right)\coloneqq B_{il}+\frac{1}{N_{l}-1}\frac{G\left(B_{il}|\boldsymbol{X}_{l},N_{l}\right)}{g\left(B_{il}|\boldsymbol{X}_{l},N_{l}\right)}.\label{eq:inverse bidding strategy}
\end{equation}

For estimation and inference, we can generate pseudo valuations from
(\ref{eq:inverse bidding strategy}) by replacing the true conditional
CDF and PDF by their kernel estimators. The GPV estimator can be defined
analogously in this general context. Asymptotically valid pointwise
confidence intervals and uniform confidence bands for $f\left(\cdot|\boldsymbol{x}\right)$
can be constructed. See Section 5 of MMS for more details. 

Following \citet{haile2003nonparametric}, we use a semi-parametric
approach to homogenize the bids.\footnote{The homogenization approach is used extensively in the literature,
e.g. \citet{athey2004csb}, \citet{liu2017nonparametric}, \citet{luo2017integrated},
and many others.} Let 
\begin{equation}
\left\{ \epsilon_{il}:i=1,...,N_{l},\,l=1,...,L\right\} \label{eq:errors}
\end{equation}
denote positive i.i.d. idiosyncratic values that are independent from
the auction-specific characteristics $\boldsymbol{X}_{l}$, $l=1,...,L$.
Let $F_{\epsilon}$ be its CDF and $\left[\underline{\epsilon},\overline{\epsilon}\right]$
be its support. Define
\[
\widetilde{B}_{il}=s\left(\epsilon_{il},N_{l}\right)\coloneqq\epsilon_{il}-\frac{1}{F_{\epsilon}\left(\epsilon_{il}\right)^{N_{l}-1}}\int_{\underline{\epsilon}}^{\epsilon_{il}}F_{\epsilon}\left(u\right)^{N_{l}-1}\mathrm{d}u.
\]
Let
\[
\widetilde{G}\left(b|n\right)\coloneqq\mathrm{P}\left[\widetilde{B}_{il}\leq b|N_{l}=n\right]
\]
be the conditional CDF and $\widetilde{g}\left(b|n\right)$ be the
corresponding conditional PDF. Note that we have 
\begin{equation}
\epsilon_{il}=s\left(\epsilon_{il},N_{l}\right)+\frac{1}{N_{l}-1}\frac{\widetilde{G}\left(s\left(\epsilon_{il},N_{l}\right)|N_{l}\right)}{\widetilde{g}\left(s\left(\epsilon_{il},N_{l}\right)|N_{l}\right)}.\label{eq:FOC error}
\end{equation}

The approach of \citet{haile2003nonparametric} assumes that, for
some parametric function $\varUpsilon$ to be specified below, $V_{il}=\varUpsilon\left(\boldsymbol{X}_{l}\right)\epsilon_{il}$
for $i=1,...,N_{l}$ and $l=1,...,L$. It then can be shown that the
conditional CDF and PDF of $\varUpsilon\left(\boldsymbol{X}_{l}\right)\widetilde{B}_{il}$
(denoted by $\widetilde{G}_{\varUpsilon}$ and $\widetilde{g}_{\varUpsilon}$,
respectively) satisfy
\[
\widetilde{G}_{\varUpsilon}\left(b|\boldsymbol{X}_{l},N_{l}\right)\coloneqq\mathrm{P}\left[\varUpsilon\left(\boldsymbol{X}_{l}\right)\widetilde{B}_{il}\leq b|\boldsymbol{X}_{l},N_{l}\right]=\widetilde{G}\left(\frac{b}{\varUpsilon\left(\boldsymbol{X}_{l}\right)}|N_{l}\right)
\]
and 
\[
\widetilde{g}_{\varUpsilon}\left(b|\boldsymbol{X}_{l},N_{l}\right)=\widetilde{g}\left(\frac{b}{\varUpsilon\left(\boldsymbol{X}_{l}\right)}|N_{l}\right)\frac{1}{\varUpsilon\left(\boldsymbol{X}_{l}\right)}.
\]
It is clear from these results and (\ref{eq:FOC error}) that 
\[
\varUpsilon\left(\boldsymbol{X}_{l}\right)\epsilon_{il}=\varUpsilon\left(\boldsymbol{X}_{l}\right)s\left(\epsilon_{il},N_{l}\right)+\frac{1}{N_{l}-1}\frac{\widetilde{G}_{\varUpsilon}\left(\varUpsilon\left(\boldsymbol{X}_{l}\right)s\left(\epsilon_{il},N_{l}\right)|\boldsymbol{X}_{l},N_{l}\right)}{\widetilde{g}_{\varUpsilon}\left(\varUpsilon\left(\boldsymbol{X}_{l}\right)s\left(\epsilon_{il},N_{l}\right)|\boldsymbol{X}_{l},N_{l}\right)},
\]
which implies that $B_{il}=\varUpsilon\left(\boldsymbol{X}_{l}\right)\widetilde{B}_{il}$,
for $i=1,...,N_{l}$ and $l=1,...,L$.

Now, we write 
\begin{equation}
\mathrm{log}\left(B_{il}\right)=\alpha\left(N_{l}\right)+\mathrm{log}\left(\varUpsilon\left(\boldsymbol{X}_{l}\right)\right)+U_{il},\label{eq:log bid linear regression}
\end{equation}
where 
\[
\alpha\left(N_{l}\right)\coloneqq\mathrm{E}\left[\mathrm{log}\left(s\left(\epsilon_{il},N_{l}\right)\right)|N_{l}\right]\textrm{ and \ensuremath{U_{il}\coloneqq}\ensuremath{\mathrm{log}\left(s\left(\epsilon_{il},N_{l}\right)\right)}}-\alpha\left(N_{l}\right).
\]
It is easy to check that $\mathrm{E}\left[U_{il}|\boldsymbol{X}_{l},N_{l}\right]=0$.
Since $N_{l}$ is discrete, we can write 
\[
\alpha\left(N_{l}\right)=\sum_{n=\underline{n}}^{\overline{n}}\alpha_{n}\mathbbm{1}\left(N_{l}=n\right).
\]
We assume that the function $\varUpsilon$ is log-linear in parameters:
$\mathrm{log}\left(\varUpsilon\left(\boldsymbol{X}_{l}\right)\right)=\boldsymbol{X}_{l}^{\mathrm{T}}\boldsymbol{\beta}$
for some unknown $\boldsymbol{\beta}$. Now (\ref{eq:log bid linear regression})
can be written as 
\[
\mathrm{log}\left(B_{il}\right)=\sum_{n=\underline{n}}^{\overline{n}}\alpha_{n}\mathbbm{1}\left(N_{l}=n\right)+\boldsymbol{X}_{l}^{\mathrm{T}}\boldsymbol{\beta}+U_{il}.
\]

Regressing the log-bids on the covariates and the indicators for the
number of bidders yields an estimator $\widehat{\boldsymbol{\beta}}$
of $\boldsymbol{\beta}$. Then, the homogenized bids are given by
\begin{equation}
B_{il}^{0}\coloneqq\mathrm{exp}\left(\mathrm{log}\left(B_{il}\right)+\boldsymbol{x}_{0}^{\mathrm{T}}\widehat{\boldsymbol{\beta}}-\boldsymbol{X}_{l}^{\mathrm{T}}\widehat{\boldsymbol{\beta}}\right),\textrm{ for \ensuremath{i=1,...,N_{l}} and \ensuremath{l=1,...,L}}.\label{eq:homogenized bids}
\end{equation}
These bids can be interpreted as the bid that would have been submitted
by the $i$-th bidder if the covariates were equal to $\boldsymbol{x}_{0}$,
in the $l$-th auction. Suppose we are interested in inference on
$f\left(\cdot|\boldsymbol{x}_{0}\right)$ for some fixed $\boldsymbol{x}_{0}$.\footnote{For example, $\boldsymbol{x}_{0}$ can be taken to be the sample mean
$L^{-1}\sum_{l=1}^{L}\boldsymbol{X}_{l}$. See \citet{haile2003nonparametric}.} 

Let 
\begin{gather*}
\widehat{G}\left(b,n\right)\coloneqq\frac{1}{L}\sum_{l=1}^{L}\mathbbm{1}\left(N_{l}=n\right)\frac{1}{N_{l}}\sum_{i=1}^{N_{l}}\mathbbm{1}\left(B_{il}^{0}\leq b\right),\\
\widehat{g}\left(b,n\right)\coloneqq\frac{1}{L}\sum_{l=1}^{L}\mathbbm{1}\left(N_{l}=n\right)\frac{1}{N_{l}}\sum_{i=1}^{N_{l}}\frac{1}{h_{g}}K_{g}\left(\frac{B_{il}^{0}-b}{h_{g}}\right)\textrm{ and}\\
\widehat{\xi}\left(b,n\right)\coloneqq b+\frac{1}{n-1}\frac{\widehat{G}\left(b,n\right)}{\widehat{g}\left(b,n\right)}.
\end{gather*}
 $\widehat{s}\left(\cdot,n\right)$ can also be defined analogously
and $\widehat{s}^{-1}\left(\cdot,n\right)$ is its pseudo inverse.
Let $\widehat{V}_{il}^{0\dagger}\coloneqq\widehat{s}^{-1}\left(B_{il}^{0},N_{l}\right)$
be the monotonicity-constrained pseudo valuations.

A semi-parametric estimator of $f\left(\cdot|\boldsymbol{x}_{0}\right)$
is 
\[
\widehat{f}\left(v|\boldsymbol{x}_{0}\right)\coloneqq\frac{1}{L}\sum_{l=1}^{L}\frac{1}{N_{l}}\sum_{i=1}^{N_{l}}\frac{1}{h_{f}}K_{f}\left(\frac{\widehat{V}_{il}^{0\dagger}-v}{h_{f}}\right),
\]
which is asymptotically normal. Its asymptotic variance can be consistently
estimated by
\begin{eqnarray*}
\widehat{\mathrm{V}}_{RGPV}\left(v\right) & \coloneqq & \sum_{n=\underline{n}}^{\overline{n}}\frac{1}{n\left(n-1\right)^{2}h_{f}^{2}h_{g}}\\
 &  & \times\frac{1}{L\left(L-1\right)\left(L-2\right)}\sum_{l=1}^{L}\sum_{k\neq l}\sum_{k'\neq k,k'\neq l}\mathbbm{1}\left(N_{l}=n,N_{k}=n,N_{k'}=n\right)\frac{1}{N_{l}}\sum_{i=1}^{N_{l}}\eta_{il,k}\left(v\right)\eta_{il,k'}\left(v\right),
\end{eqnarray*}
where 
\begin{gather*}
\eta_{il,k}\left(v\right)\coloneqq\frac{1}{N_{k}}\sum_{j=1}^{N_{k}}K_{f}'\left(\frac{\widehat{V}_{jk}^{0\dagger}-v}{h_{f}}\right)\frac{1}{\widehat{s}'\left(\widehat{V}_{jk}^{0\dagger},N_{k}\right)}\int_{\widehat{\underline{b}}_{N_{k}}^{0}}^{\widehat{\overline{b}}_{N_{k}}^{0}}\frac{1}{h_{r}}K_{r}\left(\frac{\widehat{V}_{jk}^{0\dagger}-\widehat{\xi}\left(u,N_{k}\right)}{h_{r}}\right)\frac{\widehat{G}\left(u,N_{k}\right)}{\widehat{g}\left(u,N_{k}\right)^{2}}K_{g}\left(\frac{B_{jk}^{0}-u}{h_{g}}\right)\mathrm{d}u,\\
\widehat{s}'\left(v,n\right)\coloneqq\int_{\widehat{\underline{b}}_{n}^{0}}^{\widehat{\overline{b}}_{n}^{0}}\frac{1}{h_{r}}K_{r}\left(\frac{\widehat{\xi}\left(b,n\right)-v}{h_{r}}\right)\mathrm{d}b
\end{gather*}
and 
\[
\widehat{\overline{b}}_{n}^{0}\coloneqq\underset{\left(i,l\right):N_{l}=n}{\mathrm{max}}B_{il}^{0}\textrm{ and \ensuremath{\widehat{\underline{b}}_{n}^{0}\coloneqq\underset{\left(i,l\right):N_{l}=n}{\mathrm{min}}B_{il}^{0}}}.
\]

A pointwise confidence interval for $f\left(v|\boldsymbol{x}_{0}\right)$
that is of the same form as (\ref{eq:RGPV confidence interval}) can
be proved to be asymptotically valid. For bootstrap resampling, we
treat the homogenized bids (\ref{eq:homogenized bids}) as our observed
bids and apply the two-step resampling procedure provided by \citet{Marmer_Shneyerov_Quantile_Auctions}.
In each bootstrap replication, we first randomly draw $L$ observations
from $\left\{ N_{l}:l=1,...,L\right\} $ with replacement. Next, we
randomly draw bids with replacement from bids corresponding to the
selected number of bidders. If for the $l$-th observation in the
bootstrap sample, we have $N_{l}^{*}=N_{l'}$, then let $\left\{ B_{il}^{0*}:i=1,...,N_{l}^{*}\right\} $
be i.i.d. draws from all the bids in auctions with number of bidders
being $N_{l'}$ with replacement. Now the bootstrap sample is $\left(B_{il}^{0*},N_{l}^{*}\right)$,
$i=1,...,N_{l}^{*}$ and $l=1,...,L$. Then it is straightforward
to construct the bootstrap analogue of $\widehat{f}\left(\cdot|\boldsymbol{x}_{0}\right)$
and a bootstrap-based uniform confidence band for $f\left(\cdot|\boldsymbol{x}_{0}\right)$
can be constructed analogously. 

\section{Monte Carlo Simulations\label{sec:Monte-Carlo-Simulations}}

In this section, we assess the finite-sample performances of the uniform
confidence bands based on both the unconstrained GPV estimator and
the rearrangement-based monotonicity-constrained estimator proposed
in this paper. Our simulation design follows \citet{Marmer_Shneyerov_Quantile_Auctions}
and the DGP is described in Remark \ref{numerical }. We consider
$\theta=1$ and draw 2100 independent valuations from $F_{\theta}$.
In all these cases, the number of bidders $N$ is constant. The number
of auctions is determined by $N\cdot L=2100$. We choose $K_{r}=K_{f}$
to be the second-order triweight kernel. For estimation of the inverse
bidding strategy, we use the second-order triweight kernel in the
MCE. In this case, the kernel $K_{g}$ in the expressions of the asymptotic
variances is the fourth-order triweight kernel. 

We use the same bandwidths as in GPV. We take $h_{g}=3.72\cdot\widehat{\sigma}_{b}\cdot\left(N\cdot L\right)^{-1/5}$
when estimating the inverse bidding strategy. $\widehat{\sigma}_{b}$
is the estimated standard deviation of the observed bids. We use $h_{f}=3.15\cdot\widehat{\sigma}_{v}\cdot\left(N\cdot L\right)^{-1/5}$
as the second-step bandwidth, where $\widehat{\sigma}_{v}$ is the
estimated standard deviation of the (unconstrained) pseudo valuations.
The constants $3.72$ and $3.15$ are Silverman's rule-of-thumb constants
corresponding to fourth-order and second-order triweight kernels.
When imposing monotonicity, we take $h_{r}=h_{f}$. We consider different
numbers of bidders $N\in\left\{ 3,5,7\right\} $, and also the density
function over the ranges $v\in\left[0.2,0.8\right]$ and $v\in\left[0.3,0.7\right]$.
When computing the bootstrap-based critical values, we set the number
of bootstrap replications to 499 and use grid maximization over the
grid $\left[v_{l}:0.001:v_{u}\right]$.\footnote{We also tried a finer grid $\left[v_{l}:0.0001:v_{u}\right]$, which
produced similar results.}

The main result of this paper is that imposing monotonicity using
smooth rearrangement results in more efficient inference. Therefore,
in addition to assessing coverage accuracy, we also report the ratio
of the supremum widths of the confidence bands: 
\[
W_{GPV}\coloneqq\underset{v\in I}{\mathrm{sup}}\,2\cdot\zeta_{GPV,\alpha}\sqrt{\frac{\widehat{\mathrm{V}}_{GPV}\left(v\right)}{Lh_{f}^{2}h_{g}}}\textrm{ and }W_{RGPV}\coloneqq\underset{v\in I}{\mathrm{sup}}\,2\cdot\zeta_{RGPV,\alpha}\sqrt{\frac{\widehat{\mathrm{V}}_{RGPV}\left(v\right)}{Lh_{f}^{2}h_{g}}}.
\]

\begin{table}[t]
\caption{Coverage probabilities and the relative supremum width $W_{GPV}/W_{RPGV}$
of the bootstrap-based uniform confidence bands around the GPV and
rearrangement estimators, $\theta=1$\label{tab:Results_1}}

\centering{}%
\begin{tabular}{>{\centering}p{1.3cm}>{\centering}p{1.3cm}>{\centering}p{1.3cm}>{\centering}p{1.3cm}>{\centering}p{0.15cm}>{\centering}p{1.2cm}>{\centering}p{1.2cm}>{\centering}p{1.2cm}>{\centering}p{0.15cm}>{\centering}p{1.3cm}>{\centering}p{1.3cm}>{\centering}p{1.3cm}}
\toprule 
 & \multicolumn{3}{c}{GPV} &  & \multicolumn{3}{c}{Rearrangement} &  & \multicolumn{3}{c}{$W_{GPV}/W_{RPGV}$}\tabularnewline
 & 0.90 & 0.95 & 0.99 &  & 0.90 & 0.95 & 0.99 &  & 0.90 & 0.95 & 0.99\tabularnewline
\midrule
\uline{\mbox{$v\in[0.3,0.7]$}} & \multicolumn{3}{c}{} &  & \tabularnewline
$N=3$ & 0.908 & 0.956 & 0.990 &  & 0.918 & 0.956 & 0.992 &  & 1.334 & 1.326 & 1.315\tabularnewline
$N=5$ & 0.882 & 0.940 & 0.996 &  & 0.888 & 0.936 & 0.994 &  & 1.099 & 1.094 & 1.093\tabularnewline
$N=7$ & 0.890 & 0.946 & 0.988 &  & 0.872 & 0.942 & 0.986 &  & 1.069 & 1.068 & 1.072\tabularnewline
 &  &  &  &  &  &  &  &  &  &  & \tabularnewline
\uline{\mbox{$v\in[0.2,0.8]$}} & \multicolumn{3}{c}{} &  & \tabularnewline
$N=3$ & 0.910 & 0.948 & 0.990 &  & 0.882 & 0.938 & 0.990 &  & 1.316 & 1.305 & 1.288\tabularnewline
$N=5$ & 0.918 & 0.966 & 0.996 &  & 0.906 & 0.956 & 0.992 &  & 1.132 & 1.133 & 1.132\tabularnewline
$N=7$ & 0.900 & 0.944 & 0.984 &  & 0.900 & 0.952 & 0.986 &  & 1.129 & 1.132 & 1.132\tabularnewline
\bottomrule
\end{tabular}
\end{table}

Table \ref{tab:Results_1} reports the coverage probabilities as well
as the relative supremum width of the confidence bands based on the
GPV and our rearrangement estimators. The coverage probabilities of
both methods are similar and accurate. However, our approach can produce
considerably smaller confidence with the reduction in supremum width
ranging from 6.8\% to 33.4\%. More substantial reductions in supremum
width are obtained for smaller numbers of bidders. This is due to
the inverse relationship between the number of bidders and the asymptotic
variance of the estimators.

\section{Conclusion}

The GPV nonparametric identification and estimation approach is an
indispensable working tool in structural econometrics of auctions.
This paper contributes to the literature by showing how one can reduce
the asymptotic variance of GPV-type estimators by incorporating the
monotonicity constraint in estimation. While monotonicity-constrained
estimators have been previously considered in the auction literature,
to the best of our knowledge ours is the first to obtain a reduction
in asymptotic variance in this context. Our method is simple to implement,
and as a by-product, it also produces a simple estimator for the bidding
function. We also discuss construction of uniform confidence bands
for the density of valuations. In a simulation study, we show that
by applying our approach one can increase the precision of the confidence
bands without sacrificing their coverage in finite samples. 

\bibliographystyle{chicago}
\bibliography{\string~/Dropbox/VGPV/Revision/vgpv_art,btw_selection,\string~/Dropbox/VGPV/Revision/Revision_V12}

\appendix

\part*{Appendix}

Let $\apprle$ denote an inequality up to a universal constant that
does not depend on the sample size $L$. For a sequence of classes
of functions $\mathscr{F}_{L}$ (that may depend on the sample size)
defined on $\left[\underline{b},\overline{b}\right]^{d}$, for some
$d\geq1$, let $N\left(\epsilon,\mathscr{F}_{L},\left\Vert \cdot\right\Vert _{Q,2}\right)$
denote the $\epsilon-$covering number, i.e., the smallest integer
$m$ such that there are $m$ balls of radius $\epsilon$ centered
at points in $\mathscr{F}_{L}$, with respect to the metric induced
by the norm $\left\Vert \cdot\right\Vert _{Q,2}$, where $\left\Vert f\right\Vert _{Q,2}\coloneqq\left(\int\left|f\right|^{2}\mathrm{d}Q\right)^{1/2}$,
$f\in\mathscr{F}_{L}$. A function $F_{L}:\left[\underline{b},\overline{b}\right]^{d}\rightarrow\mathbb{R}_{+}$
is an envelope of $\mathscr{F}_{L}$ if $F_{L}\geq\underset{f\in\mathscr{F}_{L}}{\mathrm{sup}}\left|f\right|$.
We say that $\mathscr{F}_{L}$ is a (uniform) Vapnik-Chervonenkis-type
(VC-type) class with respect to the envelope $F_{L}$ (see, e.g.,
\citealp[Definition 2.1]{chernozhukov2014gaussian}) if there exist
some positive constant $C_{1}$ and $C_{2}$ that are independent
of $L$ such that
\[
N\left(\epsilon\left\Vert F_{L}\right\Vert _{Q,2},\mathscr{F}_{L},\left\Vert \cdot\right\Vert _{Q,2}\right)\leq\left(\frac{C_{1}}{\epsilon}\right)^{C_{2}},\,\textrm{for all \ensuremath{\epsilon\in\left(0,1\right]},}
\]
for all finitely discrete probability measure $Q$ on $\left[\underline{b},\overline{b}\right]^{d}$.
Note that the all function classes that appear later are dependent
on $L$. We suppress the dependence for notational simplicity.

``With probability approaching 1'' is abbreviated as ``w.p.a.1''.
For notational simplicity, in the proofs, $\underset{i,l}{\mathrm{max}}$
is understood as $\underset{\left(i,l\right)\in\left\{ 1,...,N\right\} \times\left\{ 1,...,L\right\} }{\mathrm{max}}$.
$\underset{\left(2\right)}{\sum}$ is understood as $\sum_{\left(j,k\right)\neq\left(i,l\right)}$
and $\underset{\left(3\right)}{\sum}$ is understood as 
\[
\sum_{i,l}\sum_{\left(j,k\right)\neq\left(i,l\right)}\sum_{\left(j',k'\right)\neq\left(i,l\right),\,\left(j',k'\right)\neq\left(j,k\right)},
\]
i.e., summing over all distinct indices. $\left(N\cdot L\right)_{2}$
is understood as $\left(N\cdot L\right)\left(N\cdot L-1\right)$ and
$\left(N\cdot L\right)_{3}$ is understood as $\left(N\cdot L\right)\left(N\cdot L-1\right)\left(N\cdot L-2\right)$. 

Fix 
\[
\delta_{0}\coloneqq\mathrm{min}\left\{ \left(\overline{v}-v_{u}\right)/2,\left(v_{l}-\underline{v}\right)/2\right\} .
\]

\section{Proofs of the Main Results}

\begin{proof}[Proof of Theorem \ref{thm:normality}]It follows from
Lemma \ref{lem:Lemma 7} that
\[
\left(Lh_{f}^{2}h_{g}\right)^{1/2}\left(\widehat{f}_{GPV}\left(v\right)-f\left(v\right)\right)=\frac{1}{N^{1/2}\left(N-1\right)}\frac{1}{\left(N\cdot L\right)^{1/2}}\sum_{i,l}h_{f}h_{g}^{1/2}\left(\mathcal{M}_{2}\left(B_{il};v\right)-\mu_{\mathcal{M}}\left(v\right)\right)+o_{p}\left(1\right).
\]

Denote 
\begin{gather*}
\mathcal{M}^{\ddagger}\left(b',b;v\right)\coloneqq-\frac{1}{h_{f}^{2}}K_{f}'\left(\frac{\xi\left(b'\right)-v}{h_{f}}\right)\xi'\left(b'\right)\int_{\underline{b}}^{\overline{b}}\frac{1}{h_{r}}K_{r}\left(\frac{\xi\left(b'\right)-\xi\left(u\right)}{h_{r}}\right)\frac{G\left(u\right)}{g\left(u\right)^{2}}\frac{1}{h_{g}}K_{g}\left(\frac{b-u}{h_{g}}\right)\mathrm{d}u\\
\mathcal{M}_{2}^{\ddagger}\left(b;v\right)\coloneqq-\int\frac{1}{h_{f}^{2}}K_{f}'\left(\frac{\xi\left(b'\right)-v}{h_{f}}\right)\xi'\left(b'\right)\int_{\underline{b}}^{\overline{b}}\frac{1}{h_{r}}K_{r}\left(\frac{\xi\left(b'\right)-\xi\left(u\right)}{h_{r}}\right)\frac{G\left(u\right)}{g\left(u\right)^{2}}\frac{1}{h_{g}}K_{g}\left(\frac{b-u}{h_{g}}\right)\mathrm{d}u\mathrm{d}G\left(b'\right)\\
\textrm{and }\mu_{\mathcal{M}^{\ddagger}}\left(v\right)\coloneqq\int\int\mathcal{M}^{\ddagger}\left(b,b';v\right)\mathrm{d}G\left(b\right)\mathrm{d}G\left(b'\right)=\int\mathcal{M}_{2}^{\ddagger}\left(b;v\right)\mathrm{d}G\left(b\right).
\end{gather*}
Then it is clear that 
\[
\mathcal{M}_{2}\left(B_{il};v\right)-\mu_{\mathcal{M}}\left(v\right)=\mathcal{M}_{2}^{\ddagger}\left(B_{il};v\right)-\mu_{\mathcal{M}^{\ddagger}}\left(v\right)\textrm{ for all \ensuremath{i=1,...,N} and \ensuremath{L=1,...,L}}
\]
and 
\[
\mu_{\mathcal{M}^{\ddagger}}\left(v\right)=\mu_{\mathcal{M}}\left(v\right)-\int\int_{\underline{b}}^{\overline{b}}\frac{1}{h_{f}^{2}}K_{f}'\left(\frac{\xi\left(b'\right)-v}{h_{f}}\right)\xi'\left(b'\right)\frac{1}{h_{f}}K_{r}\left(\frac{\xi\left(b'\right)-\xi\left(u\right)}{h_{f}}\right)\frac{G\left(u\right)}{g\left(u\right)}\mathrm{d}u\mathrm{d}G\left(b'\right).
\]

Note that we assumed $h_{r}=h_{f}$. By change of variables,
\begin{align*}
 & \int\int_{\underline{b}}^{\overline{b}}K_{f}'\left(\frac{\xi\left(b'\right)-v}{h_{f}}\right)\xi'\left(b'\right)\frac{1}{h_{f}}K_{r}\left(\frac{\xi\left(b'\right)-\xi\left(u\right)}{h_{f}}\right)\frac{G\left(u\right)}{g\left(u\right)}\mathrm{d}u\mathrm{d}G\left(b'\right)\\
= & h_{f}\int_{\frac{\underline{v}-v}{h_{f}}}^{\frac{\overline{v}-v}{h_{f}}}\int_{\frac{\underline{v}-v}{h_{f}}}^{\frac{\overline{v}-v}{h_{f}}}K_{f}'\left(w\right)K_{r}\left(w-z\right)\frac{G\left(s\left(h_{f}z+v\right)\right)}{g\left(s\left(h_{f}z+v\right)\right)}s'\left(h_{f}z+v\right)g\left(s\left(h_{f}w+v\right)\right)\mathrm{d}z\mathrm{d}w\\
= & h_{f}\int_{\frac{\underline{v}-v}{h_{f}}}^{\frac{\overline{v}-v}{h_{f}}}\int_{\frac{\underline{v}-v}{h_{f}}}^{\frac{\overline{v}-v}{h_{f}}}K_{f}'\left(w\right)K_{r}\left(w-z\right)\left\{ \frac{G\left(s\left(v\right)\right)s'\left(v\right)}{g\left(s\left(v\right)\right)}+\frac{\left\{ g\left(s\left(\ddot{v}\right)\right)s'\left(\ddot{v}\right)^{2}+G\left(s\left(\ddot{v}\right)\right)s''\left(\ddot{v}\right)\right\} g\left(s\left(\ddot{v}\right)\right)}{g\left(s\left(\ddot{v}\right)\right)^{2}}h_{f}z\right.\\
 & \left.-\frac{G\left(s\left(\ddot{v}\right)\right)s'\left(\ddot{v}\right)^{2}g'\left(s\left(\ddot{v}\right)\right)}{g\left(s\left(\ddot{v}\right)\right)^{2}}h_{f}z\right\} \left\{ g\left(s\left(v\right)\right)+g'\left(s\left(\dot{v}\right)\right)s'\left(\dot{v}\right)h_{f}w\right\} \mathrm{d}z\mathrm{d}w,
\end{align*}
where $\dot{v}$ and $\ddot{v}$ are mean values that are dependent
on $w$ and $z$ with $\left|\dot{v}-v\right|\leq h_{f}\left|w\right|$
and $\left|\ddot{v}-v\right|\leq h_{f}\left|z\right|$.

Since when $h_{f}$ is small enough,
\[
\int_{\frac{\underline{v}-v}{h_{f}}}^{\frac{\overline{v}-v}{h_{f}}}\int_{\frac{\underline{v}-v}{h_{f}}}^{\frac{\overline{v}-v}{h_{f}}}K_{f}'\left(w\right)K_{r}\left(w-z\right)\mathrm{d}z\mathrm{d}w=0,
\]
it follows that
\[
\int\int_{\underline{b}}^{\overline{b}}K_{f}'\left(\frac{\xi\left(b'\right)-v}{h_{f}}\right)\xi'\left(b'\right)\frac{1}{h_{f}}K_{r}\left(\frac{\xi\left(b'\right)-\xi\left(u\right)}{h_{f}}\right)\frac{G\left(u\right)}{g\left(u\right)}\mathrm{d}u\mathrm{d}G\left(b'\right)=O\left(h^{2}\right),
\]
uniformly in $v\in I$. $\underset{v\in I}{\mathrm{sup}}\left|\mu_{\mathcal{M}^{\ddagger}}\left(v\right)\right|=O\left(1\right)$
follows from this result and Lemma \ref{lem:Lemma 6}.

Define
\[
U_{il}\left(v\right)\coloneqq\frac{1}{N^{1/2}\left(N-1\right)}\frac{1}{\left(N\cdot L\right)^{1/2}}h_{f}h_{g}^{1/2}\left(\mathcal{M}_{2}^{\ddagger}\left(B_{il};v\right)-\mu_{\mathcal{M}^{\ddagger}}\left(v\right)\right),
\]
and
\begin{equation}
\sigma\left(v\right)\coloneqq\left(\sum_{i,l}\mathrm{E}\left[U_{il}\left(v\right)^{2}\right]\right)^{1/2}=\left(\frac{1}{N\left(N-1\right)^{2}}h_{f}^{2}h_{g}\mathrm{E}\left[\left(\mathcal{M}_{2}^{\ddagger}\left(B_{11};v\right)-\mu_{\mathcal{M}^{\ddagger}}\left(v\right)\right)^{2}\right]\right)^{\nicefrac{1}{2}}.\label{eq:sigma_L definition}
\end{equation}

By change of variables,
\begin{eqnarray*}
\mathrm{E}\left[h_{f}^{2}h_{g}\mathcal{M}_{2}^{\ddagger}\left(B_{11};v\right)^{2}\right] & = & \int\frac{1}{h_{g}}\left\{ \int_{\underline{b}}^{\overline{b}}\int_{\underline{b}}^{\overline{b}}\frac{1}{h_{f}}K'\left(\frac{\xi\left(b'\right)-v}{h_{f}}\right)\frac{1}{s'\left(\xi\left(b'\right)\right)}\frac{1}{h_{f}}K\left(\frac{\xi\left(b'\right)-\xi\left(u\right)}{h_{f}}\right)\frac{G\left(u\right)}{g\left(u\right)^{2}}\right.\\
 &  & \left.\times K\left(\frac{b-u}{h_{g}}\right)g\left(b'\right)\mathrm{d}u\mathrm{d}b'\right\} ^{2}\mathrm{d}G\left(b\right)\\
 & = & \int_{\frac{\underline{b}-s\left(v\right)}{h_{g}}}^{\frac{\overline{b}-s\left(v\right)}{h_{g}}}\rho_{h}\left(y\right)^{2}g\left(h_{g}y+s\left(v\right)\right)\mathrm{d}y.
\end{eqnarray*}
where
\begin{eqnarray}
\rho_{h}\left(y\right) & \coloneqq & \int_{\frac{\underline{v}-v}{h_{f}}}^{\frac{\overline{v}-v}{h_{f}}}\int_{\frac{\underline{v}-v}{h_{f}}}^{\frac{\overline{v}-v}{h_{f}}}K_{f}'\left(w\right)K_{r}\left(w-z\right)\frac{G\left(s\left(h_{f}z+v\right)\right)}{g\left(s\left(h_{f}z+v\right)\right)^{2}}K_{g}\left(y-\frac{s\left(h_{f}z+v\right)-s\left(v\right)}{h_{g}}\right)\nonumber \\
 &  & \times s'\left(h_{f}z+v\right)g\left(s\left(h_{f}w+v\right)\right)\mathrm{d}z\mathrm{d}w.\label{eq:rho_h definition}
\end{eqnarray}
We can easily verify that the dominated convergence theorem applies
to the integral on the right hand side of (\ref{eq:rho_h definition})
(as $h\downarrow0$) for each $y\in\mathbb{R}$:
\[
\underset{h\downarrow0}{\mathrm{lim}}\rho_{h}\left(y\right)=\frac{G\left(s\left(v\right)\right)s'\left(v\right)}{g\left(s\left(v\right)\right)}\int\int K_{f}'\left(w\right)K_{r}\left(w-z\right)K_{g}\left(y-s'\left(v\right)\frac{\lambda_{f}}{\lambda_{g}}z\right)\mathrm{d}z\mathrm{d}w.
\]
Since $K_{f}'$ and $K_{r}$ are compactly supported on $\left[-1,1\right]$,
by reverse triangle inequality, 
\begin{eqnarray*}
\left|\rho_{h}\left(y\right)\right| & \apprle & \int_{\frac{\underline{v}-v}{h_{f}}}^{\frac{\overline{v}-v}{h_{f}}}\int_{\frac{\underline{v}-v}{h_{f}}}^{\frac{\overline{v}-v}{h_{f}}}\left|K'\left(w\right)\right|\left|\mathbbm{1}\left(\left|z\right|\leq1+\left|w\right|\right)\right|\mathbbm{1}\left(\left|y\right|\leq1+\left|\frac{s\left(h_{f}z+v\right)-s\left(v\right)}{h_{g}}\right|\right)\mathrm{d}z\mathrm{d}w\\
 & \apprle & \mathbbm{1}\left(\left|y\right|\leq1+2\frac{\lambda_{f}}{\lambda_{g}}\overline{C}_{s'}\right),
\end{eqnarray*}
where $\overline{C}_{s'}\coloneqq\underset{v\in\left[\underline{v},\overline{v}\right]}{\mathrm{sup}}s'\left(v\right)$. 

Now the dominated convergence theorem applies and 
\begin{align*}
 & \underset{h\downarrow0}{\mathrm{lim}}\int_{\frac{\underline{b}-s\left(v\right)}{h_{g}}}^{\frac{\overline{b}-s\left(v\right)}{h_{g}}}\rho_{h}\left(y\right)^{2}g\left(h_{g}y+s\left(v\right)\right)\mathrm{d}y\\
= & \frac{G\left(s\left(v\right)\right)^{2}s'\left(v\right)^{2}}{g\left(s\left(v\right)\right)}\int\left\{ \int\int K'\left(w\right)K\left(w-z\right)K\left(y-s'\left(v\right)\frac{\lambda_{f}}{\lambda_{g}}z\right)\mathrm{d}z\mathrm{d}w\right\} ^{2}\mathrm{d}y\\
= & \frac{F\left(v\right)^{2}f\left(v\right)^{2}}{g\left(s\left(v\right)\right)^{3}}\int\left\{ \int\int K'\left(w\right)K\left(w-z\right)K\left(y-s'\left(v\right)\frac{\lambda_{f}}{\lambda_{g}}z\right)\mathrm{d}z\mathrm{d}w\right\} ^{2}\mathrm{d}y
\end{align*}
where the second equality follows from the relation $f\left(v\right)=g\left(s\left(v\right)\right)s'\left(v\right)$.
It follows from this result and $\underset{v\in I}{\mathrm{sup}}\left|\mu_{\mathcal{M}^{\ddagger}}\left(v\right)\right|=O\left(1\right)$
that
\begin{align}
 & \mathrm{E}\left[h_{f}^{2}h_{g}\left(\mathcal{M}_{2}^{\ddagger}\left(B_{11};v\right)-\mu_{\mathcal{M}^{\ddagger}}\left(v\right)\right)^{2}\right]\nonumber \\
= & \mathrm{E}\left[h_{f}^{2}h_{g}\mathcal{M}_{2}^{\ddagger}\left(B_{11};v\right)^{2}\right]-h_{f}^{2}h_{g}\mu_{\mathcal{M}^{\ddagger}}\left(v\right)^{2}\nonumber \\
= & \frac{F\left(v\right)^{2}f\left(v\right)^{2}}{g\left(s\left(v\right)\right)^{3}}\int\left\{ \int\int K_{f}'\left(w\right)K_{r}\left(w-z\right)K_{g}\left(y-s'\left(v\right)\frac{\lambda_{f}}{\lambda_{g}}z\right)\mathrm{d}z\mathrm{d}w\right\} ^{2}\mathrm{d}y+o\left(1\right)\label{eq:Em_2^2}
\end{align}
and thus 
\begin{equation}
\sigma\left(v\right)=\frac{1}{N^{1/2}\left(N-1\right)}\left\{ \frac{F\left(v\right)^{2}f\left(v\right)^{2}}{g\left(s\left(v\right)\right)^{3}}\int\left\{ \int\int K_{f}'\left(w\right)K_{r}\left(w-z\right)K_{g}\left(y-s'\left(v\right)\frac{\lambda_{f}}{\lambda_{g}}z\right)\mathrm{d}z\mathrm{d}w\right\} ^{2}\mathrm{d}y\right\} ^{1/2}+o\left(1\right).\footnotemark\label{eq:s_L limit}
\end{equation}
\footnotetext{It is easy to adapt the proofs to show that the remainder terms are indeed uniform in $v\in I$. See the proof of Theorem 2.1 of MMS.}

By the $c_{r}$ inequality (see, e.g., \citealp[9.28]{Davidson_Stochastic_Limit}),
we have
\begin{eqnarray}
\sum_{i,l}\mathrm{E}\left[\left|\frac{U_{il}\left(v\right)}{\sigma\left(v\right)}\right|^{3}\right] & = & \sigma\left(v\right)^{-3}\left(N-1\right)^{-3}\left(N\cdot L\right)^{-1/2}\mathrm{E}\left[h_{f}^{3}h_{g}^{3/2}\left|\left(\mathcal{M}_{2}^{\ddagger}\left(B_{11};v\right)-\mu_{\mathcal{M}^{\ddagger}}\left(v\right)\right)\right|^{3}\right]\nonumber \\
 & \apprle & \sigma\left(v\right)^{-3}\left(N\cdot L\right)^{-1/2}\left(h_{f}^{3}h_{g}^{3/2}\mathrm{E}\left[\left|\mathcal{M}_{2}^{\ddagger}\left(B_{11};v\right)\right|^{3}\right]+h_{f}^{3}h_{g}^{3/2}\left|\mu_{\mathcal{M}^{\ddagger}}\left(v\right)\right|^{3}\right).\label{eq:Liapunov's condition sum}
\end{eqnarray}
Then it is easy to verify that the Lyapunov's condition holds:
\begin{equation}
\underset{L\uparrow\infty}{\mathrm{lim}}\sum_{i,l}\mathrm{E}\left[\left|\frac{U_{il}\left(v\right)}{\sigma\left(v\right)}\right|^{3}\right]=0.\label{eq:Liapunov's condition}
\end{equation}
By Lyapunov's central limit theorem,
\begin{equation}
\sum_{i,l}\frac{U_{il}\left(v\right)}{\sigma\left(v\right)}\rightarrow_{d}\mathrm{N}\left(0,1\right),\textrm{ as \ensuremath{L\uparrow\infty}}.\label{eq:convergence of sum of U_il}
\end{equation}

For the second part, it suffices to show 
\[
\int\left\{ \int\int K_{f}'\left(u\right)K_{r}\left(u-z\right)K_{g}\left(w-\frac{\lambda_{f}}{\lambda_{g}}s'\left(v\right)z\right)\mathrm{d}z\mathrm{d}u\right\} ^{2}\mathrm{d}w\leq\int\left\{ \int K_{f}'\left(u\right)K_{g}\left(w-s'\left(v\right)\frac{\lambda_{f}}{\lambda_{g}}u\right)\mathrm{d}u\right\} ^{2}\mathrm{d}w.
\]
For each $w\in\mathbb{R}$, since $K_{r}$ and $K_{f}'$ are assumed
to be bounded and compactly supported, the Fubini-Tonelli theorem
applies and therefore,
\[
\int\int K_{f}'\left(u\right)K_{r}\left(u-z\right)K_{g}\left(w-\frac{\lambda_{f}}{\lambda_{g}}s'\left(v\right)z\right)\mathrm{d}z\mathrm{d}u=\int\left(\int K_{f}'\left(u\right)K_{r}\left(u-z\right)\mathrm{d}u\right)K_{g}\left(w-\frac{\lambda_{f}}{\lambda_{g}}s'\left(v\right)z\right)\mathrm{d}z.
\]
Since $K_{r}$ and $K_{f}'$ are supported on $\left[-1,1\right]$,
by integration by parts,
\begin{equation}
\int_{-1}^{1}K_{f}'\left(u\right)K_{r}\left(u-z\right)\mathrm{d}u-\int_{-1}^{1}K_{f}\left(u\right)K_{r}'\left(u-z\right)\mathrm{d}u=0.\label{eq:theorem 2 part b inter_1}
\end{equation}

Now,
\begin{eqnarray}
\left\{ \int\int K_{f}'\left(u\right)K_{r}\left(u-z\right)K_{g}\left(w-\frac{\lambda_{f}}{\lambda_{g}}s'\left(v\right)z\right)\mathrm{d}z\mathrm{d}u\right\} ^{2} & = & \left\{ \int\left(\int K_{f}\left(u\right)K_{r}'\left(u-z\right)\mathrm{d}u\right)K_{g}\left(w-\frac{\lambda_{f}}{\lambda_{g}}s'\left(v\right)z\right)\mathrm{d}z\right\} ^{2}\nonumber \\
 & = & \left\{ \int\int K_{f}\left(u\right)K_{r}'\left(u-z\right)K_{g}\left(w-\frac{\lambda_{f}}{\lambda_{g}}s'\left(v\right)z\right)\mathrm{d}z\mathrm{d}u\right\} ^{2}\nonumber \\
 & \leq & \int K_{f}\left(u\right)\left\{ \int K_{r}'\left(u-z\right)K_{g}\left(w-\frac{\lambda_{f}}{\lambda_{g}}s'\left(v\right)z\right)\mathrm{d}z\right\} ^{2}\mathrm{d}u,\label{eq:theorem 2 part b inter_2}
\end{eqnarray}
for each $w\in\mathbb{R}$, where the first equality follows from
(\ref{eq:theorem 2 part b inter_1}), the second equality follows
from the Fubini-Tonelli theorem and the inequality follows from Jensen's
inequality since $K_{f}$ is a probability density function. 

Since the inequality (\ref{eq:theorem 2 part b inter_2}) holds for
all $w\in\mathbb{R}$, by the Fubini-Tonelli theorem, 
\begin{align}
 & \int\left\{ \int\int K_{f}'\left(u\right)K_{r}\left(u-z\right)K_{g}\left(w-\frac{\lambda_{f}}{\lambda_{g}}s'\left(v\right)z\right)\mathrm{d}z\mathrm{d}u\right\} ^{2}\mathrm{d}w\nonumber \\
\leq & \int K_{f}\left(u\right)\int\left\{ \int K_{r}'\left(u-z\right)K_{g}\left(w-\frac{\lambda_{f}}{\lambda_{g}}s'\left(v\right)z\right)\mathrm{d}z\right\} ^{2}\mathrm{d}w\mathrm{d}u.\label{eq:theorem 2 part b inter_7}
\end{align}
Now for any fixed $\left(u,w\right)\in\mathbb{R}^{2}$, by change
of variables, 
\[
\int K_{r}'\left(u-z\right)K_{g}\left(w-\frac{\lambda_{f}}{\lambda_{g}}s'\left(v\right)z\right)\mathrm{d}z=-\int K_{r}'\left(y\right)K_{g}\left(w-\frac{\lambda_{f}}{\lambda_{g}}s'\left(v\right)\left(y+u\right)\right)\mathrm{d}y.
\]
Then,
\begin{align}
 & \int K_{f}\left(u\right)\int\left\{ \int K_{r}'\left(u-z\right)K_{g}\left(w-\frac{\lambda_{f}}{\lambda_{g}}s'\left(v\right)z\right)\mathrm{d}z\right\} ^{2}\mathrm{d}w\mathrm{d}u\nonumber \\
= & \int K_{f}\left(u\right)\int\left\{ \int K_{r}'\left(y\right)K_{g}\left(w-\frac{\lambda_{f}}{\lambda_{g}}s'\left(v\right)\left(y+u\right)\right)\mathrm{d}y\right\} ^{2}\mathrm{d}w\mathrm{d}u.\label{eq:theorem 2 part b inter_6}
\end{align}
It follows from change of variables that 
\[
\int\left\{ \int K_{r}'\left(y\right)K_{g}\left(w-\frac{\lambda_{f}}{\lambda_{g}}s'\left(v\right)\left(y+u\right)\right)\mathrm{d}y\right\} ^{2}\mathrm{d}w=\int\left\{ \int K_{r}'\left(y\right)K_{g}\left(w-\frac{\lambda_{f}}{\lambda_{g}}s'\left(v\right)y\right)\mathrm{d}y\right\} ^{2}\mathrm{d}w,
\]
for all $u\in\mathbb{R}$. The conclusion follows from this result,
(\ref{eq:theorem 2 part b inter_6}) and the assumption $K_{r}=K_{f}$.\end{proof}

\begin{proof}[Proof of Theorem \ref{thm:variance estimator}]The proof
is very similar to that of Theorem 3.1 of MMS. Let 

\begin{align*}
\widetilde{\mathrm{V}}\left(v\right)\coloneqq & \frac{1}{N\left(N-1\right)^{2}h_{f}h_{g}^{2}}\frac{1}{\left(N\cdot L\right)_{3}}\sum_{\left(3\right)}\mathbb{T}_{jk}\widetilde{\eta}_{il,jk}(v)\mathbb{T}_{j'k'}\widetilde{\eta}_{il,j'k'}(v),\\
\text{with }\widetilde{\eta}_{il,jk}(v)\coloneqq & K_{f}'\left(\frac{\widehat{V}_{jk}^{\dagger}-v}{h_{f}}\right)\frac{1}{s'\left(\widehat{V}_{jk}^{\dagger}\right)}\int_{\underline{b}}^{\overline{b}}\frac{1}{h_{r}}K_{r}\left(\frac{\widehat{V}_{jk}^{\dagger}-\xi\left(u\right)}{h_{r}}\right)\frac{G\left(u\right)}{g\left(u\right)^{2}}K_{g}\left(\frac{B_{il}-u}{h_{g}}\right)\mathrm{d}u
\end{align*}
and 
\begin{align*}
\overline{\mathrm{V}}\left(v\right)\coloneqq & \frac{1}{N\left(N-1\right)^{2}h_{f}h_{g}^{2}}\frac{1}{\left(N\cdot L\right)_{3}}\sum_{\left(3\right)}\overline{\eta}_{il,jk}(v)\overline{\eta}_{il,j'k'}(v),\\
\text{with }\overline{\eta}_{il,jk}(v)\coloneqq & K_{f}'\left(\frac{V_{jk}^{\dagger}-v}{h_{f}}\right)\frac{1}{s'\left(V_{jk}^{\dagger}\right)}\int_{\underline{b}}^{\overline{b}}\frac{1}{h_{r}}K_{r}\left(\frac{V_{jk}^{\dagger}-\xi\left(u\right)}{h_{r}}\right)\frac{G\left(u\right)}{g\left(u\right)^{2}}K_{g}\left(\frac{B_{il}-u}{h_{g}}\right)\mathrm{d}u.
\end{align*}

By the arguments used in the proof of Lemma \ref{lem:Lemma 3}, 
\[
\widehat{\mathrm{V}}_{RGPV}\left(v\right)=\frac{1}{N\left(N-1\right)^{2}h_{f}h_{g}^{2}}\frac{1}{\left(N\cdot L\right)_{3}}\sum_{\left(3\right)}\mathbb{T}_{jk}\widehat{\eta}_{il,jk}(v)\mathbb{T}_{j'k'}\widehat{\eta}_{il,j'k'}(v),\textrm{ for all \ensuremath{v\in I}, w.p.a.1.}
\]
By using Lemma \ref{lem:Lemma 1}, Lemma \ref{lem:Lemma 2}, (\ref{eq:V_dag - V sup rate}),
Taylor expansion, tedious algebra and empirical process techniques
invoked in the proof of Theorem 3.1 of MMS, one can show that 
\[
\underset{v\in I}{\mathrm{sup}}\left|\frac{1}{N\left(N-1\right)^{2}h_{f}h_{g}^{2}}\frac{1}{\left(N\cdot L\right)_{3}}\sum_{\left(3\right)}\mathbb{T}_{jk}\widehat{\eta}_{il,jk}(v)\mathbb{T}_{j'k'}\widehat{\eta}_{il,j'k'}(v)-\widetilde{\mathrm{V}}\left(v\right)\right|=O_{p}\left(\left(\frac{\mathrm{log}\left(L\right)}{Lh^{3}}\right)^{1/2}+h^{2}\right)
\]
and
\[
\underset{v\in I}{\mathrm{sup}}\left|\overline{\mathrm{V}}\left(v\right)-\widetilde{\mathrm{V}}\left(v\right)\right|=O_{p}\left(\left(\frac{\mathrm{log}\left(L\right)}{Lh^{3}}\right)^{1/2}+h^{2}\right).
\]

Now note that $\left\{ \overline{\mathrm{V}}\left(v\right):v\in I\right\} $
is a U process with $\mathrm{V}_{\mathcal{M}}\left(v\right)=\mathrm{E}\left[\overline{\mathrm{V}}\left(v\right)\right]$.
One can apply Hoeffding decomposition to $\overline{\mathrm{V}}\left(v\right)$
and apply techniques invoked in the proof of Theorem 3.1 of MMS from
empirical process and U process theory to derive the uniform rate
of convergence of $\overline{\mathrm{V}}\left(v\right)-\mathrm{V}_{\mathcal{M}}\left(v\right)$.
Then the conclusion follows.\end{proof}

\begin{proof}[Proof of Theorem \ref{thm:uniform confidence band}]First,
it can be verified by standard arguments that the function class $\left\{ \mathcal{M}_{2}^{\ddagger}\left(\cdot;v\right):v\in I\right\} $
is (uniformly) VC-type with respect to a constant envelope that is
a multiple of $h_{f}^{-2}h_{g}^{-1}$. Then it essentially follows
from Lemma \ref{lem:Lemma 7} and Theorem \ref{thm:variance estimator}
that the process
\begin{equation}
Z\left(v\right)\coloneqq\frac{\widehat{f}_{RGPV}\left(v\right)-f\left(v\right)}{\left(Lh_{f}^{2}h_{g}\right)^{-1/2}\widehat{\mathrm{V}}_{RGPV}\left(v\right)^{1/2}},\,v\in I\label{eq:Z definition}
\end{equation}
can be approximated by 
\[
\varGamma\left(v\right)\coloneqq\frac{1}{\left(N\cdot L\right)^{1/2}}\sum_{i,l}\frac{\mathcal{M}_{2}^{\ddagger}\left(B_{il};v\right)-\mu_{\mathcal{M}^{\ddagger}}\left(v\right)}{\mathrm{Var}\left[\mathcal{M}_{2}^{\ddagger}\left(B_{11};v\right)\right]^{1/2}},\,v\in I
\]
uniformly in $v\in I$, with an estimated rate of uniform approximation
error. See the proof of Lemma B.4 for details. 

By adapting the proofs of Lemmas B.5 - B.9 of MMS, we can show that
the bootstrap process $\left\{ Z^{*}\left(v\right):v\in I\right\} $
can be approximated by the bootstrap analogue of $\varGamma\left(v\right)$
uniformly in $v\in I$. The rest of the proof is identical to that
of Corollary 4.3 of MMS. We can show that the difference between the
distribution of $\left\Vert Z\right\Vert _{I}$ and that of $\left\Vert \varGamma_{G}\right\Vert _{I}$,
where $\left\{ \varGamma_{G}\left(v\right):v\in I\right\} $ is an
intermediate Gaussian process that has the same covariance structure
as that of $\varGamma$ , converges to zero uniformly. See Theorem
4.3 of MMS and its proof. Then it can be shown that the difference
between the distribution of $\left\Vert Z^{*}\right\Vert _{I}$ and
that of $\left\Vert \varGamma_{G}\right\Vert _{I}$ converges to zero
uniformly in the bootstrap world. See Theorem 4.4 of MMS and its proof.
The conclusion follows easily from these observations. See the proof
of Corollaries 4.2 and 4.3 of MMS.\end{proof}

\section{Lemmas}
\begin{lem}
\label{lem:Lemma 1}Suppose that Assumptions \ref{assu:DGP} - \ref{assu:bandwidth}
are satisfied. Then we have
\[
\underset{b\in\left[\underline{b},\overline{b}\right]}{\mathrm{sup}}\left|\widehat{\xi}\left(b\right)-\xi\left(b\right)\right|=O_{p}\left(\left(\frac{\mathrm{log}\left(L\right)}{Lh}\right)^{1/2}+h^{3}\right)
\]
and
\[
\underset{b\in\left[\underline{b},\overline{b}\right]}{\mathrm{sup}}\left|\widehat{\xi}\left(b\right)-\xi\left(b\right)+\frac{1}{N-1}\frac{G\left(b\right)}{g\left(b\right)^{2}}\left(\widehat{g}\left(b\right)-g\left(b\right)\right)-\frac{1}{N-1}\frac{\widehat{G}\left(b\right)-G\left(b\right)}{g\left(b\right)}\right|=O_{p}\left(\frac{\mathrm{log}\left(L\right)}{Lh}+h^{6}\right).
\]
\end{lem}
\begin{proof}[Proof of Lemma \ref{lem:Lemma 1}]Standard arguments
(see Lemma 1 of \citealp{Marmer_Shneyerov_Quantile_Auctions}) yield
\begin{equation}
\underset{b\in\left[\underline{b},\overline{b}\right]}{\mathrm{sup}}\left|\widehat{G}\left(b\right)-G\left(b\right)\right|=O_{p}\left(\left(\frac{\mathrm{log}\left(L\right)}{L}\right)^{1/2}\right).\label{eq:G_hat - G standard uniform rates}
\end{equation}
The bias of the local quadratic MCE $\mathrm{E}\left[\widehat{g}\left(b\right)\right]-g\left(b\right)$
is $O\left(h^{3}\right)$ uniformly over the entire support $b\in\left[\underline{b},\overline{b}\right]$.
The stochastic part $\widehat{g}\left(b\right)-\mathrm{E}\left[\widehat{g}\left(b\right)\right]$
can be shown to be $O_{p}\left(\mathrm{log}\left(L\right)^{1/2}\left(Lh\right)^{-1/2}\right)$
uniformly over the entire support $b\in\left[\underline{b},\overline{b}\right]$,
by using standard arguments (see, e.g., \citealp{Newey_Kernel_ET_1994}).
Therefore, 
\begin{align}
\underset{b\in\left[\underline{b},\overline{b}\right]}{\mathrm{sup}}\left|\widehat{g}\left(b\right)-g\left(b\right)\right|=O_{p}\left(\left(\frac{\mathrm{log}\left(L\right)}{Lh}\right)^{1/2}+h^{3}\right).\label{eq:g_hat - g standard uniform rates}
\end{align}

Applying the identity $\frac{a}{b}=\frac{a}{c}-\frac{a\left(b-c\right)}{c^{2}}+\frac{a\left(b-c\right)^{2}}{bc^{2}}$,
\begin{eqnarray}
\widehat{\xi}\left(b\right)-\xi\left(b\right) & = & \frac{1}{N-1}\left\{ -\frac{G\left(b\right)\left(\widehat{g}\left(b\right)-g\left(b\right)\right)}{g\left(b\right)^{2}}+\frac{\widehat{G}\left(b\right)-G\left(b\right)}{g\left(b\right)}+\frac{\widehat{G}\left(b\right)}{\widehat{g}\left(b\right)}\frac{\left(\widehat{g}\left(b\right)-g\left(b\right)\right)^{2}}{g\left(b\right)^{2}}\right.\nonumber \\
 &  & \left.-\frac{\left(\widehat{G}\left(b\right)-G\left(b\right)\right)\left(\widehat{g}\left(b\right)-g\left(b\right)\right)}{g\left(b\right)^{2}}\right\} .\label{eq:ksi_hat - ksi decomposition}
\end{eqnarray}
By using (\ref{eq:G_hat - G standard uniform rates}), (\ref{eq:g_hat - g standard uniform rates})
and (\ref{eq:density of bids bounded  from 0}), 
\begin{equation}
\underset{b\in\left[\underline{b},\overline{b}\right]}{\mathrm{sup}}\left|\frac{G\left(b\right)\left(\widehat{g}\left(b\right)-g\left(b\right)\right)}{g\left(b\right)^{2}}\right|=O_{p}\left(\left(\frac{\mathrm{log}\left(L\right)}{Lh}\right)^{1/2}+h^{2}\right)\label{eq:ksi_hat - ksi uniform rate inter_1}
\end{equation}
and
\begin{equation}
\underset{b\in\left[\underline{b},\overline{b}\right]}{\mathrm{sup}}\left|\frac{\widehat{G}\left(b\right)-G\left(b\right)}{g\left(b\right)}\right|=O_{p}\left(\left(\frac{\mathrm{log}\left(L\right)}{L}\right)^{1/2}\right).\label{eq:ksi_hat - ksi uniform rate inter_2}
\end{equation}
Since $\underset{b\in\left[\underline{b},\overline{b}\right]}{\mathrm{sup}}\left|\widehat{g}\left(b\right)-g\left(b\right)\right|=o_{p}\left(1\right)$
w.p.a.1, $\underset{b\in\left[\underline{b},\overline{b}\right]}{\mathrm{sup}}\widehat{g}\left(b\right)^{-1}<\left(\underline{C}_{g}/2\right)^{-1}$,
w.p.a.1 and consequently, 
\begin{eqnarray*}
\underset{b\in\left[\underline{b},\overline{b}\right]}{\mathrm{sup}}\left|\frac{\widehat{G}\left(b\right)}{\widehat{g}\left(b\right)}\frac{\left(\widehat{g}\left(b\right)-g\left(b\right)\right)^{2}}{g\left(b\right)^{2}}\right| & \apprle & \underset{b\in\left[\underline{b},\overline{b}\right]}{\mathrm{sup}}\left(\widehat{g}\left(b\right)-g\left(b\right)\right)^{2}\\
 & = & O_{p}\left(\frac{\mathrm{log}\left(L\right)}{Lh}+h^{6}\right).
\end{eqnarray*}
The conclusion follows from this result, (\ref{eq:ksi_hat - ksi decomposition}),
(\ref{eq:ksi_hat - ksi uniform rate inter_1}) and (\ref{eq:ksi_hat - ksi uniform rate inter_2}).\end{proof}
\begin{lem}
\label{lem:Lemma 2}Suppose that Assumptions \ref{assu:DGP} - \ref{assu:bandwidth}
are satisfied. Then, (a). $\widehat{s}$ is strictly increasing on
$\left[\xi\left(\widehat{\underline{b}}\right),\xi\left(\widehat{\overline{b}}\right)\right]$,
w.p.a.1; (b). 
\[
\widehat{s}\left(\xi\left(\widehat{\underline{b}}\right)+h_{r}\right)-\widehat{\underline{b}}=O_{p}\left(h\right)\textrm{ and \ensuremath{\widehat{\overline{b}}-\widehat{s}\left(\xi\left(\widehat{\overline{b}}\right)-h_{r}\right)=O_{p}\left(h\right);}}
\]
(c). 
\[
\underset{z\in\left[\xi\left(\widehat{\underline{b}}\right)+h_{r},\xi\left(\widehat{\overline{b}}\right)-h_{r}\right]}{\mathrm{sup}}\left|\widehat{s}\left(z\right)-s\left(z\right)\right|=O_{p}\left(\left(\frac{\mathrm{log}\left(L\right)}{Lh}\right)^{1/2}+h^{2}\right);
\]
(d). 
\[
\underset{z\in\left[\xi\left(\widehat{\underline{b}}\right)+h_{r},\xi\left(\widehat{\overline{b}}\right)-h_{r}\right]}{\mathrm{sup}}\left|\widehat{s}'\left(z\right)-s'\left(z\right)\right|=O_{p}\left(\left(\frac{\mathrm{log}\left(L\right)}{Lh^{3}}\right)^{1/2}+h^{2}\right);
\]
(e). For any inner closed sub-interval $\left[b_{l},b_{u}\right]$
of $\left[\underline{b},\overline{b}\right]$, 
\[
\underset{b\in\left[b_{l},b_{u}\right]}{\mathrm{sup}}\left|\widehat{s}^{-1}\left(b\right)-\xi\left(b\right)\right|=O_{p}\left(\left(\frac{\mathrm{log}\left(L\right)}{Lh}\right)^{1/2}+h^{2}\right).
\]
\end{lem}
\begin{proof}[Proof of Lemma \ref{lem:Lemma 2}]By the definition
of $\widehat{s}$, we have
\[
\widehat{s}'\left(t\right)=\int_{\widehat{\underline{b}}}^{\widehat{\overline{b}}}\frac{1}{h_{r}}K_{r}\left(\frac{\widehat{\xi}\left(b\right)-t}{h_{r}}\right)\mathrm{d}b.
\]
For any $t\in\left[\xi\left(\widehat{\underline{b}}\right),\xi\left(\widehat{\overline{b}}\right)\right]$,
$\widehat{s}'\left(t\right)>0$ if the measurable set $\left\{ b\in\left[\widehat{\underline{b}},\widehat{\overline{b}}\right]:\left|\widehat{\xi}\left(b\right)-t\right|\leq h_{r}\right\} $
has positive measure. Let 
\[
r_{\xi}\coloneqq\underset{b\in\left[\underline{b},\overline{b}\right]}{\mathrm{sup}}\left|\widehat{\xi}\left(b\right)-\xi\left(b\right)\right|.
\]
Clearly,
\[
\left\{ b\in\left[\widehat{\underline{b}},\widehat{\overline{b}}\right]:\left|\xi\left(b\right)-t\right|+r_{\xi}\leq h_{r}\right\} \subseteq\left\{ b\in\left[\widehat{\underline{b}},\widehat{\overline{b}}\right]:\left|\widehat{\xi}\left(b\right)-t\right|\leq h_{r}\right\} .
\]
 By Lemma \ref{lem:Lemma 1}, $r_{\xi}=o_{p}\left(h\right)$. Consequently,
$r_{\xi}\leq h_{r}/2$ w.p.a.1, and
\begin{equation}
\left\{ b\in\left[\widehat{\underline{b}},\widehat{\overline{b}}\right]:\left|\xi\left(b\right)-t\right|\leq h_{r}/2\right\} \subseteq\left\{ b\in\left[\widehat{\underline{b}},\widehat{\overline{b}}\right]:\left|\widehat{\xi}\left(b\right)-t\right|\leq h_{r}\right\} ,\textrm{for all \ensuremath{t\in\left[\xi\left(\widehat{\underline{b}}\right),\xi\left(\widehat{\overline{b}}\right)\right]}, w.p.a.1.}\label{eq:inverse image 2}
\end{equation}
Since $\xi$ is continuous and strictly increasing on $\left[\widehat{\underline{b}},\widehat{\overline{b}}\right]$,
$\left\{ b\in\left[\widehat{\underline{b}},\widehat{\overline{b}}\right]:\left|\widehat{\xi}\left(b\right)-t\right|\leq h_{r}\right\} $
has positive measure.

For Part (b), by change of variables and Fubini-Tonelli theorem, we
have
\begin{align*}
\widehat{s}\left(\xi\left(\widehat{\underline{b}}\right)+h_{r}\right)-\widehat{\underline{b}}= & \int_{\widehat{\underline{b}}}^{\widehat{\overline{b}}}\int_{-\infty}^{\infty}\mathbbm{1}\left(w\leq\frac{\xi\left(\widehat{\underline{b}}\right)+h_{r}-\widehat{\xi}\left(b\right)}{h_{r}}\right)K_{r}\left(w\right)\mathrm{d}w\mathrm{d}b\\
\leq & \int_{-\infty}^{\infty}\int_{\widehat{\underline{b}}}^{\widehat{\overline{b}}}\mathbbm{1}\left(\xi\left(b\right)\leq r_{\xi}+\xi\left(\widehat{\underline{b}}\right)+\left(1-w\right)h_{r}\right)\mathrm{d}bK_{r}\left(w\right)\mathrm{d}w\\
= & \int_{-\infty}^{\infty}\mathrm{max}\left\{ s\left(r_{\xi}+\xi\left(\widehat{\underline{b}}\right)+\left(1-w\right)h_{r}\right)-\widehat{\underline{b}},0\right\} K_{r}\left(w\right)\mathrm{d}w.
\end{align*}
By a mean value expansion, we have 
\begin{align*}
 & \int_{-\infty}^{\infty}\mathrm{max}\left\{ s\left(\xi\left(\widehat{\underline{b}}\right)+\left(1-w\right)h_{r}+r_{\xi}\right)-\widehat{\underline{b}},0\right\} K\left(w\right)\mathrm{d}w\\
\leq & \int_{-\infty}^{\infty}\left|s\left(\xi\left(\widehat{\underline{b}}\right)+\left(1-w\right)h_{r}+r_{\xi}\right)-\widehat{\underline{b}}\right|K\left(w\right)\mathrm{d}w\\
\apprle & \int_{-\infty}^{\infty}\left|\left(1-w\right)h_{r}+r_{\xi}\right|K\left(w\right)\mathrm{d}w\\
= & O_{p}\left(h\right),
\end{align*}
where the last equality holds since $r_{\xi}=o_{p}\left(h\right)$.
Therefore, $\widehat{s}\left(\xi\left(\widehat{\underline{b}}\right)+h_{r}\right)-\widehat{\underline{b}}=O_{p}\left(h\right)$.
The proof of $\widehat{\overline{b}}-\widehat{s}\left(\xi\left(\widehat{\overline{b}}\right)-h_{r}\right)=O_{p}\left(h\right)$
is similar. 

For Part (c), by the triangle inequality and a second-order Taylor
expansion, we have
\begin{eqnarray}
\left|\widehat{s}\left(z\right)-s\left(z\right)\right| & \leq & \left|\int_{\widehat{\underline{b}}}^{\widehat{\overline{b}}}\widetilde{K}_{r}\left(\frac{z-\widehat{\xi}\left(b\right)}{h_{r}}\right)\mathrm{d}b-\int_{\widehat{\underline{b}}}^{\widehat{\overline{b}}}\widetilde{K}_{r}\left(\frac{z-\xi\left(b\right)}{h_{r}}\right)\mathrm{d}b\right|+\left|\int_{\widehat{\underline{b}}}^{\widehat{\overline{b}}}\widetilde{K}_{r}\left(\frac{z-\xi\left(b\right)}{h_{r}}\right)\mathrm{d}b+\widehat{\underline{b}}-s\left(z\right)\right|.\nonumber \\
 & \leq & \left|\int_{\widehat{\underline{b}}}^{\widehat{\overline{b}}}\frac{1}{h_{r}}K_{r}\left(\frac{z-\xi\left(b\right)}{h_{r}}\right)\left(\widehat{\xi}\left(b\right)-\xi\left(b\right)\right)\mathrm{d}b\right|+\left|\frac{1}{2}\int_{\widehat{\underline{b}}}^{\widehat{\overline{b}}}\frac{1}{h_{r}^{2}}K_{r}\left(\frac{z-\dot{\xi}\left(b\right)}{h_{r}}\right)\left(\widehat{\xi}\left(b\right)-\xi\left(b\right)\right)^{2}\mathrm{d}b\right|\nonumber \\
 &  & +\left|\int_{\widehat{\underline{b}}}^{\widehat{\overline{b}}}\widetilde{K}_{r}\left(\frac{z-\xi\left(b\right)}{h_{r}}\right)\mathrm{d}b+\widehat{\underline{b}}-s\left(z\right)\right|,\label{eq:K_til(ksi_hat) - K_til(ksi) integral bound inter_1}
\end{eqnarray}
where $\dot{\xi}\left(b\right)$ is the mean value satisfying $\left|\dot{\xi}\left(b\right)-\xi\left(b\right)\right|\leq\left|\widehat{\xi}\left(b\right)-\xi\left(b\right)\right|$
for each $b\in\left[\widehat{\underline{b}},\widehat{\overline{b}}\right]$.
Then we have
\begin{eqnarray}
\underset{z\in\left[\xi\left(\widehat{\underline{b}}\right)+h_{r},\xi\left(\widehat{\overline{b}}\right)-h_{r}\right]}{\mathrm{sup}}\left|\int_{\widehat{\underline{b}}}^{\widehat{\overline{b}}}\frac{1}{h_{r}}K_{r}\left(\frac{z-\xi\left(b\right)}{h_{r}}\right)\left(\widehat{\xi}\left(b\right)-\xi\left(b\right)\right)\mathrm{d}b\right| & \leq & \left\{ \underset{z\in\left[\xi\left(\widehat{\underline{b}}\right)+h_{r},\xi\left(\widehat{\overline{b}}\right)-h_{r}\right]}{\mathrm{sup}}\int_{\widehat{\underline{b}}}^{\widehat{\overline{b}}}\frac{1}{h_{r}}K_{r}\left(\frac{z-\xi\left(b\right)}{h_{r}}\right)\mathrm{d}b\right\} r_{\xi}\nonumber \\
 & \leq & \left\{ \underset{z\in\left[\xi\left(\widehat{\underline{b}}\right)+h_{r},\xi\left(\widehat{\overline{b}}\right)-h_{r}\right]}{\mathrm{sup}}\int_{\widehat{\underline{b}}}^{\widehat{\overline{b}}}\frac{1}{h_{r}}\mathbbm{1}\left(\left|z-\xi\left(b\right)\right|\leq h_{r}\right)\mathrm{d}b\right\} r_{\xi}\nonumber \\
 & = & O_{p}\left(\left(\frac{\mathrm{log}\left(L\right)}{Lh}\right)^{1/2}+h^{3}\right),\label{eq:K_til(ksi_hat) - K_til(ksi) integral bound inter_2}
\end{eqnarray}
where the equality follows from Lemma \ref{lem:Lemma 1} . 

For all $z\in\left[\xi\left(\widehat{\underline{b}}\right)+h_{r},\xi\left(\widehat{\overline{b}}\right)-h_{r}\right]$,
\begin{eqnarray}
\left|\int_{\widehat{\underline{b}}}^{\widehat{\overline{b}}}\frac{1}{h_{r}^{2}}K\left(\frac{z-\dot{\xi}\left(b\right)}{h_{r}}\right)\left(\hat{\xi}\left(b\right)-\xi\left(b\right)\right)^{2}\mathrm{d}b\right| & \leq & \left\{ \int_{\widehat{\underline{b}}}^{\widehat{\overline{b}}}\frac{1}{h_{r}^{2}}K_{r}\left(\frac{z-\dot{\xi}\left(b\right)}{h_{r}}\right)\mathrm{d}b\right\} r_{\xi}^{2}\nonumber \\
 & \apprle & \left\{ \int_{\widehat{\underline{b}}}^{\widehat{\overline{b}}}\frac{1}{h_{r}^{2}}\mathbbm{1}\left(\left|z-\xi\left(b\right)\right|\leq h_{r}+r_{\xi}\right)\mathrm{d}b\right\} r_{\xi}^{2}\nonumber \\
 & \leq & \left\{ \int_{\widehat{\underline{b}}}^{\widehat{\overline{b}}}\frac{1}{h_{r}^{2}}\mathbbm{1}\left(\left|z-\xi\left(b\right)\right|\leq2h_{r}\right)\mathrm{d}b\right\} r_{\xi}^{2},\label{eq:K_til(ksi_hat) - K_til(ksi) integral bound inter_3}
\end{eqnarray}
where the last inequality holds w.p.a.1 since $r_{\xi}=o_{p}\left(h\right)$.
Now by (\ref{eq:K_til(ksi_hat) - K_til(ksi) integral bound inter_1}),
(\ref{eq:K_til(ksi_hat) - K_til(ksi) integral bound inter_2}) and
(\ref{eq:K_til(ksi_hat) - K_til(ksi) integral bound inter_3}),
\begin{equation}
\underset{z\in\left[\xi\left(\widehat{\underline{b}}\right)+h_{r},\xi\left(\widehat{\overline{b}}\right)-h_{r}\right]}{\mathrm{sup}}\left|\int_{\widehat{\underline{b}}}^{\widehat{\overline{b}}}\widetilde{K}_{r}\left(\frac{z-\widehat{\xi}\left(b\right)}{h_{r}}\right)\mathrm{d}b-\int_{\widehat{\underline{b}}}^{\widehat{\overline{b}}}\widetilde{K}_{r}\left(\frac{z-\xi\left(b\right)}{h_{r}}\right)\mathrm{d}b\right|=O_{p}\left(\left(\frac{\mathrm{log}\left(L\right)}{Lh}\right)^{1/2}+h^{3}\right).\label{eq:K_til(ksi_hat) - K_til(ksi) integral uniform bound}
\end{equation}

For all $z\in\left[\xi\left(\widehat{\underline{b}}\right)+h_{r},\xi\left(\widehat{\overline{b}}\right)-h_{r}\right]$,
since $\widetilde{K}_{r}\left(u\right)=\int_{-\infty}^{u}K_{r}\left(t\right)\mathrm{d}t$
and $K_{r}$ is supported on $\left[-1,1\right]$,
\begin{eqnarray*}
\left|\int_{\widehat{\underline{b}}}^{\widehat{\overline{b}}}\widetilde{K}_{r}\left(\frac{z-\xi\left(b\right)}{h_{r}}\right)\mathrm{d}b+\widehat{\underline{b}}-s\left(z\right)\right| & = & \left|\int_{\widehat{\underline{b}}}^{s\left(z-h_{r}\right)}\widetilde{K}_{r}\left(\frac{z-\xi\left(b\right)}{h_{r}}\right)\mathrm{d}b+\int_{s\left(z-h_{r}\right)}^{s\left(z+h_{r}\right)}\widetilde{K}_{r}\left(\frac{z-\xi\left(b\right)}{h_{r}}\right)\mathrm{d}b+\widehat{\underline{b}}-s\left(z\right)\right|\\
 & = & \left|s\left(z-h_{r}\right)+\int_{s\left(z-h_{r}\right)}^{s\left(z+h_{r}\right)}\widetilde{K}_{r}\left(\frac{z-\xi\left(b\right)}{h_{r}}\right)\mathrm{d}b-s\left(z\right)\right|.
\end{eqnarray*}
By change of variables and integration by parts,
\begin{eqnarray*}
\int_{s\left(z-h_{r}\right)}^{s\left(z+h_{r}\right)}\widetilde{K}_{r}\left(\frac{z-\xi\left(b\right)}{h_{r}}\right)\mathrm{d}b & = & \int_{-1}^{1}\widetilde{K}_{r}\left(-u\right)h_{r}s'\left(h_{r}u+z\right)\mathrm{d}u\\
 & = & -s\left(z-h_{r}\right)+\int_{-1}^{1}s\left(z+h_{r}u\right)K_{r}\left(u\right)\mathrm{d}u.
\end{eqnarray*}
Now it follows that 
\begin{eqnarray}
\underset{z\in\left[\xi\left(\widehat{\underline{b}}\right)+h_{r},\xi\left(\widehat{\overline{b}}\right)-h_{r}\right]}{\mathrm{sup}}\left|\int_{\widehat{\underline{b}}}^{\widehat{\overline{b}}}\widetilde{K}_{r}\left(\frac{z-\xi\left(b\right)}{h_{r}}\right)\mathrm{d}b+\widehat{\underline{b}}-s\left(z\right)\right| & = & \underset{z\in\left[\xi\left(\widehat{\underline{b}}\right)+h_{r},\xi\left(\widehat{\overline{b}}\right)-h_{r}\right]}{\mathrm{sup}}\left|\int_{-1}^{1}K_{r}\left(u\right)s\left(z+h_{r}u\right)\mathrm{d}u-s\left(z\right)\right|\nonumber \\
 & = & O_{p}\left(h^{2}\right),\label{eq:K_til(ksi) - b_lower_bar - s uniform bound}
\end{eqnarray}
where the second equality follows from Taylor expansion and the fact
$\int uK_{r}\left(u\right)\mathrm{d}u=0$. The conclusion of Part
(c) follows from (\ref{eq:K_til(ksi_hat) - K_til(ksi) integral bound inter_1}),
(\ref{eq:K_til(ksi_hat) - K_til(ksi) integral uniform bound}) and
(\ref{eq:K_til(ksi) - b_lower_bar - s uniform bound}).

For Part (d), note 
\[
\left|\widehat{s}'\left(z\right)-s'\left(z\right)\right|\leq\left|\int_{\widehat{\underline{b}}}^{\widehat{\overline{b}}}\frac{1}{h_{r}}K_{r}\left(\frac{\widehat{\xi}\left(b\right)-z}{h_{r}}\right)\mathrm{d}b-\int_{\widehat{\underline{b}}}^{\widehat{\overline{b}}}\frac{1}{h_{r}}K_{r}\left(\frac{\xi\left(b\right)-z}{h_{r}}\right)\mathrm{d}b\right|+\left|\int_{\widehat{\underline{b}}}^{\widehat{\overline{b}}}\frac{1}{h_{r}}K_{r}\left(\frac{\xi\left(b\right)-z}{h_{r}}\right)\mathrm{d}b-s'\left(z\right)\right|.
\]
By arguments that are similar to those used to prove (\ref{eq:K_til(ksi_hat) - K_til(ksi) integral uniform bound}),
\[
\underset{z\in\left[\xi\left(\widehat{\underline{b}}\right)+h_{r},\xi\left(\widehat{\overline{b}}\right)-h_{r}\right]}{\mathrm{sup}}\left|\int_{\widehat{\underline{b}}}^{\widehat{\overline{b}}}\frac{1}{h_{r}}K_{r}\left(\frac{\widehat{\xi}\left(b\right)-z}{h_{r}}\right)\mathrm{d}b-\int_{\widehat{\underline{b}}}^{\widehat{\overline{b}}}\frac{1}{h_{r}}K_{r}\left(\frac{\xi\left(b\right)-z}{h_{r}}\right)\mathrm{d}b\right|=O_{p}\left(\left(\frac{\mathrm{log}\left(L\right)}{Lh^{3}}\right)^{1/2}+h^{2}\right).
\]
By integration by parts and Taylor expansion, 
\begin{eqnarray*}
\underset{z\in\left[\xi\left(\widehat{\underline{b}}\right)+h_{r},\xi\left(\widehat{\overline{b}}\right)-h_{r}\right]}{\mathrm{sup}}\left|\int_{\widehat{\underline{b}}}^{\widehat{\overline{b}}}\frac{1}{h_{r}}K_{r}\left(\frac{\xi\left(b\right)-z}{h_{r}}\right)\mathrm{d}b-s'\left(z\right)\right| & = & \underset{z\in\left[\xi\left(\widehat{\underline{b}}\right)+h_{r},\xi\left(\widehat{\overline{b}}\right)-h_{r}\right]}{\mathrm{sup}}\left|\int_{-1}^{1}K_{r}\left(u\right)s'\left(z+h_{r}u\right)\mathrm{d}u-s'\left(z\right)\right|\\
 & = & O_{p}\left(h^{2}\right).
\end{eqnarray*}
The conclusion of Part (d) therefore follows.

For Part (e), Part (b) implies that $\left[b_{l},b_{u}\right]$ is
contained in the interior of $\left[\xi\left(\widehat{\underline{b}}\right)+h_{r},\xi\left(\widehat{\overline{b}}\right)-h_{r}\right]$
w.p.a.1. Then Part (a) implies that 
\begin{equation}
\widehat{s}^{-1}\left(b\right)\in\left[\xi\left(\widehat{\underline{b}}\right)+h_{r},\xi\left(\widehat{\overline{b}}\right)-h_{r}\right]\textrm{ and }b=\widehat{s}\left(\widehat{s}^{-1}\left(b\right)\right),\textrm{ for all \ensuremath{b\in\left[b_{l},b_{u}\right],} w.p.a.1}.\label{eq:s_hat^-1}
\end{equation}
Therefore, for any $b\in\left[b_{l},b_{u}\right]$, 
\[
s\left(\widehat{s}^{-1}\left(b\right)\right)-s\left(\xi\left(b\right)\right)=s'\left(\dot{v}\right)\left(\widehat{s}^{-1}\left(b\right)-\xi\left(b\right)\right)
\]
for some mean value $\dot{v}$ with $\left|\dot{v}-\xi\left(b\right)\right|\leq\left|\widehat{s}^{-1}\left(b\right)-\xi\left(b\right)\right|$.
Then, since $s'$ is bounded away from zero (see Lemma A1 of GPV),
\begin{eqnarray*}
\underset{b\in\left[b_{l},b_{u}\right]}{\mathrm{sup}}\left|\widehat{s}^{-1}\left(b\right)-\xi\left(b\right)\right| & \apprle & \underset{b\in\left[b_{l},b_{u}\right]}{\mathrm{sup}}\left|s\left(\widehat{s}^{-1}\left(b\right)\right)-s\left(\xi\left(b\right)\right)\right|\\
 & \leq & \underset{z\in\left[\xi\left(\widehat{\underline{b}}\right)+h_{r},\xi\left(\widehat{\overline{b}}\right)-h_{r}\right]}{\mathrm{sup}}\left|s\left(z\right)-\widehat{s}\left(z\right)\right|\\
 & = & O_{p}\left(\left(\frac{\mathrm{log}\left(L\right)}{Lh}\right)^{1/2}+h^{2}\right),
\end{eqnarray*}
where the second inequality follows from (\ref{eq:s_hat^-1}) and
holds w.p.a.1. 

\end{proof}
\begin{lem}
\label{lem:Lemma 3}Suppose that Assumptions \ref{assu:DGP} - \ref{assu:bandwidth}
hold. Let $\mathbb{T}_{il}\coloneqq\mathbbm{1}\left(V_{il}\in\left[v-\delta_{0},v+\delta_{0}\right]\right)$.
Then we have
\begin{eqnarray*}
\widehat{f}_{RGPV}\left(v\right)-f\left(v\right) & = & \frac{1}{N\cdot L}\sum_{i,l}\mathbb{T}_{il}\frac{1}{h_{f}^{2}}K_{f}'\left(\frac{V_{il}-v}{h_{f}}\right)\left(\widehat{V}_{il}^{\dagger}-V_{il}\right)+\frac{1}{2}f''\left(v\right)\left(\int K_{f}\left(u\right)u^{2}\mathrm{d}u\right)h_{f}^{2}\\
 &  & +O_{p}\left(\frac{\mathrm{log}\left(L\right)}{Lh^{3}}+\left(\frac{\mathrm{log}\left(L\right)}{Lh}\right)^{1/2}+h^{2}\right),
\end{eqnarray*}
where the remainder term is uniform in $v\in I$.
\end{lem}
\begin{proof}[Proof of Lemma \ref{lem:Lemma 3}]Write
\[
\widehat{f}_{RGPV}\left(v\right)=\frac{1}{N\cdot L}\sum_{i,l}\left\{ \mathbb{T}_{il}\frac{1}{h_{f}}K_{f}\left(\frac{\widehat{V}_{il}^{\dagger}-v}{h_{f}}\right)+\left(1-\mathbb{T}_{il}\right)\frac{1}{h_{f}}K_{f}\left(\frac{\widehat{V}_{il}^{\dagger}-v}{h_{f}}\right)\right\} .
\]
Now for any $v\in\left[\xi\left(\widehat{\underline{b}}\right)+h_{f}+h_{r},\xi\left(\widehat{\overline{b}}\right)-h_{f}-h_{r}\right]$,
\begin{align*}
 & \left|\frac{1}{N\cdot L}\sum_{i,l}\left(1-\mathbb{T}_{il}\right)\frac{1}{h_{f}}K_{f}\left(\frac{\widehat{V}_{il}^{\dagger}-v}{h_{f}}\right)\right|\\
\leq & \frac{1}{N\cdot L}\sum_{i,l}h_{f}^{-1}\left(1-\mathbb{T}_{il}\right)\mathbbm{1}\left(\left|\widehat{V}_{il}^{\dagger}-v\right|\leq h_{f}\right)\\
\leq & \frac{1}{N\cdot L}\sum_{i,l}h_{f}^{-1}\mathbbm{1}\left(B_{il}>s\left(v+\delta_{0}\right)\right)\mathbbm{1}\left(B_{il}\in\left[\widehat{s}\left(v-h_{f}\right),\widehat{s}\left(v+h_{f}\right)\right]\right)\\
 & +\frac{1}{N\cdot L}\sum_{i,l}h_{f}^{-1}\mathbbm{1}\left(B_{il}<s\left(v-\delta_{0}\right)\right)\mathbbm{1}\left(B_{il}\in\left[\widehat{s}\left(v-h_{f}\right),\widehat{s}\left(v+h_{f}\right)\right]\right)\\
\leq & \frac{1}{N\cdot L}\sum_{i,l}h_{f}^{-1}\mathbbm{1}\left(B_{il}>s\left(v+\delta_{0}\right)\right)\mathbbm{1}\left(B_{il}\in\left[s\left(v-h_{f}\right)-r_{s},s\left(v+h_{f}\right)+r_{s}\right]\right)\\
 & +\frac{1}{N\cdot L}\sum_{i,l}h_{f}^{-1}\mathbbm{1}\left(B_{il}<s\left(v-\delta_{0}\right)\right)\mathbbm{1}\left(B_{il}\in\left[s\left(v-h_{f}\right)-r_{s},s\left(v+h_{f}\right)+r_{s}\right]\right),
\end{align*}
where $r_{s}\coloneqq\underset{z\in\left[\xi\left(\widehat{\underline{b}}\right)+h_{r},\xi\left(\widehat{\overline{b}}\right)-h_{r}\right]}{\mathrm{sup}}\left|\widehat{s}\left(z\right)-s\left(z\right)\right|$
and the second inequality holds w.p.a.1. Therefore,
\[
\widehat{f}_{RGPV}\left(v\right)=\frac{1}{N\cdot L}\sum_{i,l}\mathbb{T}_{il}\frac{1}{h_{f}}K_{f}\left(\frac{\widehat{V}_{il}^{\dagger}-v}{h_{f}}\right),\textrm{ for all \ensuremath{v\in I}, w.p.a.1.}
\]

Let $\widetilde{f}$ denote the infeasible estimator that uses the
unobserved true valuations:
\[
\widetilde{f}\left(v\right)=\frac{1}{N\cdot L}\sum_{i,l}\frac{1}{h_{f}}K_{f}\left(\frac{V_{il}-v}{h_{f}}\right).
\]
It now follows that
\[
\widehat{f}_{RGPV}\left(v\right)-\widetilde{f}\left(v\right)=\frac{1}{N\cdot L}\sum_{i,l}\mathbb{T}_{il}\frac{1}{h_{f}}\left(K_{f}\left(\frac{\widehat{V}_{il}^{\dagger}-v}{h_{f}}\right)-K_{f}\left(\frac{V_{il}-v}{h_{f}}\right)\right),\textrm{ for all \ensuremath{v\in I}, w.p.a.1.}
\]
By a second-order Taylor expansion of the right-hand side of the above
equality,
\begin{equation}
\widehat{f}_{RGPV}\left(v\right)-\widetilde{f}\left(v\right)=\frac{1}{N\cdot L}\sum_{i,l}\mathbb{T}_{il}\frac{1}{h_{f}^{2}}K_{f}'\left(\frac{V_{il}-v}{h_{f}}\right)\left(\widehat{V}_{il}^{\dagger}-V_{il}\right)+\frac{1}{2}\cdot\frac{1}{N\cdot L}\sum_{i,l}\mathbb{T}_{il}\frac{1}{h_{f}^{3}}K_{f}''\left(\frac{\dot{V}_{il}-v}{h_{f}}\right)\left(\widehat{V}_{il}^{\dagger}-V_{il}\right)^{2},\label{eq:difference f_hat f_til 1}
\end{equation}
for some mean value $\dot{V}_{il}$ that lies on the line joining
$\widehat{V}_{il}^{\dagger}$ and $V_{il}$. 

Lemma \ref{lem:Lemma 2}(e) implies that 
\begin{equation}
\underset{v\in I}{\mathrm{sup}}\,\underset{i,l}{\mathrm{max}}\,\mathbb{T}_{il}\left|\widehat{V}_{il}^{\dagger}-V_{il}\right|=O_{p}\left(\left(\frac{\mathrm{log}\left(L\right)}{Lh}\right)^{1/2}+h^{2}\right).\label{eq:V_dag - V sup rate}
\end{equation}
Since $K_{f}''$ is compactly supported on $\left[-1,1\right]$ and
bounded, by the triangle inequality, 
\begin{align}
 & \underset{v\in I}{\mathrm{sup}}\left|\frac{1}{N\cdot L}\sum_{i,l}\mathbb{T}_{il}\frac{1}{h_{f}^{3}}K_{f}''\left(\frac{\dot{V}_{il}-v}{h_{f}}\right)\left(\widehat{V}_{il}^{\dagger}-V_{il}\right)^{2}\right|\nonumber \\
\apprle & \left\{ \underset{v\in I}{\mathrm{sup}}\,\frac{1}{N\cdot L}\sum_{i,l}\mathbb{T}_{il}h_{f}^{-3}\mathbbm{1}\left(\left|\dot{V}_{il}-v\right|\leq h_{f}\right)\right\} \left\{ \underset{v\in I}{\mathrm{sup}}\,\underset{i,l}{\mathrm{max}}\,\mathbb{T}_{il}\left(\widehat{V}_{il}^{\dagger}-V_{il}\right)^{2}\right\} \nonumber \\
\leq & \left\{ \underset{v\in I}{\mathrm{sup}}\,\frac{1}{N\cdot L}\sum_{i,l}\mathbb{T}_{il}h_{f}^{-3}\mathbbm{1}\left(\left|V_{il}-v\right|\leq2h_{f}\right)\right\} \left\{ \underset{v\in I}{\mathrm{sup}}\,\underset{i,l}{\mathrm{max}}\,\mathbb{T}_{il}\left(\widehat{V}_{il}^{\dagger}-V_{il}\right)^{2}\right\} ,\label{eq:K_double_prime * (V_hat - V)^2 bound}
\end{align}
where the last inequality holds w.p.a.1, since $\underset{v\in I}{\mathrm{sup}}\,\underset{i,l}{\mathrm{max}}\,\mathbb{T}_{il}\left|\dot{V}_{il}-V_{il}\right|=o_{p}\left(h\right)$.
It follows that
\begin{equation}
\underset{v\in I}{\mathrm{sup}}\left|\frac{1}{N\cdot L}\sum_{i,l}\mathbb{T}_{il}\frac{1}{h_{f}^{3}}K_{f}''\left(\frac{\dot{V}_{il}-v}{h_{f}}\right)\left(\widehat{V}_{il}^{\dagger}-V_{il}\right)^{2}\right|=O_{p}\left(\frac{\mathrm{log}\left(L\right)}{Lh^{3}}+h^{2}\right).\label{eq:difference f_hat f_til 2}
\end{equation}

By standard arguments for kernel density estimation,
\begin{equation}
\widetilde{f}\left(v\right)-\mathrm{E}\left[\widetilde{f}\left(v\right)\right]=O_{p}\left(\left(\frac{\mathrm{log}\left(L\right)}{Lh}\right)^{1/2}\right)\textrm{ and }\mathrm{E}\left[\widetilde{f}\left(v\right)\right]-f\left(v\right)=\frac{1}{2}f''\left(v\right)\left(\int K_{f}\left(u\right)u^{2}\mathrm{d}u\right)h_{f}^{2}+o\left(h^{2}\right),\label{eq:infeasible PDF estimate error rate}
\end{equation}
where the remainder terms are uniform in $v\in I$. The conclusion
follows. \end{proof}
\begin{lem}
\label{lem:Lemma 4}Suppose that Assumptions \ref{assu:DGP} - \ref{assu:bandwidth}
hold. Let
\[
\widetilde{s}\left(t\right)\coloneqq\int_{\underline{b}}^{\overline{b}}\int_{-\infty}^{t}\frac{1}{h_{r}}K_{r}\left(\frac{\xi\left(b\right)-u}{h_{r}}\right)\mathrm{d}u\mathrm{d}b+\underline{b},\;t\in\mathbb{R}.
\]
Then,
\begin{eqnarray*}
\widehat{f}_{RGPV}\left(v\right)-f\left(v\right) & = & -\frac{1}{N\cdot L}\sum_{i,l}\frac{1}{h_{f}^{2}}\mathbb{T}_{il}K_{f}'\left(\frac{V_{il}-v}{h_{f}}\right)\frac{\widehat{s}\left(V_{il}\right)-\widetilde{s}\left(V_{il}\right)}{s'\left(V_{il}\right)}+\frac{1}{2}f''\left(v\right)\left(\int K_{f}\left(u\right)u^{2}\mathrm{d}u\right)h_{f}^{2}\\
 &  & +\frac{1}{2}\frac{\left(s'''\left(v\right)f\left(v\right)+s''\left(v\right)f'\left(v\right)\right)s'\left(v\right)-s''\left(v\right)f\left(v\right)s''\left(v\right)}{s'\left(v\right)^{2}}\left(\int K_{r}\left(u\right)u^{2}\mathrm{d}u\right)h_{r}^{2}\\
 &  & +O_{p}\left(\frac{\mathrm{log}\left(L\right)}{Lh^{3}}+\left(\frac{\mathrm{log}\left(L\right)}{Lh}\right)^{1/2}+h^{2}\right),
\end{eqnarray*}
where the remainder term is uniform in $v\in I$.
\end{lem}
\begin{proof}[Proof of Lemma \ref{lem:Lemma 4}]It follows from Lemma
\ref{lem:Lemma 2}(b) that $\left[s\left(v_{l}-\delta_{0}\right),s\left(v_{u}+\delta_{0}\right)\right]$
is an inner closed sub-interval of $\left[\widehat{s}\left(\xi\left(\widehat{\underline{b}}\right)+h_{r}\right),\widehat{s}\left(\xi\left(\widehat{\overline{b}}\right)-h_{r}\right)\right]$
w.p.a.1. By Lemma \ref{lem:Lemma 2}(a), $\widehat{s}$ is strictly
increasing on $\left[\xi\left(\widehat{\underline{b}}\right)+h_{r},\xi\left(\widehat{\overline{b}}\right)-h_{r}\right]$
w.p.a.1 and for all $B_{il}$ satisfying $B_{il}\in\left[s\left(v_{l}-\delta_{0}\right),s\left(v_{u}+\delta_{0}\right)\right]$,
we have the following expansion by \citet[Lemma A.1]{dette2006simple}:
\[
\widehat{V}_{il}^{\dagger}-V_{il}=-\left(\frac{\widehat{s}-s}{s'}\right)\circ\xi\left(B_{il}\right)+\chi_{1,il}+\chi_{2,il}
\]
where
\[
\chi_{1,il}\coloneqq-2\left(\frac{\widehat{s}-s}{s'+\lambda_{il}\left(\widehat{s}'-s'\right)}\cdot\frac{\widehat{s}'-s'}{s'+\lambda_{il}\left(\widehat{s}'-s'\right)}\right)\circ\left(s+\lambda_{il}\left(\widehat{s}-s\right)\right)^{-1}\left(B_{il}\right)
\]
and
\[
\chi_{2,il}\coloneqq\left\{ \frac{\widehat{s}-s}{s'+\lambda_{il}\left(\widehat{s}'-s'\right)}\cdot\frac{\left(\widehat{s}-s\right)\left(s''+\lambda_{il}\left(\widehat{s}''-s''\right)\right)}{\left(s'+\lambda_{il}\left(\widehat{s}'-s'\right)\right)^{2}}\right\} \circ\left(s+\lambda_{il}\left(\widehat{s}-s\right)\right)^{-1}\left(B_{il}\right)
\]
for some $\lambda_{il}\in\left(0,1\right)$ that depends on $B_{il}$.
By Lemma \ref{lem:Lemma 2}(b),
\[
\left(s+\lambda_{il}\left(\widehat{s}-s\right)\right)^{-1}\left(B_{il}\right)\in\left[\xi\left(\widehat{\underline{b}}\right)+h_{r},\xi\left(\widehat{\overline{b}}\right)-h_{r}\right],\textrm{ w.p.a.1,}
\]
for all $B_{il}$ satisfying $B_{il}\in\left[s\left(v_{l}-\delta_{0}\right),s\left(v_{u}+\delta_{0}\right)\right]$.
Next, write
\begin{align}
 & \frac{1}{NL}\sum_{i,l}\mathbb{T}_{il}\frac{1}{h_{f}^{2}}K_{f}'\left(\frac{V_{il}-v}{h_{f}}\right)\left(\widehat{V}_{il}^{\dagger}-V_{il}\right)\nonumber \\
= & -\frac{1}{NL}\sum_{i,l}\frac{1}{h_{f}^{2}}\mathbb{T}_{il}K_{f}'\left(\frac{V_{il}-v}{h_{f}}\right)\frac{\widehat{s}\left(V_{il}\right)-\widetilde{s}\left(V_{il}\right)}{s'\left(V_{il}\right)}-\frac{1}{NL}\sum_{i,l}\frac{1}{h_{f}^{2}}\mathbb{T}_{il}K_{f}'\left(\frac{V_{il}-v}{h_{f}}\right)\frac{\widetilde{s}\left(V_{il}\right)-B_{il}}{s'\left(V_{il}\right)}\nonumber \\
 & +\frac{1}{NL}\sum_{i,l}\frac{1}{h_{f}^{2}}\mathbb{T}_{il}K_{f}'\left(\frac{V_{il}-v}{h_{f}}\right)\chi_{1,il}+\frac{1}{NL}\sum_{i,l}\frac{1}{h_{f}^{2}}\mathbb{T}_{il}K_{f}'\left(\frac{V_{il}-v}{h_{f}}\right)\chi_{2,il}.\label{eq:average K_prime V_hat_dagger - V}
\end{align}

By Lemma \ref{lem:Lemma 2}(c) and Lemma \ref{lem:Lemma 2}(d), 
\[
\underset{i,l}{\mathrm{sup}}\mathbb{T}_{il}\left|\chi_{1,il}\right|=O_{p}\left(\frac{\mathrm{log}\left(L\right)}{Lh^{2}}+h^{4}\right).
\]
For $\chi_{il}^{2}$, since $K_{r}'$ is supported on $\left[-1,1\right]$,
\begin{eqnarray*}
\underset{t\in\left[\xi\left(\widehat{\underline{b}}\right)+h_{r},\xi\left(\widehat{\overline{b}}\right)-h_{r}\right]}{\mathrm{sup}}\left|\widehat{s}''\left(t\right)\right| & \apprle & \underset{t\in\left[\xi\left(\widehat{\underline{b}}\right)+h_{r},\xi\left(\widehat{\overline{b}}\right)-h_{r}\right]}{\mathrm{sup}}\int_{\widehat{\underline{b}}}^{\widehat{\overline{b}}}\frac{1}{h_{r}^{2}}\mathbbm{1}\left(\left|t-\xi\left(b\right)\right|\leq h_{r}+r_{\xi}\right)\mathrm{d}b\\
 & = & O_{p}\left(h^{-1}\right).
\end{eqnarray*}
It follows from this result, Lemma \ref{lem:Lemma 2}(c) and Lemma
\ref{lem:Lemma 2}(d) that
\[
\underset{i,l}{\mathrm{sup}}\mathbb{T}_{il}\left|\chi_{2,il}\right|=O_{p}\left(\frac{\mathrm{log}\left(L\right)}{Lh^{2}}+h^{3}\right).
\]
Now it is clear that
\begin{eqnarray}
\underset{v\in I}{\mathrm{sup}}\left|\frac{1}{N\cdot L}\sum_{i,l}\frac{1}{h_{f}^{2}}\mathbb{T}_{il}K_{f}'\left(\frac{V_{il}-v}{h_{f}}\right)\chi_{1,il}\right| & \leq & \left(\underset{v\in I}{\mathrm{sup}}\frac{1}{N\cdot L}\sum_{i,l}\left|\frac{1}{h_{f}^{2}}K_{f}'\left(\frac{V_{il}-v}{h_{f}}\right)\right|\right)\left(\underset{i,l}{\mathrm{sup}}\mathbb{T}_{il}\left|\chi_{1,il}\right|\right)\nonumber \\
 & = & O_{p}\left(\frac{\mathrm{log}\left(L\right)}{Lh^{3}}+h^{3}\right)\label{eq:eq:average K_prime V_hat_dagger - V inter_2}
\end{eqnarray}
and similarly,
\begin{equation}
\underset{v\in I}{\mathrm{sup}}\left|\frac{1}{NL}\sum_{i,l}\frac{1}{h_{f}^{2}}\mathbb{T}_{il}K_{f}'\left(\frac{V_{il}-v}{h_{f}}\right)\chi_{2,il}\right|=O_{p}\left(\frac{\mathrm{log}\left(L\right)}{Lh^{3}}+h^{2}\right).\label{eq:eq:average K_prime V_hat_dagger - V inter_3}
\end{equation}

By the definition of $\widetilde{s}$, when $h_{f}$ is small enough,
\begin{align*}
 & \frac{1}{NL}\sum_{i,l}\frac{1}{h_{f}^{2}}\mathbb{T}_{il}K_{f}'\left(\frac{V_{il}-v}{h_{f}}\right)\frac{\widetilde{s}\left(V_{il}\right)-B_{il}}{s'\left(V_{il}\right)}\\
= & \frac{1}{NL}\sum_{i,l}\frac{1}{h_{f}^{2}}K_{f}'\left(\frac{V_{il}-v}{h_{f}}\right)\frac{1}{s'\left(V_{il}\right)}\left(\int_{\underline{b}}^{\overline{b}}\widetilde{K}_{r}\left(\frac{V_{il}-\xi\left(b\right)}{h_{r}}\right)\mathrm{d}b+\underline{b}-s\left(V_{il}\right)\right).
\end{align*}
By change of variable and integration by parts, 
\begin{align*}
\int_{\underline{b}}^{\overline{b}}\widetilde{K}_{r}\left(\frac{V_{il}-\xi\left(b\right)}{h_{r}}\right)\mathrm{d}b= & \int_{\underline{v}}^{\overline{v}}\widetilde{K}_{r}\left(\frac{V_{il}-u}{h_{r}}\right)s'\left(u\right)\mathrm{d}u\\
= & \overline{b}\cdot\widetilde{K}_{r}\left(\frac{V_{il}-\overline{v}}{h_{r}}\right)-\underline{b}\cdot\widetilde{K}_{r}\left(\frac{V_{il}-\underline{v}}{h_{r}}\right)+\int_{\underline{v}}^{\overline{v}}\frac{1}{h_{r}}K_{r}\left(\frac{V_{il}-u}{h_{r}}\right)s\left(u\right)\mathrm{d}u\\
= & \int_{\underline{v}}^{\overline{v}}\frac{1}{h_{r}}K_{r}\left(\frac{V_{il}-u}{h_{r}}\right)s\left(u\right)\mathrm{d}u-\underline{b},
\end{align*}
when $h_{r}$ is small enough, for all $V_{il}$ satisfying $V_{il}\in\left[v_{l}-\delta_{0},v_{u}+\delta_{0}\right]$. 

Denote
\[
\beta_{s}\left(w\right)\coloneqq\int_{\underline{v}}^{\overline{v}}\frac{1}{h_{r}}K_{r}\left(\frac{w-u}{h_{r}}\right)s\left(u\right)\mathrm{d}u-s\left(w\right)
\]
and write
\begin{align}
 & \frac{1}{N\cdot L}\sum_{i,l}\frac{1}{h_{f}^{2}}K_{f}'\left(\frac{V_{il}-v}{h_{f}}\right)\frac{1}{s'\left(V_{il}\right)}\left(\int_{\underline{v}}^{\overline{v}}\frac{1}{h_{r}}K_{r}\left(\frac{V_{il}-u}{h_{r}}\right)s\left(u\right)\mathrm{d}u-s\left(V_{il}\right)\right)\nonumber \\
= & \left\{ \frac{1}{N\cdot L}\sum_{i,l}\frac{1}{h_{f}^{2}}K_{f}'\left(\frac{V_{il}-v}{h_{f}}\right)\frac{\beta_{s}\left(V_{il}\right)}{s'\left(V_{il}\right)}-\mathrm{E}\left[\frac{1}{h_{f}^{2}}K_{f}'\left(\frac{V_{11}-v}{h_{f}}\right)\frac{\beta_{s}\left(V_{11}\right)}{s'\left(V_{11}\right)}\right]\right\} +\mathrm{E}\left[\frac{1}{h_{f}^{2}}K_{f}'\left(\frac{V_{11}-v}{h_{f}}\right)\frac{\beta_{s}\left(V_{11}\right)}{s'\left(V_{11}\right)}\right].\label{eq:average K_prime * (integral K_til - s) decomposition}
\end{align}

By standard argument for kernel density estimation (see, e.g., \citealp{Newey_Kernel_ET_1994}),
since $s$ is three-times continuously differentiable,
\begin{gather}
\beta_{s}\left(w\right)=\frac{h_{r}^{2}}{2}s''\left(w\right)\int u^{2}K_{r}\left(u\right)\mathrm{d}u+o\left(h^{2}\right)\nonumber \\
\beta_{s}'\left(w\right)=\frac{h_{r}^{2}}{2}s'''\left(w\right)\int u^{2}K_{r}\left(u\right)\mathrm{d}u+o\left(h^{2}\right),\label{eq:beta_s expansion}
\end{gather}
where the remainder terms are uniform in $v\in I$.

By change of variables,
\begin{align}
 & \mathrm{E}\left[\frac{1}{h_{f}^{2}}K_{f}'\left(\frac{V_{11}-v}{h_{f}}\right)\frac{\beta_{s}\left(V_{11}\right)}{s'\left(V_{11}\right)}\right]\nonumber \\
= & \int_{\frac{\underline{v}-v}{h_{f}}}^{\frac{\overline{v}-v}{h_{f}}}\frac{1}{h_{f}}K_{f}'\left(z\right)\frac{\beta_{s}\left(h_{f}z+v\right)}{s'\left(h_{f}z+v\right)}f\left(h_{f}z+v\right)\mathrm{d}z\nonumber \\
= & \int_{\frac{\underline{v}-v}{h_{f}}}^{\frac{\overline{v}-v}{h_{f}}}\frac{1}{h_{f}}K_{f}'\left(z\right)\left\{ \frac{\beta_{s}\left(v\right)f\left(v\right)}{s'\left(v\right)}+\frac{\left(\beta_{s}'\left(\dot{v}\right)f\left(\dot{v}\right)+\beta_{s}\left(\dot{v}\right)f'\left(\dot{v}\right)\right)s'\left(\dot{v}\right)-\beta_{s}\left(\dot{v}\right)f\left(\dot{v}\right)s''\left(\dot{v}\right)}{s'\left(\dot{v}\right)^{2}}h_{f}z\right\} \mathrm{d}z,\label{eq:expectation zeta_1 decomposition}
\end{align}
where $\dot{v}$ is the mean value depending on $z$ with $\left|\dot{v}-v\right|\leq h_{f}\left|z\right|$.
It is clear that for small enough $h_{f}$, 
\[
\frac{\beta_{s}\left(v\right)f\left(v\right)}{s'\left(v\right)}\left(\int_{\frac{\underline{v}-v}{h_{f}}}^{\frac{\overline{v}-v}{h_{f}}}K_{f}'\left(z\right)\mathrm{d}z\right)=0\textrm{ for all \ensuremath{v\in I}}.
\]
Now it follows from (\ref{eq:beta_s expansion}) and (\ref{eq:expectation zeta_1 decomposition})
that
\begin{equation}
\mathrm{E}\left[\frac{1}{h_{f}^{2}}K_{f}'\left(\frac{V_{11}-v}{h_{f}}\right)\frac{\beta_{s}\left(V_{11}\right)}{s'\left(V_{11}\right)}\right]=\frac{h_{r}^{2}}{2}\frac{\left(s'''\left(v\right)f\left(v\right)+s''\left(v\right)f'\left(v\right)\right)s'\left(v\right)-s''\left(v\right)f\left(v\right)s''\left(v\right)}{s'\left(v\right)^{2}}\int u^{2}K_{r}\left(u\right)\mathrm{d}u+o\left(h^{2}\right),\label{eq:S expectation}
\end{equation}
where the remainder term is uniform in $v\in I$.

Denote 
\[
\mathcal{S}\left(z;v\right)\coloneqq\frac{1}{h_{f}^{2}}K_{f}'\left(\frac{z-v}{h_{f}}\right)\frac{\beta_{s}\left(z\right)}{s'\left(z\right)}
\]
and thus
\[
\frac{1}{N\cdot L}\sum_{i,l}\frac{1}{h_{f}^{2}}K_{f}'\left(\frac{V_{il}-v}{h_{f}}\right)\frac{\beta_{s}\left(V_{il}\right)}{s'\left(V_{il}\right)}-\mathrm{E}\left[\frac{1}{h_{f}^{2}}K_{f}'\left(\frac{V_{11}-v}{h_{f}}\right)\frac{\beta_{s}\left(V_{11}\right)}{s'\left(V_{11}\right)}\right]=\frac{1}{N\cdot L}\sum_{i,l}\mathcal{S}\left(V_{il};v\right)-\mathrm{E}\left[\mathcal{S}\left(V_{11};v\right)\right].
\]
By standard arguments (see the proof of Lemma B.3 of MMS), it can
be easily verified that $\left\{ \mathcal{S}\left(\cdot;v\right):v\in I\right\} $
is (uniformly) VC-type with respect to a constant envelope $F_{\mathcal{S}}$
which satisfies $F_{\mathcal{S}}\apprle h_{f}^{-2}\underset{v\in\left[v_{l}-\delta_{0},v_{u}+\delta_{0}\right]}{\mathrm{sup}}\left|\beta_{s}\left(v\right)\right|$
when $h_{f}$ is small enough. The applying a maximal inequality (\citealp[Theorem 2.14.1]{van1996weak})
yields 
\[
\mathrm{E}\left[\left|\frac{1}{N\cdot L}\sum_{i,l}\mathcal{S}\left(V_{il};v\right)-\mathrm{E}\left[\mathcal{S}\left(V_{11};v\right)\right]\right|\right]\leq L^{-1/2}\left|F_{\mathcal{S}}\right|=O\left(L^{-1/2}\right),
\]
when $h$ is small enough. The conclusion follows from this result,
(\ref{eq:average K_prime V_hat_dagger - V}), (\ref{eq:eq:average K_prime V_hat_dagger - V inter_2}),
(\ref{eq:eq:average K_prime V_hat_dagger - V inter_3}) and (\ref{eq:S expectation}).
\end{proof}
\begin{lem}
\label{lem:Lemma 5}Suppose that Assumptions \ref{assu:DGP} - \ref{assu:bandwidth}
hold. We have
\begin{eqnarray*}
\widehat{f}_{RGPV}\left(v\right)-f\left(v\right) & = & \frac{1}{\left(N-1\right)}\frac{1}{\left(N\cdot L\right)^{2}}\sum_{i,l}\sum_{j,k}\mathcal{M}\left(B_{il},B_{jk};v\right)+\frac{1}{2}f''\left(v\right)\left(\int K_{f}\left(u\right)u^{2}\mathrm{d}u\right)h_{f}^{2}\\
 &  & +\frac{1}{2}\frac{\left(s'''\left(v\right)f\left(v\right)+s''\left(v\right)f'\left(v\right)\right)s'\left(v\right)-s''\left(v\right)^{2}f\left(v\right)}{s'\left(v\right)^{2}}\left(\int K_{r}\left(u\right)u^{2}\mathrm{d}u\right)h_{r}^{2}\\
 &  & +O_{p}\left(\frac{\mathrm{log}\left(L\right)}{Lh^{3}}+\left(\frac{\mathrm{log}\left(L\right)}{Lh}\right)^{1/2}+h^{2}\right),
\end{eqnarray*}
where the remainder term is uniform in $v\in I$.
\end{lem}
\begin{proof}[Proof of Lemma \ref{lem:Lemma 5}]First we show
\begin{align*}
 & -\frac{1}{N\cdot L}\sum_{i,l}\frac{1}{h_{f}^{2}}\mathbb{T}_{il}K_{f}'\left(\frac{V_{il}-v}{h_{f}}\right)\frac{\widehat{s}\left(V_{il}\right)-\widetilde{s}\left(V_{il}\right)}{s'\left(V_{il}\right)}\\
= & -\frac{1}{N-1}\frac{1}{N\cdot L}\sum_{i,l}\frac{1}{h_{f}^{2}}\mathbb{T}_{il}K_{f}'\left(\frac{V_{il}-v}{h_{f}}\right)\frac{1}{s'\left(V_{il}\right)}\int_{\underline{b}}^{\overline{b}}\frac{1}{h_{r}}K_{r}\left(\frac{\xi\left(b\right)-V_{il}}{h_{r}}\right)\frac{G\left(b\right)}{g\left(b\right)^{2}}\left(\widehat{g}\left(b\right)-g\left(b\right)\right)\mathrm{d}b\\
 & +O_{p}\left(\frac{\mathrm{log}\left(L\right)}{Lh^{3}}+h^{3}+\left(\frac{\mathrm{log}\left(L\right)}{Lh}\right)^{\nicefrac{1}{2}}\right).
\end{align*}

Since by the Borel-Cantelli lemma we have
\[
\left|\widehat{\underline{b}}-\underline{b}\right|=O_{p}\left(\frac{\mathrm{\mathrm{log}}\left(L\right)}{L}\right),\,\left|\widehat{\overline{b}}-\overline{b}\right|=O_{p}\left(\frac{\mathrm{log}\left(L\right)}{L}\right),
\]
therefore,
\begin{align}
 & \frac{1}{N\cdot L}\sum_{i,l}\mathbb{T}_{il}\frac{1}{h_{f}^{2}}K_{f}'\left(\frac{V_{il}-v}{h_{f}}\right)\frac{\widehat{s}\left(V_{il}\right)-\widetilde{s}\left(V_{il}\right)}{s'\left(V_{il}\right)}\nonumber \\
= & \frac{1}{N\cdot L}\sum_{i,l}\mathbb{T}_{il}\frac{1}{h_{f}^{2}}K_{f}'\left(\frac{V_{il}-v}{h_{f}}\right)\frac{1}{s'\left(V_{il}\right)}\int_{\underline{b}}^{\overline{b}}\left(\widetilde{K}_{r}\left(\frac{V_{il}-\widehat{\xi}\left(b\right)}{h_{r}}\right)-\widetilde{K}_{r}\left(\frac{V_{il}-\xi\left(b\right)}{h_{r}}\right)\right)\mathrm{d}b+O_{p}\left(\frac{\mathrm{log}\left(L\right)}{Lh}\right),\label{eq:average PSI*ksi_hat - PSI*ksi decomposition 1}
\end{align}
where the remainder term is uniform in $v\in I$. 

By a second-order Taylor expansion, we have
\begin{align}
 & \frac{1}{N\cdot L}\sum_{i,l}\mathbb{T}_{il}\frac{1}{h_{f}^{2}}K_{f}'\left(\frac{V_{il}-v}{h_{f}}\right)\frac{1}{s'\left(V_{il}\right)}\int_{\underline{b}}^{\overline{b}}\left(\widetilde{K}_{r}\left(\frac{V_{il}-\widehat{\xi}\left(b\right)}{h_{r}}\right)-\widetilde{K}_{r}\left(\frac{V_{il}-\xi\left(b\right)}{h_{r}}\right)\right)\mathrm{d}b\nonumber \\
= & \frac{1}{N-1}\frac{1}{N\cdot L}\sum_{i,l}\mathbb{T}_{il}\frac{1}{h_{f}^{2}}K_{f}'\left(\frac{V_{il}-v}{h_{f}}\right)\frac{1}{s'\left(V_{il}\right)}\int_{\underline{b}}^{\overline{b}}\frac{1}{h_{r}}K_{r}\left(\frac{V_{il}-\xi\left(b\right)}{h_{r}}\right)\frac{G\left(b\right)}{g\left(b\right)^{2}}\left(\widehat{g}\left(b\right)-g\left(b\right)\right)\mathrm{d}b\nonumber \\
 & -\frac{1}{N-1}\frac{1}{N\cdot L}\sum_{i,l}\mathbb{T}_{il}\frac{1}{h_{f}^{2}}K_{f}'\left(\frac{V_{il}-v}{h_{f}}\right)\frac{1}{s'\left(V_{il}\right)}\int_{\underline{b}}^{\overline{b}}\frac{1}{h_{r}}K_{r}\left(\frac{V_{il}-\xi\left(b\right)}{h_{r}}\right)\frac{\widehat{G}\left(b\right)-G\left(b\right)}{g\left(b\right)}\mathrm{d}b\nonumber \\
 & +I_{1}\left(v\right)+I_{2}\left(v\right)\label{eq:average PSI*ksi_hat - PSI*ksi decomposition 2}
\end{align}
where
\begin{eqnarray*}
I_{1}\left(v\right) & \coloneqq & -\frac{1}{N\cdot L}\sum_{i,l}\mathbb{T}_{il}\frac{1}{h_{f}^{2}}K_{f}'\left(\frac{V_{il}-v}{h_{f}}\right)\frac{1}{s'\left(V_{il}\right)}\int_{\underline{b}}^{\overline{b}}\frac{1}{h_{r}}K_{r}\left(\frac{V_{il}-\xi\left(b\right)}{h_{r}}\right)\\
 &  & \times\left\{ \left(\widehat{\xi}\left(b\right)-\xi\left(b\right)\right)+\frac{1}{N-1}\frac{G\left(b\right)}{g\left(b\right)^{2}}\left(\widehat{g}\left(b\right)-g\left(b\right)\right)-\frac{1}{N-1}\frac{\widehat{G}\left(b\right)-G\left(b\right)}{g\left(b\right)}\right\} \mathrm{d}b
\end{eqnarray*}
and 
\[
I_{2}\left(v\right)\coloneqq\frac{1}{N\cdot L}\sum_{i,l}\mathbb{T}_{il}\frac{1}{h_{f}^{2}}K_{f}'\left(\frac{V_{il}-v}{h_{f}}\right)\frac{1}{s'\left(V_{il}\right)}\int_{\underline{b}}^{\overline{b}}\frac{1}{h_{r}^{2}}K_{r}'\left(\frac{V_{il}-\dot{\xi}\left(b\right)}{h_{r}}\right)\left(\widehat{\xi}\left(b\right)-\xi\left(b\right)\right)^{2}\mathrm{d}b
\]
for some mean value $\dot{\xi}\left(b\right)$ with $\left|\dot{\xi}\left(b\right)-\xi\left(b\right)\right|\leq\left|\widehat{\xi}\left(b\right)-\xi\left(b\right)\right|$
for each $b\in\left[\underline{b},\overline{b}\right]$. 

Then, by Lemma \ref{lem:Lemma 1},
\begin{eqnarray}
\underset{v\in I}{\mathrm{sup}}\left|I_{1}\left(v\right)\right| & \apprle & \left\{ \underset{v\in I}{\mathrm{sup}}\frac{1}{N\cdot L}\sum_{i,l}\frac{1}{h_{f}^{2}}\left|K_{f}'\left(\frac{V_{il}-v}{h_{f}}\right)\right|\right\} \left\{ \underset{v\in\left[v_{l}-\delta_{0},v_{u}+\delta_{0}\right]}{\mathrm{sup}}\int_{\underline{b}}^{\overline{b}}\frac{1}{h_{r}}K_{r}\left(\frac{v-\xi\left(b\right)}{h_{r}}\right)\mathrm{d}b\right\} \nonumber \\
 &  & \times\underset{b\in\left[\underline{b},\overline{b}\right]}{\mathrm{sup}}\left|\left(\widehat{\xi}\left(b\right)-\xi\left(b\right)\right)+\frac{1}{N-1}\frac{G\left(b\right)}{g\left(b\right)^{2}}\left(\widehat{g}\left(b\right)-g\left(b\right)\right)-\frac{1}{N-1}\frac{\widehat{G}\left(b\right)-G\left(b\right)}{g\left(b\right)}\right|\nonumber \\
 & = & O_{p}\left(\frac{\mathrm{log}\left(L\right)}{Lh^{2}}+h^{3}\right)\label{eq:I_1 uniform bound}
\end{eqnarray}
and
\begin{eqnarray}
\underset{v\in I}{\mathrm{sup}}\left|I_{2}\left(v\right)\right| & \apprle & \left\{ \underset{v\in I}{\mathrm{sup}}\frac{1}{N\cdot L}\sum_{i,l}\frac{1}{h_{f}^{2}}\left|K_{f}'\left(\frac{V_{il}-v}{h_{f}}\right)\right|\right\} \left\{ \underset{v\in\left[v_{l}-\delta_{0},v_{u}+\delta_{0}\right]}{\mathrm{sup}}\int_{\underline{b}}^{\overline{b}}\frac{1}{h_{r}^{2}}K\left(\frac{v-\dot{\xi}\left(b\right)}{h_{r}}\right)\mathrm{d}b\right\} r_{\xi}^{2}\nonumber \\
 & = & O_{p}\left(\frac{\mathrm{log}\left(L\right)}{Lh^{3}}+h^{3}\right),\label{eq:I_2 uniform bound}
\end{eqnarray}
when the inequalities hold when $h_{r}$ is small enough. 

Define 
\[
\mathcal{G}\left(b,b';v\right)\coloneqq\frac{1}{h_{f}^{2}}K_{f}'\left(\frac{\xi\left(b\right)-v}{h_{f}}\right)\frac{1}{s'\left(\xi\left(b\right)\right)}\int_{\underline{b}}^{\overline{b}}\frac{1}{h_{r}}K_{r}\left(\frac{\xi\left(b\right)-\xi\left(z\right)}{h_{r}}\right)\frac{\mathbbm{1}\left(b'\leq z\right)-G\left(z\right)}{g\left(z\right)}\mathrm{d}z.
\]
By the definition of $\mathcal{G}$, we have 
\begin{align}
 & \frac{1}{N\cdot L}\sum_{i,l}\mathbb{T}_{il}\frac{1}{h_{f}^{2}}K_{f}'\left(\frac{V_{il}-v}{h_{f}}\right)\frac{1}{s'\left(V_{il}\right)}\int_{\underline{b}}^{\overline{b}}\frac{1}{h_{r}}K_{r}\left(\frac{V_{il}-\xi\left(b\right)}{h_{r}}\right)\frac{\widehat{G}\left(b\right)-G\left(b\right)}{g\left(b\right)}\mathrm{d}b\nonumber \\
= & \frac{1}{\left(N\cdot L\right)^{2}}\sum_{\left(2\right)}\mathcal{G}\left(B_{il},B_{jk};v\right)+\frac{1}{\left(N\cdot L\right)^{2}}\sum_{i,l}\mathcal{G}\left(B_{il},B_{il};v\right),\,\textrm{for all \ensuremath{v\in I}},\label{eq:G decomposition 1}
\end{align}
when $h_{f}$ is small enough. The kernel $\mathcal{G}$ satisfies
\[
\mathcal{G}_{1}\left(b;v\right)\coloneqq\int\mathcal{G}\left(b,b';v\right)\mathrm{d}G\left(b'\right)=0\textrm{ and }\mu_{\mathcal{G}}\left(v\right)\coloneqq\int\int\mathcal{G}\left(b,b';v\right)\mathrm{d}G\left(b'\right)\mathrm{d}G\left(b\right)=0,\textrm{ for all \ensuremath{v\in I}}.
\]
Also define 
\[
\mathcal{G}_{2}\left(b;v\right)\coloneqq\int\mathcal{G}\left(b',b;v\right)\mathrm{d}G\left(b'\right).
\]
Hoeffding decomposition gives 
\begin{equation}
\frac{1}{\left(N\cdot L\right)_{2}}\sum_{\left(2\right)}\mathcal{G}\left(B_{il},B_{jk};v\right)=\frac{1}{N\cdot L}\sum_{i,l}\mathcal{G}_{2}\left(B_{il};v\right)+\frac{1}{\left(N\cdot L\right)_{2}}\sum_{\left(2\right)}\left\{ \mathcal{G}\left(B_{il},B_{jk};v\right)-\mathcal{G}_{2}\left(B_{il};v\right)\right\} .\label{eq:G decomposition 2}
\end{equation}
By standard arguments (see the proof of Lemma B.2 of MMS for details),
it can be easily verified that $\left\{ \mathcal{G}\left(\cdot,\cdot;v\right):v\in I\right\} $
is (uniformly) VC-type with respect to a constant envelope $F_{\mathcal{G}}$
which satisfies $F_{\mathcal{G}}\apprle h_{f}^{-2}$. This implies
\begin{equation}
\underset{v\in I}{\mathrm{sup}}\left|\frac{1}{\left(N\cdot L\right)^{2}}\sum_{i,l}\mathcal{G}\left(B_{il},B_{il};v\right)\right|=O_{p}\left(\left(Lh^{2}\right)^{-1}\right).\label{eq:G decomposition 3}
\end{equation}
Application of a maximal inequality (\citealp[Corollary 5.6]{Chen_Kato_U_Process})
and Markov's inequality gives 
\begin{equation}
\underset{v\in I}{\mathrm{sup}}\left|\frac{1}{\left(N\cdot L\right)_{2}}\sum_{\left(2\right)}\left\{ \mathcal{G}\left(B_{il},B_{jk};v\right)-\mathcal{G}_{2}\left(B_{il};v\right)\right\} \right|=O_{p}\left(\left(Lh^{2}\right)^{-1}\right).\label{eq:G decomposition 4}
\end{equation}

Next, we show that 
\begin{equation}
\underset{v\in I}{\mathrm{sup}}\,\mathrm{E}\left[\mathcal{G}_{2}\left(B_{11};v\right)^{2}\right]\apprle h^{-1},\label{eq:E=00005BG^2=00005D uniform bound}
\end{equation}
when $h$ is small enough. Denote 
\[
\tau\left(z\right)\coloneqq\int_{\underline{b}}^{\overline{b}}\frac{1}{h_{f}^{2}}K_{f}'\left(\frac{\xi\left(b\right)-v}{h_{f}}\right)\xi'\left(b\right)\frac{1}{h_{r}}K_{r}\left(\frac{\xi\left(b\right)-\xi\left(z\right)}{h_{r}}\right)g\left(b\right)\mathrm{d}b.
\]
Then by change of variables and the Fubini-Tonelli theorem,
\begin{align}
 & \int\mathcal{G}_{2}\left(b;v\right)^{2}\mathrm{d}G\left(b\right)\nonumber \\
\leq & \int_{\underline{b}}^{\overline{b}}\left\{ \int_{\underline{b}}^{\overline{b}}\tau\left(z\right)\frac{\mathbbm{1}\left(b\leq z\right)}{g\left(z\right)}\mathrm{d}z\right\} ^{2}g\left(b\right)\mathrm{d}b\nonumber \\
= & \int_{\underline{b}}^{\overline{b}}\int_{\underline{b}}^{\overline{b}}\frac{\tau\left(z\right)}{g\left(z\right)}\frac{\tau\left(z'\right)}{g\left(z'\right)}G\left(\mathrm{min}\left\{ z,z'\right\} \right)\mathrm{d}z\mathrm{d}z'\nonumber \\
= & h_{r}^{2}\int_{\frac{\underline{v}-v}{h_{r}}}^{\frac{\overline{v}-v}{h_{r}}}\int_{\frac{\underline{v}-v}{h_{r}}}^{\frac{\overline{v}-v}{h_{r}}}\frac{\tau\left(s\left(h_{r}w+v\right)\right)}{g\left(s\left(h_{r}w+v\right)\right)}\frac{\tau\left(s\left(h_{r}w'+v\right)\right)}{g\left(s\left(h_{r}w'+v\right)\right)}G\left(\mathrm{min}\left\{ s\left(h_{r}w+v\right),s\left(h_{r}w'+v\right)\right\} \right)s'\left(h_{r}w+v\right)s'\left(h_{r}w'+v\right)\mathrm{d}w\mathrm{d}w'\nonumber \\
= & \frac{2}{h_{f}^{2}}\int_{\frac{\underline{v}-v}{h_{r}}}^{\frac{\overline{v}-v}{h_{r}}}\frac{s'\left(h_{r}w'+v\right)}{g\left(s\left(h_{r}w'+v\right)\right)}\int_{\frac{\underline{v}-v}{h_{r}}}^{\frac{\overline{v}-v}{h_{r}}}K_{f}'\left(u'\right)K_{r}\left(\frac{\lambda_{f}}{\lambda_{r}}u'-w'\right)g\left(s\left(h_{f}u'+v\right)\right)\mathrm{d}u'\nonumber \\
 & \times\int_{\frac{\underline{v}-v}{h_{r}}}^{w'}\frac{s'\left(h_{r}w+v\right)G\left(s\left(h_{r}w+v\right)\right)}{g\left(s\left(h_{r}w+v\right)\right)}\int_{\frac{\underline{v}-v}{h_{r}}}^{\frac{\overline{v}-v}{h_{r}}}K_{f}'\left(u\right)K_{r}\left(\frac{\lambda_{f}}{\lambda_{r}}u-w\right)g\left(s\left(h_{f}u+v\right)\right)\mathrm{d}u\mathrm{d}w\mathrm{d}w',\label{eq:E=00005BG^2=00005D decomposition}
\end{align}
where the last equality follows from symmetry. It follows from integration
by parts that 
\[
\int\int_{-\infty}^{w'}\left\{ \int K_{f}'\left(u'\right)K_{r}\left(\frac{\lambda_{f}}{\lambda_{r}}u'-w'\right)\mathrm{d}u'\right\} \left\{ \int K_{f}'\left(u\right)K_{r}\left(\frac{\lambda_{f}}{\lambda_{r}}u-w\right)\mathrm{d}u\right\} \mathrm{d}w\mathrm{d}w'=0.
\]
Then (\ref{eq:E=00005BG^2=00005D uniform bound}) follows from this
result, (\ref{eq:E=00005BG^2=00005D decomposition}) and Taylor expansion. 

Since $\left\{ \mathcal{G}\left(\cdot,\cdot;v\right):v\in I\right\} $
is (uniformly) VC-type with respect to a constant envelope $F_{\mathcal{G}}$
which satisfies $F_{\mathcal{G}}\apprle h_{f}^{-2}$, it follows from
(\ref{eq:E=00005BG^2=00005D uniform bound}) and a maximal inequality
(\citealp[Corollary 5.1]{chernozhukov2014anti}) that 
\[
\underset{v\in I}{\mathrm{sup}}\left|\frac{1}{N\cdot L}\sum_{i,l}\mathcal{G}_{2}\left(B_{il};v\right)\right|=O_{p}\left(\left(\frac{\mathrm{log}\left(L\right)}{Lh}\right)^{1/2}+\frac{\mathrm{log}\left(L\right)}{Lh^{2}}\right).\footnotemark
\]
\footnotetext{See the proof of Lemma B.2 of MMS for more details.}Now
it follows from this result, (\ref{eq:G decomposition 1}), (\ref{eq:G decomposition 2}),
(\ref{eq:G decomposition 3}) and (\ref{eq:G decomposition 4}) that
\[
\frac{1}{N\cdot L}\sum_{i,l}\mathbb{T}_{il}\frac{1}{h_{f}^{2}}K_{f}'\left(\frac{V_{il}-v}{h_{f}}\right)\frac{1}{s'\left(V_{il}\right)}\int_{\underline{b}}^{\overline{b}}\frac{1}{h_{r}}K_{r}\left(\frac{V_{il}-\xi\left(b\right)}{h_{r}}\right)\frac{\widehat{G}\left(b\right)-G\left(b\right)}{g\left(b\right)}\mathrm{d}b=O_{p}\left(\left(\frac{\mathrm{log}\left(L\right)}{Lh}\right)^{1/2}+\frac{\mathrm{log}\left(L\right)}{Lh^{2}}\right),
\]
uniformly in $v\in I$. Then it follows from this result, (\ref{eq:average PSI*ksi_hat - PSI*ksi decomposition 1}),
(\ref{eq:average PSI*ksi_hat - PSI*ksi decomposition 2}), (\ref{eq:I_1 uniform bound})
and (\ref{eq:I_2 uniform bound}) that 
\begin{align*}
 & \frac{1}{N\cdot L}\sum_{i,l}\mathbb{T}_{il}\frac{1}{h_{f}^{2}}K_{f}'\left(\frac{V_{il}-v}{h_{f}}\right)\frac{\widehat{s}\left(V_{il}\right)-\widetilde{s}\left(V_{il}\right)}{s'\left(V_{il}\right)}\\
= & \frac{1}{N-1}\frac{1}{N\cdot L}\sum_{i,l}\mathbb{T}_{il}\frac{1}{h_{f}^{2}}K_{f}'\left(\frac{V_{il}-v}{h_{f}}\right)\frac{1}{s'\left(V_{il}\right)}\int_{\underline{b}}^{\overline{b}}\frac{1}{h_{r}}K_{r}\left(\frac{V_{il}-\xi\left(b\right)}{h_{r}}\right)\frac{G\left(b\right)}{g\left(b\right)^{2}}\left(\widehat{g}\left(b\right)-g\left(b\right)\right)\mathrm{d}b\\
 & +O_{p}\left(\left(\frac{\mathrm{log}\left(L\right)}{Lh}\right)^{1/2}+\frac{\mathrm{log}\left(L\right)}{Lh^{2}}+h^{3}\right),
\end{align*}
where the remainder term is uniform in $v\in I$.

By the definition of the MCE, 
\[
\widehat{g}\left(b\right)=\frac{1}{N\cdot L}\sum_{i,l}\frac{1}{h_{g}}K_{g}\left(\frac{B_{il}-b}{h_{g}}\right),\textrm{ for \ensuremath{b\in\left[\widehat{\underline{b}}+h_{g},\widehat{\overline{b}}-h_{g}\right].}}
\]
Therefore, since $K_{f}'$ and $K_{r}$ are compactly supported on
$\left[-1,1\right]$, it is easy to verify that
\begin{align*}
 & -\frac{1}{N-1}\frac{1}{N\cdot L}\sum_{i,l}\frac{1}{h_{f}^{2}}\mathbb{T}_{il}K_{f}'\left(\frac{V_{il}-v}{h_{f}}\right)\frac{1}{s'\left(V_{il}\right)}\int_{\underline{b}}^{\overline{b}}\frac{1}{h_{r}}K_{r}\left(\frac{\xi\left(b\right)-V_{il}}{h_{r}}\right)\frac{G\left(b\right)}{g\left(b\right)^{2}}\left(\widehat{g}\left(b\right)-g\left(b\right)\right)\mathrm{d}b\\
= & -\frac{1}{N-1}\frac{1}{N\cdot L}\sum_{i,l}\frac{1}{h_{f}^{2}}\mathbb{T}_{il}K_{f}'\left(\frac{V_{il}-v}{h_{f}}\right)\frac{1}{s'\left(V_{il}\right)}\int_{\underline{b}}^{\overline{b}}\frac{1}{h_{r}}K_{r}\left(\frac{\xi\left(b\right)-V_{il}}{h_{r}}\right)\frac{G\left(b\right)}{g\left(b\right)^{2}}\\
 & \times\left(\frac{1}{N\cdot L}\sum_{j,k}\frac{1}{h_{g}}K_{g}\left(\frac{B_{jk}-b}{h_{g}}\right)-g\left(b\right)\right)\mathrm{d}b,\textrm{ for all \ensuremath{v\in I}, w.p.a.1.}
\end{align*}
The conclusion follows from this result, (\ref{eq:average PSI*ksi_hat - PSI*ksi decomposition 1}),
(\ref{eq:average PSI*ksi_hat - PSI*ksi decomposition 2}), (\ref{eq:I_1 uniform bound}),
(\ref{eq:I_2 uniform bound}) and Lemma \ref{lem:Lemma 5}. \end{proof}
\begin{lem}
\label{lem:Lemma 6}Suppose that Assumptions \ref{assu:DGP} - \ref{assu:bandwidth}
hold. Let 
\begin{equation}
\beta_{g}\left(b\right)\coloneqq\int\left\{ \frac{1}{h_{g}}K_{g}\left(\frac{b'-b}{h_{g}}\right)-g\left(b\right)\right\} \mathrm{d}G\left(b'\right)\label{eq:bias of KDE of g(b)}
\end{equation}
be the bias of the kernel density estimator of $g\left(b\right)$.
Then
\[
\mu_{\mathcal{M}}\left(v\right)\coloneqq-\int\left\{ \frac{1}{h_{f}^{2}}K_{f}'\left(\frac{\xi\left(u\right)-v}{h_{f}}\right)\frac{1}{s'\left(\xi\left(u\right)\right)}\int_{\underline{b}}^{\overline{b}}\frac{1}{h_{r}}K_{r}\left(\frac{\xi\left(b\right)-\xi\left(u\right)}{h_{r}}\right)\frac{G\left(b\right)\beta_{g}\left(b\right)}{g\left(b\right)^{2}}\mathrm{d}b\right\} \mathrm{d}G\left(u\right)=o\left(h^{2}\right),
\]
uniformly in $v\in I$.
\end{lem}
\begin{proof}[Proof of Lemma \ref{lem:Lemma 6}]By change of variables,
\begin{align}
 & \int\left\{ \frac{1}{h_{f}^{2}}K_{f}'\left(\frac{\xi\left(u\right)-v}{h_{f}}\right)\frac{1}{s'\left(\xi\left(u\right)\right)}\int_{\underline{b}}^{\overline{b}}\frac{1}{h_{r}}K_{r}\left(\frac{\xi\left(b\right)-\xi\left(u\right)}{h_{r}}\right)\frac{G\left(b\right)\beta_{g}\left(b\right)}{g\left(b\right)^{2}}\mathrm{d}b\right\} \mathrm{d}G\left(u\right)\nonumber \\
= & \int_{\frac{\underline{v}-v}{h_{f}}}^{\frac{\overline{v}-v}{h_{f}}}\int_{\frac{\underline{v}-v}{h_{r}}}^{\frac{\overline{v}-v}{h_{r}}}\frac{1}{h_{f}}K_{f}'\left(w\right)K_{r}\left(z-\frac{\lambda_{f}}{\lambda_{r}}w\right)\frac{G\left(s\left(h_{r}z+v\right)\right)\beta_{g}\left(s\left(h_{r}z+v\right)\right)}{g\left(s\left(h_{r}z+v\right)\right)^{2}}s'\left(h_{r}z+v\right)g\left(s\left(h_{f}w+v\right)\right)\mathrm{d}z\mathrm{d}w.\label{eq:miu_M change of variable}
\end{align}
Let $\psi\left(z\right)\coloneqq G\left(s\left(z\right)\right)s'\left(z\right)/g\left(s\left(z\right)\right)^{2}$.
By a mean value expansion, the second line of (\ref{eq:miu_M change of variable})
is equal to
\begin{align}
 & \frac{1}{h_{f}}\int_{\frac{\underline{v}-v}{h_{f}}}^{\frac{\overline{v}-v}{h_{f}}}\int_{\frac{\underline{v}-v}{h_{r}}}^{\frac{\overline{v}-v}{h_{r}}}K_{f}'\left(w\right)K_{r}\left(z-\frac{\lambda_{f}}{\lambda_{r}}w\right)\left\{ \psi\left(v\right)\beta_{g}\left(s\left(v\right)\right)+\left(\psi'\left(\dot{v}\right)\beta_{g}\left(s\left(\dot{v}\right)\right)+\psi\left(\dot{v}\right)\beta_{g}'\left(s\left(\dot{v}\right)\right)s'\left(\dot{v}\right)\right)h_{r}z\right\} \nonumber \\
 & \times\left\{ g\left(s\left(v\right)\right)+g'\left(s\left(\ddot{v}\right)\right)s'\left(\ddot{v}\right)h_{f}w\right\} \mathrm{d}z\mathrm{d}w\nonumber \\
= & \psi\left(v\right)\beta_{g}\left(s\left(v\right)\right)g\left(s\left(v\right)\right)\frac{1}{h_{f}}\int_{\frac{\underline{v}-v}{h_{f}}}^{\frac{\overline{v}-v}{h_{f}}}\int_{\frac{\underline{v}-v}{h_{r}}}^{\frac{\overline{v}-v}{h_{r}}}K_{f}'\left(w\right)K_{r}\left(z-\frac{\lambda_{f}}{\lambda_{r}}w\right)\mathrm{d}z\mathrm{d}w\nonumber \\
 & +\psi\left(v\right)\beta_{g}\left(s\left(v\right)\right)\int_{\frac{\underline{v}-v}{h_{f}}}^{\frac{\overline{v}-v}{h_{f}}}\int_{\frac{\underline{v}-v}{h_{r}}}^{\frac{\overline{v}-v}{h_{r}}}K_{f}'\left(w\right)K_{r}\left(z-\frac{\lambda_{f}}{\lambda_{r}}w\right)g'\left(s\left(\ddot{v}\right)\right)s'\left(\ddot{v}\right)w\mathrm{d}z\mathrm{d}w\nonumber \\
 & +g\left(s\left(v\right)\right)\frac{1}{h_{f}}\int_{\frac{\underline{v}-v}{h_{f}}}^{\frac{\overline{v}-v}{h_{f}}}\int_{\frac{\underline{v}-v}{h_{r}}}^{\frac{\overline{v}-v}{h_{r}}}K_{f}'\left(w\right)K_{r}\left(z-\frac{\lambda_{f}}{\lambda_{r}}w\right)\left(\psi'\left(\dot{v}\right)\beta_{g}\left(s\left(\dot{v}\right)\right)+\psi\left(\dot{v}\right)\beta_{g}'\left(s\left(\dot{v}\right)\right)s'\left(\dot{v}\right)\right)h_{r}z\mathrm{d}z\mathrm{d}w\nonumber \\
 & +h_{r}\int_{\frac{\underline{v}-v}{h_{f}}}^{\frac{\overline{v}-v}{h_{f}}}\int_{\frac{\underline{v}-v}{h_{r}}}^{\frac{\overline{v}-v}{h_{r}}}K_{f}'\left(w\right)K_{r}\left(z\right)\left(\psi'\left(\dot{v}\right)\beta_{g}\left(s\left(\dot{v}\right)\right)+\psi\left(\dot{v}\right)\beta_{g}'\left(s\left(\dot{v}\right)\right)s'\left(\dot{v}\right)\right)g'\left(s\left(\ddot{v}\right)\right)s'\left(\ddot{v}\right)zw\mathrm{d}z\mathrm{d}w,\label{eq:miu_M expansion}
\end{align}
where $\dot{v}$ and $\ddot{v}$ are mean values that are dependent
on $z$ and $w$ with $\left|\dot{v}-v\right|\leq h_{r}\left|z\right|$
and $\left|\ddot{v}-v\right|\leq h_{f}\left|w\right|$. When $h$
is small enough,
\begin{equation}
\int_{\frac{\underline{v}-v}{h_{f}}}^{\frac{\overline{v}-v}{h_{f}}}\int_{\frac{\underline{v}-v}{h_{r}}}^{\frac{\overline{v}-v}{h_{r}}}K_{f}'\left(w\right)K_{r}\left(z-\frac{\lambda_{f}}{\lambda_{r}}w\right)\mathrm{d}z\mathrm{d}w=0,\textrm{ for all \ensuremath{v\in I}}.\label{eq:miu_M expansion first term}
\end{equation}

By standard arguments for the bias of kernel estimators for the density
(see, e.g., \citealp{Newey_Kernel_ET_1994}), since $K_{g}$ is supported
on $\left[-1,1\right]$, for each $b\in\left[s\left(v_{l}-\delta_{0}\right),s\left(v_{u}+\delta_{0}\right)\right]$,
\begin{equation}
\left|\beta_{g}\left(b\right)\right|\leq\frac{h_{g}^{3}}{6}\underset{b'\in\left[b-h_{g},b+h_{g}\right]}{\mathrm{sup}}\left\vert g'''\left(b'\right)\right\vert \int\left\vert u^{3}K_{g}\left(u\right)\right\vert \mathrm{d}u,\label{eq:beta bound}
\end{equation}
when $h_{g}$ is small enough. By change of variable and Taylor expansion,
we have
\begin{align}
\underset{b\in\left[s\left(v_{l}-\delta_{0}\right),s\left(v_{u}+\delta_{0}\right)\right]}{\mathrm{sup}}\left|\beta_{g}'\left(b\right)\right|= & \underset{b\in\left[s\left(v_{l}-\delta_{0}\right),s\left(v_{u}+\delta_{0}\right)\right]}{\mathrm{sup}}\left|\int\frac{1}{h_{g}}K_{g}\left(\frac{b'-b}{h_{g}}\right)g'\left(b'\right)\mathrm{d}b'-g'\left(b\right)\right|\nonumber \\
\leq & \underset{b\in\left[s\left(v_{l}-\delta_{0}\right),s\left(v_{u}+\delta_{0}\right)\right]}{\mathrm{sup}}\frac{h_{g}^{2}}{2}\left|\int K_{g}\left(u\right)u^{2}\left(g'''\left(\dot{b}\right)-g'''\left(b\right)\right)\mathrm{d}u\right|,\label{eq:sup beta_prime bound}
\end{align}
when $h_{g}$ is small enough, where $\dot{b}$ is the mean value
depending on $u$ with $\left|\dot{b}-b\right|\leq h_{g}\left|u\right|$.
Since $g'''$ is uniformly continuous on any inner closed subset of
$\left[\underline{b},\overline{b}\right]$, the assumption that $K_{g}$
is supported on $\left[-1,1\right]$ and (\ref{eq:sup beta_prime bound})
imply that $\beta_{g}'\left(b\right)=o\left(h^{2}\right)$ uniformly
in $b\in\left[s\left(v_{l}-\delta_{0}\right),s\left(v_{u}+\delta_{0}\right)\right]$.
The conclusion follows from these results, (\ref{eq:miu_M expansion})
and (\ref{eq:miu_M expansion first term}). \end{proof}
\begin{lem}
\label{lem:Lemma 7}Suppose that Assumptions \ref{assu:DGP} - \ref{assu:bandwidth}
hold. Then 
\begin{eqnarray*}
\widehat{f}_{RGPV}\left(v\right)-f\left(v\right) & = & \frac{1}{\left(N-1\right)}\frac{1}{\left(N\cdot L\right)^{2}}\sum_{i,l}\left(\mathcal{M}_{2}\left(B_{il};v\right)-\mu_{\mathcal{M}}\left(v\right)\right)+\frac{1}{2}f''\left(v\right)\left(\int K_{f}\left(u\right)u^{2}\mathrm{d}u\right)h_{f}^{2}\\
 &  & +\frac{1}{2}\frac{\left(s'''\left(v\right)f\left(v\right)+s''\left(v\right)f'\left(v\right)\right)s'\left(v\right)-s''\left(v\right)^{2}f\left(v\right)}{s'\left(v\right)^{2}}\left(\int K_{r}\left(u\right)u^{2}\mathrm{d}u\right)h_{r}^{2}\\
 &  & +O_{p}\left(\frac{\mathrm{log}\left(L\right)}{Lh^{3}}+\left(\frac{\mathrm{log}\left(L\right)}{Lh}\right)^{1/2}+h^{2}\right).
\end{eqnarray*}
\end{lem}
\begin{proof}[Proof of Lemma \ref{lem:Lemma 7}]Hoeffding decomposition
gives
\begin{eqnarray}
\frac{1}{\left(N\cdot L\right)^{2}}\sum_{i,l}\sum_{j,k}\mathcal{M}\left(B_{il},B_{jk};v\right) & = & \mu_{\mathcal{M}}\left(v\right)+\left\{ \frac{1}{N\cdot L}\sum_{i,l}\mathcal{M}_{1}\left(B_{il};v\right)-\mu_{\mathcal{M}}\left(v\right)\right\} +\left\{ \frac{1}{N\cdot L}\sum_{i,l}\mathcal{M}_{2}\left(B_{il};v\right)-\mu_{\mathcal{M}}\left(v\right)\right\} \nonumber \\
 &  & +\frac{1}{\left(N\cdot L\right)_{2}}\sum_{\left(2\right)}\left\{ \mathcal{M}\left(B_{il},B_{jk};v\right)-\mathcal{M}_{1}\left(B_{il};v\right)-\mathcal{M}_{2}\left(B_{jk};v\right)+\mu_{\mathcal{M}}\left(v\right)\right\} \nonumber \\
 &  & +\frac{1}{\left(N\cdot L\right)^{2}}\sum_{i,l}\mathcal{M}\left(B_{il},B_{il};v\right)-\frac{1}{\left(N\cdot L\right)\left(N\cdot L\right)_{2}}\sum_{\left(2\right)}\mathcal{M}\left(B_{il},B_{jk};v\right).\label{eq:M Hoeffding decomposition}
\end{eqnarray}
By standard arguments used in the proofs of Lemma B.2 and Lemma B.3
of MMS, it can be easily verified that the class $\left\{ \mathcal{M}\left(\cdot,\cdot;v\right):v\in I\right\} $
is (uniformly) VC-type with respect to a constant envelope $F_{\mathcal{M}}$
that satisfies $F_{\mathcal{M}}=O\left(h^{-3}\right)$. Therefore,
it is easy to check that 
\begin{equation}
\left|\frac{1}{\left(N\cdot L\right)^{2}}\sum_{i,l}\mathcal{M}\left(B_{il},B_{il};v\right)\right|=O_{p}\left(\left(Lh^{3}\right)^{-1}\right)\textrm{ and }\left|\frac{1}{\left(N\cdot L\right)\left(N\cdot L\right)_{2}}\sum_{\left(2\right)}\mathcal{M}\left(B_{il},B_{jk};v\right)\right|=O_{p}\left(\left(Lh^{3}\right)^{-1}\right),\label{eq:M decomposition 1}
\end{equation}
uniformly in $v\in I$, and 
\begin{equation}
\underset{v\in I}{\mathrm{sup}}\left|\frac{1}{\left(N\cdot L\right)_{2}}\sum_{\left(2\right)}\left\{ \mathcal{M}\left(B_{il},B_{jk};v\right)-\mathcal{M}_{1}\left(B_{il};v\right)-\mathcal{M}_{2}\left(B_{jk};v\right)+\mu_{\mathcal{M}}\left(v\right)\right\} \right|=O_{p}\left(\left(Lh^{3}\right)^{-1}\right)\label{eq:M decomposition 2}
\end{equation}
follows from the maximal inequality \citet[Corollary 5.6]{Chen_Kato_U_Process}
and Markov's inequality.

By change of variables,
\begin{eqnarray*}
\mathcal{M}_{1}\left(b;v\right) & = & \frac{1}{h_{f}^{2}}K_{f}\left(\frac{\xi\left(b\right)-v}{h_{f}}\right)\xi'\left(b\right)\int_{\underline{b}}^{\overline{b}}\frac{1}{h_{r}}K_{r}\left(\frac{\xi\left(b\right)-\xi\left(z\right)}{h_{r}}\right)\frac{G\left(z\right)}{g\left(z\right)^{2}}\beta_{g}\left(z\right)\mathrm{d}z\\
 & = & \frac{1}{h_{f}^{2}}K_{f}\left(\frac{\xi\left(b\right)-v}{h_{f}}\right)\xi'\left(b\right)\int_{\frac{\underline{v}-\xi\left(b\right)}{h_{r}}}^{\frac{\overline{v}-\xi\left(b\right)}{h_{r}}}K_{r}\left(u\right)\frac{G\left(s\left(h_{r}u+\xi\left(b\right)\right)\right)\beta_{g}\left(s\left(h_{r}u+\xi\left(b\right)\right)\right)s'\left(h_{r}u+\xi\left(b\right)\right)}{g\left(s\left(h_{r}u+\xi\left(b\right)\right)\right)^{2}}\mathrm{d}u.
\end{eqnarray*}
By the arguments used in the proof of Lemma B.3, it can be verified
that $\left\{ \mathcal{M}_{1}\left(\cdot;v\right):v\in I\right\} $
is (uniformly) VC-type with respect to a constant envelope $F_{\mathcal{M}_{1}}$
that satisfies
\[
F_{\mathcal{M}_{1}}\apprle h_{f}^{-2}\underset{b\in\left[s\left(v_{l}-\delta_{0}\right),s\left(v_{u}+\delta_{0}\right)\right]}{\mathrm{sup}}\left|\beta_{g}\left(b\right)\right|,
\]
when $h$ is small enough. The maximal inequality \citet[Theorem 2.14.1]{van1996weak}
yields 
\[
\mathrm{E}\left[\underset{v\in I}{\mathrm{sup}}\left|\frac{1}{N\cdot L}\sum_{i,l}\mathcal{M}_{1}\left(B_{il};v\right)-\mu_{\mathcal{M}}\left(v\right)\right|\right]\apprle L^{-1/2}F_{\mathcal{M}_{1}}=O\left(L^{-1/2}h\right).
\]
The conclusion follows from this result, Markov's inequality, (\ref{eq:M Hoeffding decomposition}),
(\ref{eq:M decomposition 1}), (\ref{eq:M decomposition 2}) and Lemma
\ref{lem:Lemma 6}.\end{proof}
\end{document}